\documentclass[journal]{IEEEtran}

\usepackage{hyperref}
\usepackage{cite}

\ifCLASSINFOpdf
  \usepackage[pdftex]{graphicx}
   \DeclareGraphicsExtensions{.pdf,.jpeg,.png}
\else
\fi

\usepackage{amsmath, amssymb, bbm}
\usepackage{url}

\ifCLASSOPTIONcompsoc
  \usepackage[caption=false,font=normalsize,labelfont=sf,textfont=sf]{subfig}
\else
  \usepackage[caption=false,font=footnotesize]{subfig}
\fi

\newtheorem{theorem}{Theorem}
\newtheorem{lemma}{Lemma}[section]
\newtheorem{defi}{Definition}[section]

\newtheorem{remark}{Remark}[section]
\newtheorem{prop}{Proposition}[section]
\newtheorem{corr}{Corollary}[section]

\newcommand{\abs}[1]{\left \lvert #1\right \rvert}

\DeclareMathOperator*{\argmax}{arg\,max}
\usepackage{mathtools}

\newcommand{\be}{\begin{equation}}
\newcommand{\ee}{\end{equation}}
\newcommand{\ben}{\begin{equation*}}
\newcommand{\een}{\end{equation*}}
\newcommand{\mc}{\mathcal}

\newcommand{\stdnorm}{\mathcal{N}(0,1)}

\newcommand{\iid}{\text{i.i.d.}}
\newcommand{\expec}{\mathbb{E}}
\newcommand{\prob}{\mathbb{P}}

\newcommand{\snr}{\textsf{snr}}

\renewcommand\L{L}
\newcommand{\K}{K}
\newcommand{\M}{M}
\newcommand{\Lc}{\textsf{C}}
\newcommand{\Lr}{\textsf{R}}

\newcommand{\sfr}{\textsf{r}}
\newcommand{\sfc}{\textsf{c}}

\newcommand{\A}{\boldsymbol{A}}
\newcommand{\W}{\boldsymbol{W}}
\newcommand{\bbeta}{\boldsymbol{\beta}}
\newcommand{\bb}{\boldsymbol{b}}
\newcommand{\bc}{\boldsymbol{c}}
\newcommand{\bx}{\boldsymbol{x}}
\newcommand{\by}{\boldsymbol{y}}
\newcommand{\bw}{\boldsymbol{w}}

\newcommand{\e}{\epsilon}
\newcommand\p{c}
\newcommand\imag{\textrm{j}}

\begin{document}
%
\title{Modulated Sparse Superposition Codes\\ for the Complex AWGN Channel}
%
%
%

\author{Kuan~Hsieh~and~Ramji~Venkataramanan
\thanks{This paper was supported in part by an EPSRC Doctoral Training Partnership Award. This paper was presented in part at the 2020 IEEE International Symposium on Information Theory.}
\thanks{
K. Hsieh and R. Venkataramanan are with the Department of Engineering, University of Cambridge, Cambridge CB2 1PZ, U.K. (e-mail: kh525@cam.ac.uk; rv285@cam.ac.uk).}
}

\maketitle

\begin{abstract}
This paper studies a generalization of sparse superposition codes (SPARCs) for communication over the complex additive white Gaussian noise (AWGN) channel.
In a SPARC, the codebook is defined in terms of a design matrix, and each codeword is a generated by multiplying the design matrix with a sparse message vector. In the standard SPARC construction, information is encoded in the locations of the non-zero entries of the message vector.  
In this paper we generalize the construction and consider \emph{modulated} SPARCs, where information is encoded in both the locations and the values of the non-zero entries of the message vector. We focus on the case where the non-zero entries take values from a phase-shift keying (PSK) constellation. We propose a computationally efficient approximate message passing (AMP) decoder, and obtain analytical bounds on the state evolution parameters which predict the error performance of the decoder. Using these bounds we show that PSK-modulated SPARCs are asymptotically capacity achieving for the complex AWGN channel, with either spatial coupling or power allocation. We also provide numerical simulation results to demonstrate the error performance at finite code lengths. These results show that introducing modulation to the SPARC design can significantly reduce decoding complexity without sacrificing error performance.
\end{abstract}


%
\IEEEpeerreviewmaketitle

\section{Introduction}\label{sec:intro}
%
%
%
%
\IEEEPARstart{S}{parse} superposition codes, or sparse regression codes (SPARCs), were  introduced by Joseph and Barron\cite{joseph2012least, joseph2014fast} for communication over the real-valued additive white Gaussian noise (AWGN) channel.  For any fixed rate less than the channel capacity, these codes have been shown to have near-exponential decay of error probability with various efficient decoders \cite{joseph2014fast, cho2013approximate,rush2017capacity, venkataramanan19monograph}. In this paper we propose a variant of SPARCs, called \emph{modulated} sparse regression codes, for communication over the complex AWGN channel.

In the complex AWGN channel, the output symbol $y$ is produced from (complex) input symbol $x$ according to $y = x+ w$. The noise random variable  $w$  is circularly-symmetric complex Gaussian with zero mean and variance $\sigma^2$, denoted by $\mc{CN}(0,\sigma^2)$. There is an average power constraint  $P$ on the channel input: if a codeword $\boldsymbol{x} = x_1, x_2, \ldots, x_n$ is transmitted over $n$ uses of the channel, then
\be\label{eq:average_power_constraint}
\frac{1}{n}\sum_{i=1}^n |x_i|^2 \leq P +\varepsilon,
\ee
for a suitably small tolerance  $\varepsilon >0$.
Here $|\cdot|$ denotes the modulus of a complex number.
The Shannon capacity of the channel is  $\mc{C} = \ln \left(1+\frac{P}{\sigma^2}\right)$ nats.

\begin{figure}[t]
\centering
\includegraphics[width=\columnwidth]{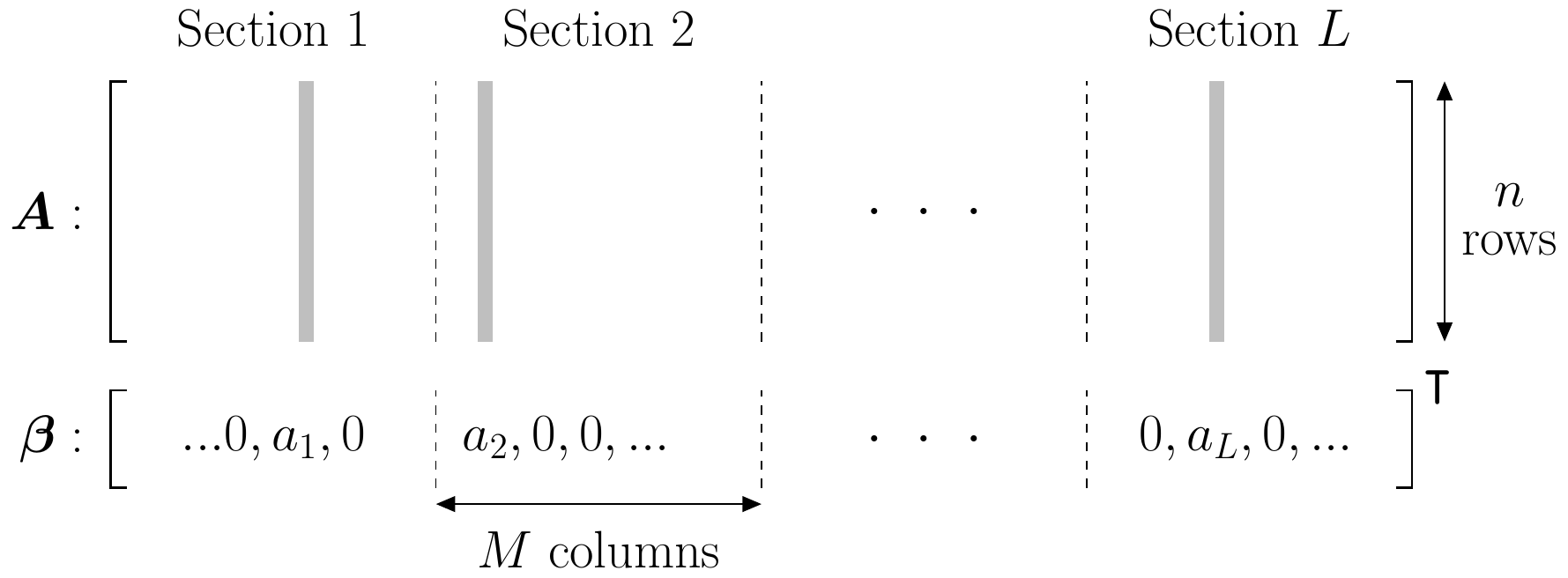}
\caption{\small $\A$ is an $n \times \L \M$ design matrix and $\bbeta$ is an $\L \M\times 1$ message vector with one non-zero entry in each of its $\L$ sections.  Codewords are of the form $\A\bbeta$. The non-zero values  $a_1, \ldots, a_\L$ are fixed a priori in standard SPARCs, but are chosen from a $\K$-ary constellation in modulated SPARCs, such as $\K$-ary PSK.}
\label{fig:sparc_code_construction}
\end{figure}

A SPARC is defined via a design matrix $\A$ of dimensions $n\times \L \M$, where $n$ is the code length and $\L$, $\M$ are integers such that $\A$ has $\L$ sections with $\M$ columns each. The construction is depicted in Fig. \ref{fig:sparc_code_construction}. Codewords are generated as linear combinations of $\L$ columns of $\A$, with one column from each section. Thus a codeword can be represented as $\A\bbeta$, with $\bbeta$ being an $\L \M\times 1$ \emph{message vector} with exactly one non-zero entry in each of its $\L$ sections.  In the standard SPARC construction, the message is indexed by the locations of the non-zero entries in $\bbeta$, with their values fixed a priori. Since each section encodes $\log\M$ bits (or $\ln \M$ nats) and there are $\L$ sections, the rate of the SPARC can be expressed as $R = \frac{\L\ln\M}{n}$ nats.

In a \emph{modulated} SPARC, information is encoded in both the locations and the values of the non-zeros entries of the message vector $\bbeta$. In particular, we allow each non-zero entry of
$\bbeta$  to take values in  a $\K$-ary constellation. Therefore, each section encodes 
$\log\K + \log \M$ bits (or $\ln(\K\M)$ nats). Since there are $\L$ sections, the rate of the modulated SPARC can be expressed as 
\be\label{eq:rate_eq}
R = \frac{\L\ln (\K\M)}{n} \ \text{nats}.
\ee
A SPARC without modulation can be seen as having modulation factor $\K=1$.

Modulation introduces an extra degree of freedom in the design of SPARCs which can be used to reduce decoding complexity without sacrificing finite-length error performance (see Section \ref{sec:numerical_results}). Adding modulation to the SPARC design may also be useful in new multiuser settings such as unsourced random access \cite{polyanskiy2017perspective,fengler2019sparcs,fengler2020unsourced,amalladinne2020approximate} and many-access channels \cite{chen2017capacity,zadik2019improved,ravi2019capacity}. 

Since we consider communication over the complex AWGN channel, the SPARC codeword $\A\bbeta$ can be complex-valued. Accordingly, the design matrix $\A$ is chosen to have independent zero-mean complex Gaussian entries. For a modulated complex SPARC, the non-zero entries of $\bbeta$ take values in a $\K$-ary complex constellation. The power constraint $P$ on the channel input implies that the variances of the entries of $\A$ and the non-zero values of $\bbeta$ should be chosen such that $\frac{1}{n} \|\A\bbeta\|^2 \leq P + \varepsilon$
with high probability, for a suitably small $\varepsilon >0$.
For example, if the values of the non-zero entries of $\bbeta$ are all chosen to be 1 (the unmodulated case) and the entries of $\A$ are chosen i.i.d. $\sim \mc{CN}(0, P/\L)$,  then
$\frac{1}{n} \mathbb{E} [\| \A \bbeta\|^2] = P$, and  the power constraint is satisfied with high probability for large $n$. 
With optimal decoding, such a design has error probability decaying exponentially in the code length for rates $R < \mc{C}$ \cite{joseph2012least}.\footnote{The result in \cite{joseph2012least} was proved for a real-valued SPARC over a real AWGN channel, but the result can be extended to the complex case by similar arguments.} However, different designs of the matrix $\A$ are required for good error performance with computationally efficient decoders, such as designs that use \textit{power allocation} \cite{joseph2014fast, cho2013approximate, rush2017capacity} or \textit{spatial coupling} \cite{barbier2015approximate, barbier2017approximate, barbier2019universal, rush2020capacity}.  

 Both power allocation and spatial coupling facilitate iterative decoding by ensuring that obtaining good estimates of some sections of the message vector makes other sections easier to decode. 
In particular, using  Approximate Message Passing (AMP) decoding, both power allocated and spatially coupled unmodulated SPARCs  have been to proven to have vanishing error probability in the large system limit, for any $R < \mc{C}$  \cite{rush2017capacity,rush2019theerror, barbier2016proof, rush2020capacity}. (Here ``large system limit'' refers to $(\L,\M,n)$ all tending to infinity such that $n R = \L \ln \M$.)  Both power allocated and spatially coupled SPARCs can be studied under a single framework, as discussed in Section \ref{sec:pa_sc}.

\subsection{Structure of the paper and main contributions}

In Section \ref{sec:complex_mod_SPARCs}, we introduce modulated SPARCs with $\K$-ary phase shift keying (PSK) constellations  and justify the choice of PSK. We describe how  the design matrix can be constructed from a base matrix. Power allocation and spatial coupling can be implemented by appropriate choices of the base matrix.

In Section \ref{sec:amp_se},  we propose an AMP decoder for  modulated  complex SPARCs and describe its state evolution recursion. The state evolution recursion predicts the mean-squared error between the true message vector and its estimate in each iteration of AMP decoding. Since SPARCs have been previously analyzed only for real-valued channels (even in the unmodulated case),  we  briefly discuss the differences between unmodulated real and complex SPARCs, comparing their AMP decoders and state evolution recursions.
 
In Section \ref{sec:err_perf_analysis}, we analyze the error performance of the AMP decoder for modulated complex  SPARCs using state evolution. The main technical result (Proposition \ref{prop:se_psi_bound}) gives upper and lower bounds on a key state evolution parameter which predicts the mean-squared error of the AMP estimate in each iteration. Using this bound, we show that in the large system limit, the state evolution recursion for complex SPARCs with $K$-ary PSK modulation  is  the \emph{same} for any fixed value of $K$, including $K=1$ (unmodulated).  
We use this result to prove that  $K$-PSK modulated SPARCs  with AMP decoding are capacity achieving for the complex AWGN channel, with either spatial coupling or an appropriately chosen power allocation (Theorems \ref{thm:sc_mod_sparcs} and \ref{thm:pa_mod_sparcs}). To show that $K$-PSK modulated SPARCs with a specified design are capacity achieving in the large system limit, two steps are required:
\begin{enumerate}
\item Prove that  for any rate $R < \mc{C}$, state evolution predicts vanishing probability of decoder error in the large system limit.
\item Prove that  the error rate of the AMP decoder is accurately tracked by the state evolution parameters for sufficiently large code length.
\end{enumerate}

In Section \ref{sec:err_perf_analysis}, we carry out Step 1. Step 2 was proved in \cite[Theorems 1 and 2]{rush2020capacity} for the case of unmodulated real-valued SPARCs (including spatially coupled and power allocated ones), where it was shown that the normalized mean-squared error of the AMP decoder concentrates on the state evolution prediction. Step 2 for modulated complex-valued SPARCs can be proved along the same lines. We do not detail the proof in this paper as it is a straightforward extension of the analysis in \cite{rush2020capacity}: the proof uses the same induction argument and sequence of steps as \cite{rush2020capacity}, with modifications to account for $\A$ and $\bbeta$ being complex valued. The key conceptual difference between PSK-modulated complex SPARCs and unmodulated real-valued SPARCs is in the state evolution and its analysis, which is discussed in Sections \ref{sec:amp_se} and \ref{sec:err_perf_analysis}.

In Section \ref{sec:numerical_results}, we evaluate the finite length error performance of  modulated complex SPARCs with AMP decoding via numerical simulations, and compare their performance with coded modulation schemes using LDPC codes from the DVB-S2 standard \cite{dvb-s2}. With a  DFT-based implementation, the per-iteration complexity of the AMP decoder  is $\mc{O}(\L\M (\K  + \log (\L\M)))$. For $\K \ll \log (\L\M)$, our numerical results  demonstrate that modulation allows one to significantly reduce the decoder complexity without sacrificing error performance.

SPARCs for the real AWGN channel with binary PSK (BPSK) modulation ($K=2$) and power allocation were discussed in Greig's PhD thesis \cite[Chapter~5]{greig2017thesis}. That work motivated the current study of  PSK-modulated SPARCs for the complex AWGN channel. Power allocated and spatially coupled designs are treated in a unified way using the framework of base matrices which we describe in Section \ref{sec:complex_mod_SPARCs}.

\emph{Related work}: Compressed coding is a technique recently proposed in \cite{liang2016ISTC,liang2020cc} for efficient communication over AWGN channels. Here a coded modulation scheme (e.g., a binary code + BPSK) is first used to produce a code sequence $\bc$, which is then used to generate the channel input sequence $\bx = \A \bc$ via a Gaussian matrix $\A$. An AMP algorithm is used to recover the code sequence $\bc$ from the noisy channel output $\by = \A\bc + \bw$. It is shown in \cite{liang2020cc} that the rate of this scheme can asymptotically approach the channel capacity provided a certain curve matching condition is satisfied. 

Though the  decoders for compressed coding and SPARC are both based on estimating a vector from a noisy linear observation,  the SPARC decoder is tailored to the sparse, section-wise structure of the message vector $\bbeta$. Moreover, the SPARC construction does not use an outer code. In compressed coding the coded sequence $\bc$  is not sparse as each of its entries is drawn from a constellation. Consequently, the  underlying code design and the analysis of the compressed coding scheme (based on curve matching principles) is very different from that of both unmodulated and modulated SPARCs.

\subsection{Notation}
We use $\log$ and $\ln$ to denote the base 2 logarithm and natural logarithm, respectively. 
The real-valued Gaussian distribution with mean $\mu$ and variance $\sigma^2$ is denoted by $\mc{N}(\mu, \sigma^2)$.
The indicator function of an event $\mc{A}$ is denoted by $\mathbbm{1}\{\mc{A}\}$.
For a positive integer $m$, we use $[m]$ to denote the set  $\{1, \dots, m \}$.
Throughout the paper, we use bold lower case letters or Greek symbols for vectors, and bold upper case for matrices. We use plain font for scalars, and subscripts denote entries of a vector or matrix. For example, $\boldsymbol{x}$ denotes a vector, with $x_i$ being the $i^{th}$ element of $\boldsymbol{x}$. Similarly, $\boldsymbol{X}$ is a matrix, and its $(i,j)^{th}$ entry is denoted by $X_{ij}$. 
The real and imaginary parts of a complex number $x$ are denoted by $\Re(x)$ and $\Im(x)$ whilst its complex conjugate is denoted by $\overline{x}$. The conjugate transpose of a matrix $\boldsymbol{X}$ is denoted by $\boldsymbol{X}^*$. 
The squared $\ell^2$-norm of a complex vector $\boldsymbol{x}$ is denoted by $\|\boldsymbol{x}\|^2 := \sum_i |x_i|^2$.

\section{Modulated SPARCs}\label{sec:complex_mod_SPARCs}

As described in Section \ref{sec:intro} and illustrated in Fig. \ref{fig:sparc_code_construction}, a SPARC is defined by a design matrix $\A$ of dimension $n \times \L\M$ with independent Gaussian entries. The codeword is $\boldsymbol{x} = \A\bbeta$, where the message vector $\bbeta$ has exactly one non-zero entry in each of its $\L$ sections.

\subsection{Modulation}\label{sec:modulation}

In a modulated SPARC, the value of the non-zero entry in each section of $\bbeta$ is chosen from a $\K$-ary constellation, and the information is encoded in both the location and the value of the non-zero entry. Since each section has $\M$ entries and there are $\K$ possible values for the non-zero entry, each section encodes $\log \K + \log \M$ bits.

\begin{figure}[!t]
\centering
\includegraphics[width=\columnwidth]{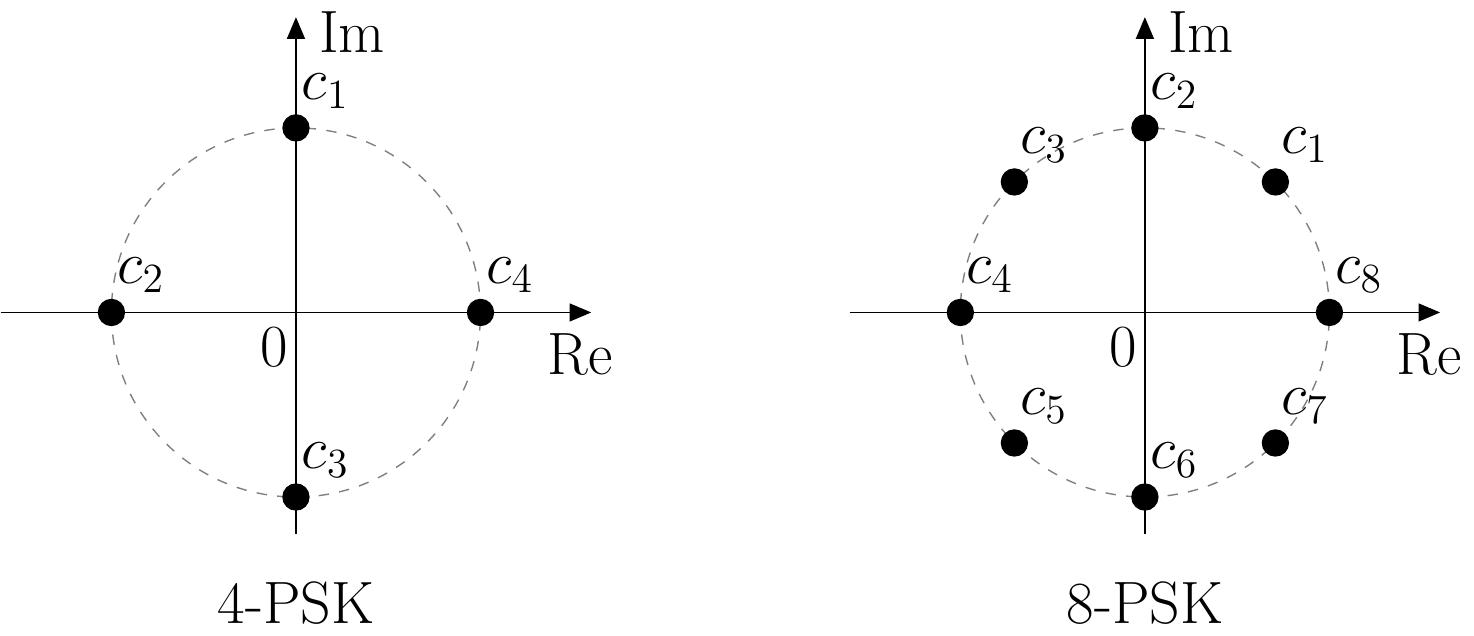}
\caption{\small Labeling of PSK constellation symbols for $\K=4$ and $\K=8$.}
\label{fig:4PSK}
\end{figure}

In this paper, we consider the phase shift keying (PSK) constellation, where the $\K$ constellation symbols are equally spaced on the unit circle of the complex plane. The constellation symbols are denoted by $\{\p_k\}_{k=1,\ldots,\K}$, where $\p_k = e^{\imag 2\pi k/\K}$ and $\imag \coloneqq \sqrt{-1}$. (Fig. \ref{fig:4PSK} illustrates the labeling of the constellation points.)
PSK is chosen rather than other modulation techniques such as quadrature amplitude modulation (QAM) because PSK has constellation symbols of equal magnitude. Unequal symbol magnitudes will counteract the effect of power allocation and spatial coupling, and simulations show that errors are much more likely to occur in the sections where symbols of smaller magnitude are chosen.

Note that when $\K=2$, the PSK constellation symbols are $+1$ and $-1$, which can be used to modulate real-valued SPARCs. For $K >2$, the information bits can be mapped to constellation symbols via Gray coding to minimise the bit error in decoding.

\subsection{Power allocation and spatial coupling}\label{sec:pa_sc}

The main idea behind both power allocation and spatial coupling is to change the variances of the (independent Gaussian) entries of $\A$ so that certain parts of the message vector are easier to decode than others. Power allocation varies the variance across the sections (columns) of $\A$, while spatial coupling varies the variance across both rows and columns such that $\A$ has a band diagonal structure. Both techniques can be described in terms of \emph{base matrices}, as described below.\footnote{In standard unmodulated SPARCs, power allocation is often described as the choice of the non-zero values of the message vector $\bbeta$ across all $\L$ sections while the design matrix $\A$ has i.i.d.\ Gaussian entries. This is equivalent to a choice of the variances of the Gaussian entries  across the $\L$ sections of the design matrix while the non-zero values of $\bbeta$ are all equal to 1. The latter perspective allows a unified treatment of power allocation and spatial coupling.}

\begin{figure}[!t]
\centering
\includegraphics[width=\columnwidth]{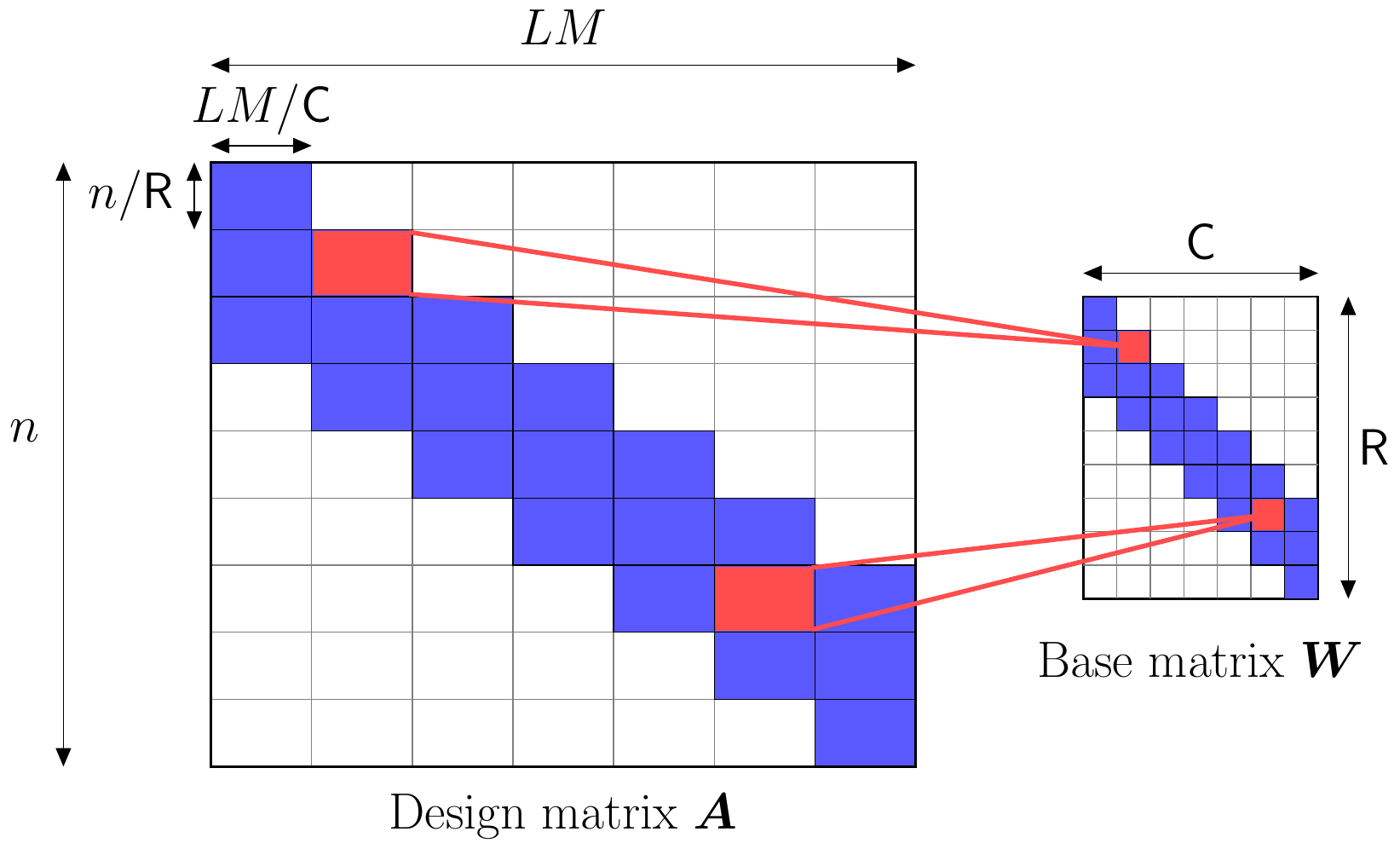}
\caption{\small An example of a spatially coupled design matrix $\A$ and corresponding base matrix $\W$. Each square in $\W$ represents a scalar entry which  specifies the variance of the entries in a block of $\A$. The white parts of $\A$ and $\W$ correspond to zeros. The design matrix $\A$ has independent complex Gaussian entries $A_{ij} \sim \mc{CN}(0, W_{\sfr(i)\sfc(j)}/\L)$.}
\label{fig:sparc_scmatrix}
\end{figure}

The design matrix $\A\in\mathbb{C}^{n\times\L\M}$ is divided into $\Lr$-by-$\Lc$ equally size \emph{blocks}. The entries within each block are i.i.d.\ complex Gaussian with zero mean and variance specified by the corresponding entry of a base matrix $\W \in \mathbb{R}^{\Lr \times \Lc}$.
The $n \times \L\M$ design matrix $\A$ is constructed by replacing each entry of the base matrix $W_{\sfr \sfc}$, for $\sfr \in[\Lr]$, $\sfc \in [\Lc]$, by an $\frac{n}{\Lr} \times \frac{\L\M}{\Lc}$ matrix with entries drawn i.i.d.\ from $\mc{CN}(0, W_{\sfr \sfc}/{L})$. See Fig. \ref{fig:sparc_scmatrix} for an example. Hence, given base matrix $\W$, the design matrix $\A$ has independent complex Gaussian entries
\be\label{eq:construct_Aij_complex}
A_{ij} \sim \mc{CN}\left(0,\frac{1}{L} W_{\sfr(i) \sfc(j)} \right),  \quad \text{for }  i \in [n], \ j\in[\L \M].
\ee
The operators $\sfr(\cdot):[n]\rightarrow[\Lr]$ and $\sfc(\cdot):[\M L]\rightarrow[\Lc]$ in \eqref{eq:construct_Aij_complex} map a particular row or column index to its corresponding \emph{row block} or \emph{column block} index. We require  $\Lc$ to divide $\L$, resulting in $\frac{\L}{\Lc}$ sections per column block.

The non-zero entries of message vector $\bbeta$ all have magnitude equal to 1 due to PSK modulation. In order to satisfy the  power constraint \eqref{eq:average_power_constraint},  the entries of the base matrix $\W$ must satisfy
\be
\label{eq:W_power_constraint}
\frac{1}{\Lr \,\Lc}\sum_{\sfr=1}^{\Lr}  \sum_{\sfc=1}^{\Lc} W_{\sfr \sfc} = P.
\ee
This ensures that for each message vector $\bbeta$, we have $ \frac{1}{n}\expec[ \| \A \bbeta  \|^2]=P$.

The trivial base matrix with $\Lr=\Lc=1$ corresponds to a design matrix with entries drawn i.i.d.\ from $\mc{CN}(0, P/L)$.
A single-row base matrix with $\Lr=1$ and $\Lc=L$ corresponds to a design matrix with power allocation. For example, the exponential power allocation used in \cite{joseph2014fast, cho2013approximate, rush2017capacity} has $W_{1\ell} \propto e^{-\mc{C}\ell/\L}$ for $\ell \in [\L]$.
In this paper we will consider two kinds of base matrices: i) the one corresponding to an exponentially decaying power allocation, and ii)  a class of base matrices called $(\omega, \Lambda, \rho)$ base matrices \cite{hsieh2018spatially, rush2020capacity}. Other spatial coupling designs for SPARCs can be found in \cite{barbier2017approximate}.

\begin{defi}
\label{def:ome_lamb}
An $(\omega , \Lambda, \rho)$ base matrix $\W$ for spatially coupled SPARCs is described by three parameters: coupling width $\omega\geq1$, coupling length $\Lambda\geq 2\omega-1$, and $\rho \in [0,1)$ which specifies the fraction of power allocated to the uncoupled entries in each column. The matrix has $\Lr=\Lambda+\omega-1$ rows,  $\Lc=\Lambda$ columns, with each column having $\omega$ identical non-zero entries. For an average power constraint $P$, the  $(\sfr,\sfc)$th entry of the base matrix, for  $\sfr \in [\Lr], \sfc\in[\Lc]$, is given by
\begin{equation}\label{eq:W_rc}
W_{\sfr \sfc} =
\begin{cases}
 	\ (1-\rho)P \cdot \frac{\Lambda+\omega-1}{\omega} \quad &\text{if} \ \sfc \leq \sfr \leq \sfc+\omega-1,\\
	\ \rho P \cdot \frac{\Lambda + \omega -1}{\Lambda-1} \quad &\text{otherwise}.
\end{cases}
\end{equation}
\end{defi}

For example, the base matrix in Fig. \ref{fig:sparc_scmatrix} has parameters $\omega=3, \Lambda=7, \rho=0$. For our simulations in Section \ref{sec:numerical_results}, we use $\rho= 0$, whereas for Theorem \ref{thm:sc_mod_sparcs} we choose $\rho$ to be a small positive value proportional to the rate gap from capacity. When $\rho=0$,  base matrix $\W$ has non-zero entries only in the band-diagonal region. 
Each non-zero $\frac{n}{\Lr} \times \frac{\L\M}{\Lc}$ block of the design matrix $\A$ can be viewed as a standard (non spatially coupled) design matrix with $\frac{\L}{\Lc}$ sections, code length $\frac{n}{\Lr}$, and rate $R_\text{inner} = \frac{(\L/\Lc) \ln (\K\M)}{n/\Lr}$ nats. Using \eqref{eq:rate_eq}, the overall rate of the spatially coupled SPARC is related to $R_\text{inner}$ according to
\be\label{eq:R_Rsparc}
R = \frac{\Lc}{\Lr} \,  R_\text{inner} = \frac{\Lambda}{\Lambda + \omega -1} \, R_\text{inner}.
\ee
The coupling width $\omega$ is an integer greater than 1, and the difference $R_\text{inner} -R = \frac{\omega-1}{\Lambda} R$ is often referred to as a rate loss. The rate loss  becomes negligible when $\Lambda$ is much larger than $\omega$.

We mention that the idea of constructing a power allocated or spatially coupled design matrix from a base matrix is inspired by the construction of a irregular or spatially coupled low-density parity-check (LDPC) code from a protograph \cite{luby2001improved, thorpe03, divsalar2009capacity, mitchell2015spatially}.

In the remainder of the paper, we use subscripts in sans-serif font ($\sfr$ or $\sfc$) to denote row or column block indices. For example, $\bbeta_\sfc \in \mathbb{C}^{\frac{\L\M}{\Lc}}$ denotes the $\sfc$-th column block of $\bbeta \in \mathbb{C}^{\L\M}$ for $\sfc \in [\Lc]$, and $\boldsymbol{y}_\sfr \in \mathbb{C}^{\frac{n}{\Lr}}$ denotes the $\sfr$-th row block of $\boldsymbol{y} \in \mathbb{C}^{n}$ for $\sfr \in [\Lr]$.


\section{AMP decoding and state evolution}\label{sec:amp_se}
The decoder aims to recover the message vector $\bbeta$ from the channel output sequence $\boldsymbol{y} \in\mathbb{C}^n$, given by
\begin{equation}\label{eq:linear_model}
\boldsymbol{y} = \A\bbeta + \boldsymbol{w},
\end{equation}
where $w_1,\ldots, w_n \stackrel{\text{i.i.d.}}{\sim} \mc{CN}(0, \sigma^2)$. The design matrix $\A$ and base matrix $\W$ are available to the decoder.

\subsection{AMP decoder}\label{sec:amp}

We propose a decoding algorithm based on approximate message passing (AMP), a technique that has previously been used for decoding real-valued SPARCs without modulation \cite{rush2017capacity, rush2019theerror, barbier2016proof, barbier2017approximate, rush2020capacity}.
AMP  refers to a class of iterative algorithms that are approximations of loopy belief propagation for certain high-dimensional estimation problems where the underlying factor graph is dense, e.g., estimation in random linear models \cite{donoho2009message,krzakala2012statistical, bayati2011thedynamics}, generalized linear models \cite{ rangan2010generalized}, and low-rank matrix estimation \cite{FletcherRanganIMA18,deshpande2014information}. 
Complex-valued versions of AMP have been proposed for the linear model in \eqref{eq:linear_model} where the matrix $\A$, as well as $\bbeta, \boldsymbol{w}$ can be complex \cite{maleki2013asymptotic, anitori2013design,jeon2015optimality}.
However, the complex AMP cannot be directly used to decode SPARCs because it does not incorporate spatial coupling and is based on an i.i.d.\ prior for $\bbeta$; in a SPARC the message vector $\bbeta$ is only section-wise i.i.d.
Hence, we start from the generalized approximate message passing algorithm (GAMP) \cite[Alg.~1]{rangan2010generalized} and derive the following AMP decoder for complex SPARCs with PSK modulation. 

Given the channel output sequence $\boldsymbol{y}$, the AMP decoder generates successive estimates of the message vector, denoted by $\bbeta^t \in \mathbb{C}^{LM}$, for $t=0,1,\ldots$. It initialises $\bbeta^0$ to the all-zero vector, and for $t \geq 0$, iteratively computes
\begin{align}\label{eq:amp_decoder}
\begin{split}
\boldsymbol{z}^t &= \boldsymbol{y} - \A\bbeta^t + \widetilde{\boldsymbol{\upsilon}}^t \odot \boldsymbol{z}^{t-1},\\
\boldsymbol{\beta}^{t+1} &= \eta \big(\bbeta^t + ( \widetilde{\boldsymbol{Q}}^t \odot \A)^*  \boldsymbol{z}^t, \; \widetilde{\boldsymbol{\tau}}^t \big).
\end{split}
\end{align}
Here $\odot$ denotes the Hadamard (entry-wise) product,  
the conjugate transpose of $\A$ is denoted by $\A^*$,
and quantities with negative time indices are set to zero.
The vectors $\widetilde{\boldsymbol{\upsilon}}^t \in \mathbb{R}^n$, $\widetilde{\boldsymbol{\tau}}^t \in \mathbb{R}^{\L\M}$ and the matrix $\widetilde{\boldsymbol{Q}}^t \in \mathbb{R}^{n\times \L\M}$ are deterministic quantities defined in terms of state evolution parameters (given in the next subsection, Section \ref{sec:se}).
The function $\eta = (\eta_1, \ldots, \eta_{\L\M}): \mathbb{C}^{\L\M} \times \mathbb{R}^{\L\M} \rightarrow  \mathbb{C}^{\L\M}$ is defined as follows. 
For index $j$ in section $\ell \in [\L]$, 
denoted by $j\in\text{sec}(\ell)\coloneqq\{(\ell-1)\M+1,\ldots,\ell \M\}$,
\be\label{eq:eta_function}
\eta_j(\boldsymbol{s}, \widetilde{\boldsymbol{\tau}})
= \frac{\sum_{k=1}^{\K} \p_k \cdot e^{\Re(\overline{s_j} \p_k)/\widetilde{\tau}_j}}{\sum_{j'\in \text{sec}(\ell)} \sum_{k'=1}^{\K} e^{\Re(\overline{s_{j'}} \p_{k'})/\widetilde{\tau}_{j'}}},
\ee
where 
$\Re(z)$ and $\overline{z}$ denote the real part and complex conjugate of a complex number $z$, respectively,
and $\{\p_k = e^{\imag 2\pi k/\K} \}_{k\in[\K]}$ is the set of PSK symbols. 
Notice that $\eta_j(\boldsymbol{s}, \widetilde{\boldsymbol{\tau}})$ depends on all the entries of $\boldsymbol{s}$ and $\widetilde{\boldsymbol{\tau}}$ in the section containing $j$.

When the change in $\bbeta^t$ across successive iterations falls below a pre-specified tolerance, or the decoder reaches the maximum number of iterations allowed, the decoder terminates. Let $T$ denote the final AMP iteration (in which  $\boldsymbol{z}^{T-1}$ and  $\bbeta^T$ are computed). After iteration $T$, the decoder produces a hard-decision estimate of the message vector, denoted by $\widehat{\bbeta}^T$, as follows.  Using $\boldsymbol{s}^{T-1} = \bbeta^{T-1} + ( \widetilde{\boldsymbol{Q}}^{T-1} \odot \A)^*  \boldsymbol{z}^{T-1}$, the $\ell$-th section of $\widehat{\bbeta}^T$ is computed as 
\be\label{eq:amp_hard_decision}
\widehat{\bbeta}_{\text{sec}(\ell)}^T
= \argmax_{\boldsymbol{b} \in \mc{B}_{\M,\K}} \; \Re\Big[ \big(\boldsymbol{s}^{T-1}_{\text{sec}(\ell)} \big)^* \boldsymbol{b}  \Big], \quad \text{for } \ell \in [\L],
\ee
where $\mc{B}_{\M,\K}$ denotes the set of all possible length $\M$ vectors with a single non-zero entry whose value belongs to the PSK constellation $\{\p_k\}_{k\in\{\K\}}$.

\textit{Interpretation of the AMP decoder:}
Consider the function $\eta(\cdot, \widetilde{\boldsymbol{\tau}}^t )$ in \eqref{eq:amp_decoder}, which produces the updated  message vector estimate $\bbeta^{t+1}$.  The first input to this function, denoted by $\boldsymbol{s}^t$, can be viewed as a noisy version of the true message vector $\bbeta$. In particular,  $\boldsymbol{s}^t$ is approximately distributed as $\bbeta + \sqrt{\widetilde{\boldsymbol{\tau}}^t} \odot \boldsymbol{u}$, with $\boldsymbol{u} = u_1,\ldots,u_{\L\M} \stackrel{\text{i.i.d.}}{\sim} \mc{CN}(0,2)$ independent of $\bbeta$. Under the above distributional assumption, the function $\eta_j$ in \eqref{eq:eta_function} is the minimum mean squared error (MMSE) estimator of $\beta_{j}$. That is, for $j \in [\L\M]$,
\be
\eta_j(\boldsymbol{s}, \widetilde{\boldsymbol{\tau}})=\mathbb{E}\Big[\beta_j \, \big| \,\boldsymbol{s} = \bbeta + \sqrt{\widetilde{\boldsymbol{\tau}}} \odot \boldsymbol{u} \Big],
\label{eq:eta_function_interpret}
\ee
where the expectation is calculated over $\bbeta$ and $\boldsymbol{u}$, with the location and value of the non-zero entry in each section of $\bbeta$ being uniformly distributed among the possible choices.
Under the same distributional assumption, the final hard-decision step of the algorithm \eqref{eq:amp_hard_decision} is the maximum a posteriori (MAP) estimator for $\bbeta_{\text{sec}(\ell)}$, i.e.,
\be
\widehat{\bbeta}^T
= \argmax_{\boldsymbol{b}} \; \mathbb{P} \Big(\boldsymbol{b} \, \big| \, \boldsymbol{s}^{T-1} = \bb + \sqrt{\widetilde{\boldsymbol{\tau}}^{T-1}} \odot \boldsymbol{u} \Big),
\label{eq:hard_dec_est}
\ee
where the $\argmax$ is over the set of all valid message vectors, i.e., over all $\bb \in \mathbb{C}^{\L\M}$ that have exactly one $K$-PSK entry in each of the $L$ sections.

The vector $\boldsymbol{z}^t$ in \eqref{eq:amp_decoder} consists of a residual term $\boldsymbol{y} - \A\bbeta^t$ and an `Onsager' term $\widetilde{\boldsymbol{\upsilon}}^t \odot \boldsymbol{z}^{t-1}$ which arises naturally in the derivation of the AMP algorithm.
The Onsager term is crucial for the distributional property of $\boldsymbol{s}^t$ mentioned above, and also ensures the entries of $\boldsymbol{z}^t$ are approximately Gaussian and independent. For intuition about the role of the Onsager term, see \cite[Sec.~I-C]{bayati2011thedynamics} and \cite[Sec.~VI]{donoho2013information}.

\subsection{State evolution}\label{sec:se}
Under the distributional property described above, the normalized mean square error (NMSE) between the true message vector and its AMP estimate at each iteration can be predicted using a deterministic recursion called state evolution.
For a modulated complex SPARC defined by base matrix $\W\in\mathbb{R}^{\Lr\times\Lc}$, and a complex AWGN channel with noise variance $\sigma^2$, state evolution (SE) iteratively defines vectors $\boldsymbol{\gamma}^t, \boldsymbol{\phi}^t\in\mathbb{R}^{\Lr}$ and $\boldsymbol{\tau}^t, \boldsymbol{\psi}^t\in\mathbb{R}^{\Lc}$ as follows. Initialize $\psi_\sfc^0 = 1$ for $\sfc \in [\Lc]$, and for $t=0,1,\ldots$, compute
\begin{align}
	\gamma_\sfr^t &= \frac{1}{\Lc}\sum_{\sfc=1}^{\Lc}W_{\sfr \sfc} \psi_\sfc^t, 
	\ \quad \
	\phi_\sfr^t = \sigma^2 + \gamma_\sfr^t,
	\ \quad \text{ for } \sfr\in[\Lr], \label{eq:se_phi}\\[0.3em]
	\tau_\sfc^t &= \frac{R/2}{\ln(\K\M)} \bigg[\frac{1}{\Lr}\sum_{\sfr=1}^{\Lr}\frac{W_{\sfr \sfc}}{\phi_\sfr^t}\bigg]^{-1},
	\, \quad
	\psi_\sfc^{t+1} = 1 - \mc{E}(\tau_\sfc^t),
	\nonumber\\
	&\hspace{16.4em} \text{for } \sfc\in [\Lc], \label{eq:se_psi}
\end{align}
where $\mathcal{E}(\tau)$ is defined as follows for $\tau > 0$:
\begin{align}\label{eq:se_E}
	\mathcal{E}(\tau) 
	&= \mathbb{E}\left[\frac{\sum_{k=1}^{\K} \Re[\p_k]\cdot e^{\frac{1}{\tau}\Re\left[(1+\sqrt{\tau}U_1)\overline{\p_k}\right]}}{\sum_{a=1}^{\K} e^{\frac{1}{\tau}\Re\left[(1+\sqrt{\tau}U_1) \overline{\p_a}\right]} + \sum_{j=2}^{\M}\sum_{b=1}^{\K} e^{\frac{\Re[U_j \overline{\p_b}]}{\sqrt{\tau}} }}\right],
\end{align}
with $U_1, \ldots,U_\M \stackrel{\text{i.i.d.}}{\sim} \mathcal{CN}(0,2)$.

\begin{figure}[t]
\centering
\includegraphics[width=\columnwidth]{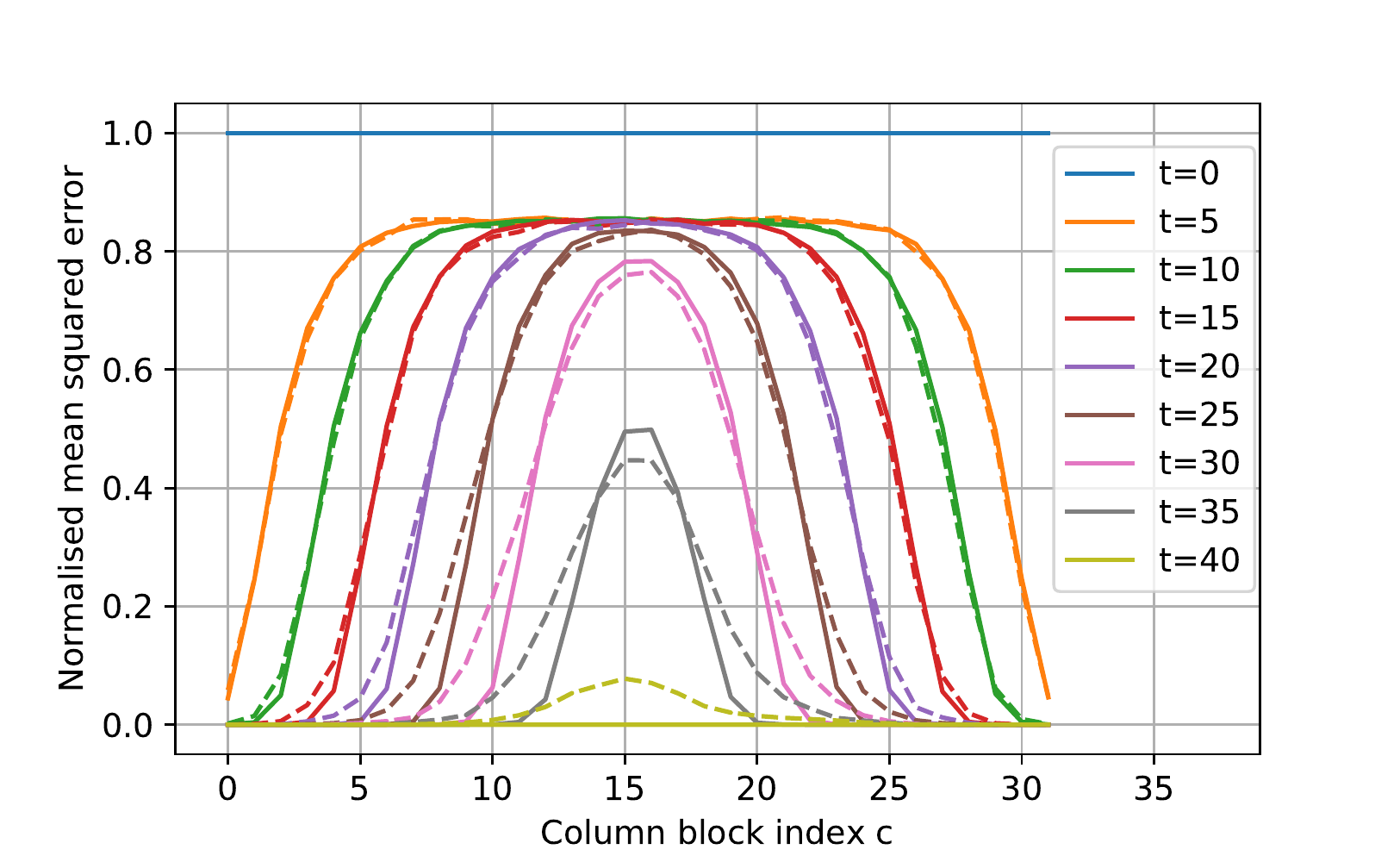}
\caption{ \small NMSE $\frac{\|\bbeta_{\sfc}^t-\bbeta_{\sfc}\|^2}{\L/\Lc}$ vs. column block index $\sfc \in [\Lc]$ for several iteration numbers. The complex SPARC is defined via an $(\omega=6 ,\Lambda=32, \rho=0)$ base matrix and has parameters $R=3.1$ bits, $\mc{C}=4$ bits, $M=256$, $L=2048$ and $n=5291$. The solid lines are the SE predictions from \eqref{eq:se_psi}, and the dashed lines are the average NMSE over 100 instances of AMP decoding.}
\label{fig:ser_waveprop}
\vspace{-5pt}
\end{figure}

As illustrated in Fig. \ref{fig:ser_waveprop}, the state evolution parameters $\{\psi_\sfc^t\}_{\sfc\in[\Lc]}$ closely track the NMSE of each block of the message vector, i.e.,  $\psi^t_\sfc \approx \frac{\|\bbeta_{\sfc}^t-\bbeta_{\sfc}\|^2}{\L/\Lc}$ for $\sfc\in[\Lc]$.
Similarly, the parameters $\{\phi_\sfr^t\}_{\sfr\in[\Lr]}$ closely track the block-wise variance of the modified residual vector $\boldsymbol{z}^t$, i.e., $\phi_\sfr^t \approx \frac{\|\boldsymbol{z}_\sfr^t\|^2}{n/\Lr}$ for $\sfr\in[\Lr]$.

For $t\geq 0$, the vectors $\widetilde{\boldsymbol{\upsilon}}^t \in \mathbb{R}^n$ and $\widetilde{\boldsymbol{\tau}}^t \in \mathbb{R}^{\L\M}$ in the AMP decoder \eqref{eq:amp_decoder} both have a block-wise structure (as indicated by the tilde), with their entries defined as follows. For $i\in[n]$ and $j\in[\L\M]$,
\begin{align}
	\widetilde{\upsilon}_i^t = \frac{\gamma_{\sfr(i)}^t}{\phi_{\sfr(i)}^{t-1}},  \qquad
	\widetilde{\tau}_j^t &= \tau_{\sfc(j)}^t, 
	\label{eq:se_tilde_tau}
\end{align}
where we recall that $\sfr(i)$ and $\sfc(j)$ denote the row and column block index of the $i$-th row entry and $j$-th column entry, respectively. The vector $\widetilde{\boldsymbol{\upsilon}}^0$ is defined to be all-zeros.
Similarly, the matrix $\widetilde{\boldsymbol{Q}}^t \in \mathbb{R}^{n\times \L\M}$ in \eqref{eq:amp_decoder} has a block-wise structure, with entries defined as follows. For $i\in[n]$ and $j\in[\L\M]$,
\be
	\widetilde{Q}_{ij}^t = \frac{\tau_{\sfc(j)}^t}{\phi_{\sfr(i)}^t}. \label{eq:se_tilde_Q}
\ee


\textit{Online estimation of SE parameters:} In the AMP decoder \eqref{eq:amp_decoder}, instead of computing the SE parameters \eqref{eq:se_phi}--\eqref{eq:se_tilde_Q} offline they can be estimated online (at runtime) using intermediate outputs from the AMP. In particular, the SE parameters $\{\gamma_\sfr^t\}_{\sfr\in[\Lr]}, \{\phi_\sfr^t\}_{\sfr\in[\Lr]}$ and $\{\tau_\sfc^t\}_{\sfc\in[\Lc]}$, which are needed to compute $\widetilde{\boldsymbol{\upsilon}}^t$, $\widetilde{\boldsymbol{\tau}}^t$ and $\widetilde{\boldsymbol{Q}}^t$, can be estimated online as follows. For $\sfr \in[\Lr]$ and $\sfc \in[\Lc]$,
\begin{align}
	\widehat{\gamma}_\sfr^t &= \frac{1}{\Lc} \sum_{\sfc=1}^{\Lc} W_{\sfr\sfc} \bigg(1 - \frac{\|\bbeta^t_{\sfc}\|^2}{\L/\Lc}\bigg), \label{eq:amp_decoder_gamma} \\
	\widehat{\phi}_\sfr^t \, &= \,
	\begin{cases}
		\sigma^2 + \widehat{\gamma}_\sfr^t \quad &\text{if the decoder knows } \sigma^2,\\
		\frac{\|\boldsymbol{z}_{\sfr}^t\|^2}{n/\Lr} \quad &\text{otherwise}, \label{eq:amp_decoder_phi}
	\end{cases}\\
	\widehat{\tau}_\sfc^t &= \frac{R/2}{\ln(\K\M)} \left[\frac{1}{\Lr}\sum_{\sfr=1}^{\Lr}\frac{W_{\sfr \sfc}}{\widehat{\phi}_\sfr^t}\right]^{-1}.  \label{eq:amp_decoder_tau}
\end{align}
These online estimates are used for the numerical simulations in Section \ref{sec:numerical_results}. The justification for these estimates comes from \cite[Lemma 7.6]{rush2020capacity}, which shows that in the case of unmodulated real-valued SPARCs, the estimates \eqref{eq:amp_decoder_gamma}--\eqref{eq:amp_decoder_tau} concentrate on their respective state evolution parameters. The arguments of \cite[Lemma 7.6]{rush2020capacity} can be extended to show similar concentration of the online estimates for modulated complex SPARCs, but we do not pursue this in the current paper. 

\subsection{Comparison of AMP for real-valued and complex-valued (unmodulated) SPARCs} \label{subsec:real_vs_complex}

The AMP decoder for complex SPARCs in \eqref{eq:amp_decoder}--\eqref{eq:eta_function} simplifies to the AMP decoder for unmodulated real-valued SPARCs in \cite{rush2020capacity} if $\K=1$ and a real-valued design matrix $\A$ is used.
Moreover, the state evolution for complex SPARCs in \eqref{eq:se_phi}--\eqref{eq:se_E} simplifies to the state evolution for unmodulated real-valued SPARCs in \cite{rush2020capacity} 
if $\K=1$, except that in the definition of $\tau_\sfc^t$ in \eqref{eq:se_psi}, the ``$R/2$'' term is replaced with ``$R$'' for a rate $R$ unmodulated real-valued SPARC.%
\footnote{One can also define $\tau_\sfc^t$ in \eqref{eq:se_psi} using ``$R$'' instead ``$R/2$'' for complex SPARCs and change the definitions of function $\eta$ in \eqref{eq:eta_function} and function $\mc{E}(\tau)$ in \eqref{eq:se_E} accordingly.
With this alternative definition of $\tau_\sfc^t$, the interpretation of the arguments to the function $\eta$ in \eqref{eq:eta_function_interpret} would be slightly different:  the first argument $\boldsymbol{s}$ would be approximately distributed as $\bbeta + \sqrt{\widetilde{\boldsymbol{\tau}}} \odot \boldsymbol{u}$, with $\boldsymbol{u} = u_1,\ldots,u_{\L\M} \stackrel{\text{i.i.d.}}{\sim} \mc{CN}(0,1)$ (instead of $\stackrel{\text{i.i.d.}}{\sim} \mc{CN}(0,2)$).
The definition of $\tau_\sfc^t$ in \eqref{eq:se_psi} using ``$R/2$'' makes the comparison of the AMP and SE equations between complex and real-valued SPARCs straightforward.}
This implies that the predicted NMSE of the AMP decoder for a rate $R$ unmodulated complex SPARC is the same as that for a rate $R/2$ unmodulated real-valued SPARC.
%
This matches our intuition since a transmission rate $R$ over a complex AWGN channel with signal-to-ratio $P/\sigma^2$ corresponds to a rate of $R/2$ over each of two independent (real) AWGN channels with signal-to-noise ratio $P/\sigma^2$.


\section{Error performance analysis}\label{sec:err_perf_analysis}

A natural error criterion for a hard decision SPARC decoder is the section error rate (SER), which is the fraction of sections decoded in error. Denoting  the final hard decision estimate of the message vector by $\widehat{\bbeta}^T$, the section error rate is defined as 
\begin{equation}\label{eq:ser_def}
\text{SER} := \frac{1}{\L}\sum_{\ell=1}^L \mathbbm{1}\{\widehat{\bbeta}^T_{\text{sec}(\ell)} \neq \bbeta_{\text{sec}(\ell)} \}.
\end{equation}
(For the AMP decoder, the hard decision estimate $\widehat{\bbeta}^T$ is given in \eqref{eq:amp_hard_decision}.)
In a modulated SPARC, a section is decoded in error when \emph{either} the location or value of the single non-zero entry is estimated incorrectly. Recall that each section corresponds to $\log\K + \log\M$ bits of information. 

We can also measure decoding performance via the bit error rate (BER). Recall that $\log \M$ bits determine the location of the non-zero entry in a section of the message vector $\bbeta$, and $\log \K$ bits determine its value.
If the location of the non-zero entry is estimated incorrectly, on average half of the $\log \M$ bits will be decoded wrongly due to the uniform mapping of bits to location.
If the value is estimated incorrectly, on average less than half of the $\log \K$ bits will be decoded wrongly because Gray coding is used to map bits to constellation symbols.

A third performance measure is the frame error rate (FER), which is based on whether the entire SPARC codeword is decoded correctly. Since an error in decoding any of the sections leads to a codeword error, the FER is at least as large as the SER. In particular, we have FER $\geq$ SER $\geq$ BER.


In this section, we first show that the SER of the AMP decoder is  upper bounded by a constant times its normalized mean squared error (NMSE). This result (Lemma \ref{lem:ser_nmse}) implies that an upper bound on the SER can be obtained from an upper bound on the state evolution parameters $\{\psi_\sfc^t\}_{\sfc\in[\Lc]}$ (which predict the block-wise NMSE). Our main technical result (Proposition \ref{prop:se_psi_bound}) provides such an upper bound on $\{\psi_\sfc^t\}_{\sfc\in[\Lc]}$. Using this bound, we show that in the large system limit, the state evolution recursion for complex SPARCs with $\K$-ary PSK modulation (Corollary \ref{corr:asymp_se}) is the \textit{same} for any fixed value of $\K$, including $\K=1$ (unmodulated).
Therefore, the same arguments that prove that unmodulated SPARCs are capacity achieving with AMP decoding also imply that $\K$-PSK modulated SPARCs are capacity achieving with AMP decoding, for any fixed $\K$ (Theorems \ref{thm:sc_mod_sparcs} and  \ref{thm:pa_mod_sparcs}).


\begin{lemma}[Bounding SER in terms of NMSE]\label{lem:ser_nmse}
Consider a complex SPARC with $\K$-ary PSK modulation, with $\K$ being a power of 2. Let the AMP decoder  be run for $T$ iterations, with $\bbeta^T$ being the final ``soft-decision'' estimate produced according to \eqref{eq:amp_decoder}, and $\widehat{\bbeta}^T$ the final hard-decision  estimate produced according to \eqref{eq:amp_hard_decision}. Let the section error rate (SER) corresponding to 
$\widehat{\bbeta}^T$ be defined as in \eqref{eq:ser_def}. Then the SER  can be bounded in terms of the normalised mean squared error $\frac{1}{\L}\| \bbeta^T - \bbeta \|^2$ as follows:
\begin{equation}
\label{eq:ser_leq_nmse}
	\text{SER} \, \leq \,
	\begin{cases}
		4 \cdot \frac{\|\bbeta^T - \bbeta \|^2}{\L} \quad &\text{if} \ \K=1,2, 4,\\
		\sin^{-4}(\frac{\pi}{\K}) \cdot \frac{\|\bbeta^T - \bbeta \|^2}{\L} \quad &\text{if} \ \K \geq 8.
	\end{cases}
\end{equation}
\end{lemma}
The proof is given in Appendix \ref{appendix:proof_lemma_se_nmse}.

Since the NMSE $\frac{1}{\L}\|\bbeta^t - \bbeta \|^2$ is predicted by the state evolution quantity $\frac{1}{\Lc}\sum_\sfc \psi_\sfc^t$, Lemma \ref{lem:ser_nmse} implies  that an upper bound on the parameters $\{\boldsymbol{\psi}_\sfc^t\}_{\sfc\in[\Lc]}$  would give an upper bound on the SER after iteration $t$. The following proposition gives such an upper bound for the state evolution predicted NMSE.

\begin{prop}[Bounding the state evolution predicted NMSE]\label{prop:se_psi_bound}
Consider any base matrix $\W$ that satisfies $\xi_1 \leq \frac{1}{\Lr}\sum_{\sfr'} W_{\sfr' \sfc} \leq \xi_2$ and $\xi_1 \leq \frac{1}{\Lc}\sum_{\sfc'} W_{\sfr\sfc'} \leq \xi_2$ for some universal positive constants $\xi_1, \xi_2$, for all $\sfr\in[\Lr], \sfc\in[\Lc]$. Let
\begin{equation}\label{eq:se_nu}
	\nu_\sfc^t \, \coloneqq \, \frac{1}{\tau_\sfc^t \cdot \ln(KM)} \, = \, \frac{2}{R\cdot\Lr}\sum_{\sfr=1}^{\Lr}\frac{W_{\sfr \sfc}}{\phi_\sfr^t}, \quad \text{for} \ \sfc\in [\Lc].
\end{equation}
Then,
\begin{equation}
\frac{2}{R} \cdot \frac{\xi_1}{\sigma^2 + 2\xi_2} \ \leq \ \nu_\sfc^t \ \leq \ \frac{2}{R} \cdot \frac{\xi_2}{\sigma^2}, \label{eq:se_nu_lb_ub}
\end{equation}
and for sufficiently large $\M$ and any $\delta\in (0, \frac{1}{2})$, $\tilde{\delta} \in (0, 1)$,
we have the following bounds on $\psi_\sfc^{t+1}$ for $\sfc \in [\Lc]$:
\begin{align}
	\text{(a)} \quad \psi_\sfc^{t+1} &\leq  f_{K,M} \cdot \mathbbm{1}\{\nu_\sfc^t > 2+\delta\} \nonumber\\  
	&\, \quad +  (1+h_{K,M}) \cdot \mathbbm{1}\{\nu_\sfc^t \leq 2+\delta\}, \label{eq:psi_c_ub}\\[0.3em]
	\text{(b)} \quad \psi_\sfc^{t+1}  &\geq  \big(1 - \M^{-\alpha_{K}\tilde{\delta}^2}\big) \cdot \mathbbm{1}\{\nu_\sfc^t < 2- \tilde{\delta}\}.\label{eq:psi_c_lb}
\end{align}
The non-negative scalars $f_{\K,\M}$ and $h_{\K,\M}$ satisfy
\be\label{eq:f1_f2_v0}
f_{\K,\M} \leq \frac{(KM)^{- \alpha_{1K} \delta^2}}{\delta \sqrt{\ln(KM)}}, \quad
h_{\K,\M} \leq \frac{(KM)^{-\alpha_{2K} \nu_\sfc^t}}{\sqrt{\nu_\sfc^t\ln(KM)}},
\ee
where the positive constants $\alpha_{K}$, 
$\alpha_{1K}$, and $\alpha_{2K}$ depend only on $\K$ and the bounds on $\nu_\sfc^t$ in \eqref{eq:se_nu_lb_ub}.
Exact expressions for $f_{\K,\M}, h_{\K,\M}$  are given in Remark \ref{rem:fKhK_exact} below.
\end{prop}

\begin{remark}
The variable $\nu_\sfc^t$ can be regarded as a measure of the ``signal-to-noise ratio'' of column block $\sfc$ after iteration $t$. 
When $\nu_\sfc^t$ exceeds $2+\delta$, the predicted NMSE of column block $\sfc$ after iteration $t+1$, denoted by $\psi_\sfc^{t+1}$, is bounded above by $f_{\K,\M}$, which can be made arbitrarily small as $\M\to\infty$ (Remark \ref{rem:f1f2_to_zero}).
On the other hand, if $\nu_\sfc^t$ is less than $2-\tilde{\delta}$, 
then $\psi_\sfc^{t+1}$ is bounded above by $1+h_{\K,\M}$ and  bounded below by $1 - \M^{-\alpha_{K}\tilde{\delta}^2}$, 
both of which tend to 1 as $\M\to\infty$.
\end{remark}

\begin{remark} \label{rem:fKhK_exact} The expressions for the scalars $f_{\K, \M}$ and $h_{\K, \M}$  when $K$ is a power of $2$, are as follows. For $\K=1,2$ and $4$,
\begin{equation}\label{eq:f1_f2}
	f_{\K, \M}, \ h_{\K,\M} = 
  	\begin{cases}
    		\frac{M^{-\kappa_1\delta^2}}{\delta \sqrt{\ln M}}, \hspace{1.5cm} 0, & \text{if } \K=1 \\[0.5em]
    		\frac{(2M)^{- \kappa_2 \delta^2}}{\delta\sqrt{\ln(2M)}}, \ 
			\frac{(2M)^{-\nu_\sfc^t/2}}{\sqrt{2\pi \nu_\sfc^t \ln(2M)}}, & \text{if } \K=2 \\[0.8em]
    		\frac{(4M)^{- \kappa_3 \delta^2}}{\delta\sqrt{\ln(4M)}}, \
			\frac{2(4M)^{-\nu_\sfc^t/2}}{\sqrt{2\pi \nu_\sfc^t \ln(4M)}}, & \text{if } \K=4,
  	\end{cases}
\end{equation}
and for $\K\geq 8$,
\begin{align}
f_{\K,\M} &= \frac{(1+\cot(\frac{2\pi}{\K})) (KM)^{- \frac{\kappa_4 \delta^2}{(1+\cot(\frac{2\pi}{\K}))^2}}}{\delta \sqrt{\ln(KM)}}\nonumber\\[0.1em]
&\hspace{4.7em}+ K (KM)^{- 2(2+\delta^2)\sin^2(\frac{\pi}{K}) }, \\[0.3em]
h_{\K,\M} &= \frac{2(1+\cot(\frac{2\pi}{K}))(KM)^{-\frac{\nu_\sfc^t}{2(1+\cot(\frac{2\pi}{K}))^2}}}{\sqrt{2\pi\nu_\sfc^t\ln(KM)}}, \label{eq:f2_K8}
\end{align}
where $\kappa_1$ to $\kappa_4$ are universal positive constants, not depending on $K$ or $M$.
\end{remark}

\begin{remark}\label{rem:f1f2_to_zero}
From \eqref{eq:f1_f2}--\eqref{eq:f2_K8},  it is clear that for any fixed $\K$ and $\delta\in(0,\frac{1}{2})$, both $f_{\K,\M}$ and $h_{\K,\M}$ approach 0 as $\M\to\infty$.  On the other hand, for any fixed $\M$ and $\delta\in(0,\frac{1}{2})$, both $f_{\K,\M}$ and $h_{\K,\M}$ increase with $\K$, for $\K \geq 8$. Therefore,  the upper bound on the predicted block-wise NMSE $\psi_\sfc^{t+1}$ in \eqref{eq:psi_c_ub} and the upper bound on the SER  in \eqref{eq:ser_leq_nmse} both increase with $\K$, when $\K\geq 8$. This is consistent with the fact that when $M$ is fixed,  the amount of information transmitted per section increases with $K$.

If we consider increasing values of $\K$, one can ask how fast $\K$ can grow 
with $\M$ such that $f_{\K,\M}$ and $h_{\K,\M}$  approach 0  as $\K,\M$ both approach infinity? Using the expressions in \eqref{eq:f1_f2}--\eqref{eq:f2_K8}, we  deduce that $f_{\K,\M}$ and $h_{\K,\M}$ approach 0 for any fixed $\delta\in(0,\frac{1}{2})$ if 
$\lim_{\K,\M\to\infty} |\frac{M}{g(\K)}| = \infty$ (i.e., $\M$ dominates $g(\K)$ asymptotically), where
\begin{align}\label{eq:gK}
		g(\K) = \frac{1}{\K} \, \max \bigg\{
		 &\Big[1+\cot\big(\frac{2\pi}{\K} \big)\Big]^{\frac{[1+\cot(\frac{2\pi}{\K})]^2}{\kappa_4 \delta^2}}, \nonumber\\
		 &\Big[1+\cot\big(\frac{2\pi}{\K}\big)\Big]^{\frac{2[1+\cot(\frac{2\pi}{\K})]^2}{\min_\sfc \nu_\sfc^t}}, \nonumber\\[0.3em]
		 & \K^{\frac{1}{2(2+\delta^2)\sin^2(\frac{\pi}{K})}} 
		 \bigg\}.
\end{align}
\end{remark}


The proof of Proposition \ref{prop:se_psi_bound} is given in Section \ref{sec:proof_prop_se_psi_bound}. A proof sketch is presented here to highlight the main ideas.

\begin{IEEEproof}[Proof Sketch]
Recall from \eqref{eq:se_psi} that $\psi_\sfc^{t+1} = 1 - \mathcal{E}(\tau_\sfc^t)$.
The upper bound on $\psi_\sfc^{t+1}$  in \eqref{eq:psi_c_ub} is proved by obtaining two lower bounds for $\mathcal{E}(\tau_\sfc^t)$, one which holds for all values of $\nu_\sfc^t>0$, and the other which holds for $\nu_\sfc^t > 2$. In particular, we show that
\begin{equation}\label{eq:se_E_lb_nu_two}
	 \mathcal{E}(\tau_\sfc^t) \geq
  	\begin{cases}
    		-h_{\K,\M}, & \text{for } \nu_\sfc^t > 0,\\
    		1-f_{\K,\M}, & \text{for } \nu_\sfc^t > 2+\delta.
  	\end{cases}
\end{equation}
Similarly, the lower bound on $\psi_\sfc^{t+1}$  in \eqref{eq:psi_c_lb} is proved by obtaining the following upper bound for $\mathcal{E}(\tau_\sfc^t)$:
\be\label{eq:se_E_ub_nu_one}
\mathcal{E}(\tau_\sfc^t) \leq \M^{-\alpha_{K}\tilde{\delta}^2} \quad \text{for } \nu_\sfc^t < 2-\tilde{\delta}.
\ee
For the case of $\K=1$ (no modulation), \eqref{eq:se_E_lb_nu_two} was proved in \cite[Appendix~A]{rush2019theerror} and \eqref{eq:se_E_ub_nu_one} in \cite[Appendix~A]{rush2020capacity}. 
The proof for $\K\geq2$ is significantly more challenging as the $\mathcal{E}(\tau_\sfc^t)$ term defined in \eqref{eq:se_E} can be negative. 
We discuss the main ideas in the proof of \eqref{eq:se_E_lb_nu_two} for the $\K\geq2$ case below.

From \eqref{eq:se_E} , we observe that $\mathcal{E}(\tau)$ is an expectation over $\M$ i.i.d.\ $\mc{CN}(0,2)$ random variables $U_1,\ldots, U_\M$. Using the tower property we write
\begin{align}
	&\mathcal{E}(\tau) = \nonumber\\
	&\mathbb{E}_{U_1}  \mathbb{E}\Bigg[\frac{\sum_{k=1}^{\K} \Re[\p_k]\cdot e^{\frac{1}{\tau}\Re\left[(1+\sqrt{\tau}U_1) \overline{\p_k}\right]}}{\sum_{a=1}^{\K} e^{\frac{1}{\tau}\Re\left[(1+\sqrt{\tau}U_1) \overline{\p_a}\right]} + \sum_{j=2}^{\M}\sum_{b=1}^{\K} e^{\frac{\Re[U_j \overline{\p_b}]}{\sqrt{\tau}}}} \, \bigg | \, U_1 \Bigg],
\end{align}
where the inner expectation is over $U_2,\ldots, U_\M$, which only appear in the denominator. The outer expectation over $U_1$ is split into four terms, each of which integrate over different ranges of $U_1^R$ and $U_1^I$, the (independent) real and imaginary parts of $U_1$.
\begin{align}
	\mathcal{E}(\tau) 
	&=\int_{\underline{u}}^{\infty} \int_{\underline{u}}^{\infty} p(u^R) \, p(u^I) \, \mathbb{E}_{U_2,\ldots,U_\M} [\ldots]\, du^R \, du^I \nonumber\\
	&\quad + \int_{-\infty}^{\underline{u}} \int_{-\infty}^{\underline{u}} p(u^R) \, p(u^I) \, \mathbb{E}_{U_2,\ldots,U_\M} [\ldots ] \, du^R \, du^I \nonumber \\
	&\quad + \int_{\underline{u}}^{\infty} \int_{-\infty}^{\underline{u}} p(u^R) \, p(u^I) \, \mathbb{E}_{U_2,\ldots,U_\M} [\ldots ] \, du^R \, du^I\nonumber\\
	&\quad + \int_{-\infty}^{\underline{u}} \int_{\underline{u}}^{\infty} p(u^R) \, p(u^I) \, \mathbb{E}_{U_2,\ldots,U_\M} [\ldots ] \, du^R \, du^I \nonumber \\
	&= I_1 + I_2 + I_3 + I_4, \label{eq:se_E_equals_I1_I2_I3_I4}
\end{align}
where $\underline{u}$ is an arbitrary parameter which splits the range of the integration.

We argue that for a suitably chosen value of $\underline{u}$ (which depends on $\K,\M$ but not on $\nu_\sfc^t$), the $\mathbb{E}_{U_2,\ldots,U_\M} [\ldots]$ term is non-negative for $U_1^R\geq \underline{u}$ and $U_1^I\geq \underline{u}$, and hence $I_1\geq 0$. For the terms $I_2$ to $I_4$, we first show that $\mathbb{E}_{U_2,\ldots,U_\M} [\ldots]  \geq -1$, and then use tail bounds for standard Gaussians to obtain $I_2+I_3 +I_4 \geq - h_{\K,\M}$. Thus we obtain the lower bound $\mathcal{E}(\tau_\sfc^t) \geq -h_{\K,\M}$ for all $\nu_\sfc^t > 0$.

For $\nu_\sfc^t > 2$, we again split the integral into the ranges in \eqref{eq:se_E_equals_I1_I2_I3_I4}, but choose $\underline{u}$ to depend on $\nu_\sfc^t$ as well as $\K,\M$. We again use $\mathbb{E}_{U_2,\ldots,U_\M} [\ldots]  \geq -1$ and tail bounds for standard Gaussians to obtain a lower bound of the form $I_2+I_3+I_4 \geq - B_1 \cdot f_{\K,\M}$ for sufficiently large $\M$ and positive constant $B_1$.
We then obtain a lower bound for $I_1$ which takes the form $I_1 \geq 1 - B_2 \cdot f_{\K,\M}$ for sufficiently large $\M$ and positive constant $B_2$. Combining the results, we obtain the lower bound on $\mathcal{E}(\tau_\sfc^t)$ for the $\nu_\sfc^t>2$ case. We therefore have both the lower bounds in \eqref{eq:se_E_lb_nu_two}.
\end{IEEEproof}


Proposition \ref{prop:se_psi_bound}  immediately yields the following asymptotic characterization of the state evolution recursion for $M \to \infty$.

\begin{corr}[Asymptotic state evolution]\label{corr:asymp_se}
For any base matrix $\W$ satisfying the conditions in Proposition \ref{prop:se_psi_bound}, the state evolution recursion in \eqref{eq:se_phi}--\eqref{eq:se_E} simplifies to the following  as $M \to \infty$, for any fixed $K \geq 1$. Initialize $\bar{\psi}_\sfc^0 = 1$ for $\sfc \in [\Lc]$, and for $t=0,1,\ldots$, compute
\begin{align}
	\bar{\phi}_\sfr^t \ &= \ \sigma^2 + \frac{1}{\Lc}\sum_{\sfc=1}^{\Lc}W_{\sfr \sfc} \bar{\psi}_\sfc^t, \hspace{1.2cm}\qquad \sfr\in[\Lr], \label{eq:se_phi_asymp}\\
	\bar{\psi}_\sfc^{t+1} \ &= \ 1 - \mathbbm{1}\left\{\frac{1}{\Lr}\sum_{\sfr=1}^{\Lr}\frac{W_{\sfr \sfc}}{\bar{\phi}_\sfr^t} > R\right\}, \qquad \sfc\in [\Lc], \label{eq:se_psi_asymp}
\end{align}
where $\bar{\phi}, \bar{\psi}$ indicate asymptotic values.
\end{corr}

\begin{remark}\label{rem:mod_unmod_asymp_se}
Though the state evolution parameters in \eqref{eq:se_phi}--\eqref{eq:se_E} depend on the modulation factor $\K$, the asymptotic values of these parameters in \eqref{eq:se_phi_asymp}--\eqref{eq:se_psi_asymp} do not.
Therefore, as $\M\to\infty$, the predicted per-iteration NMSE of the AMP decoder for complex SPARCs with $\K$-ary PSK modulation is the same for any finite $\K$, including $\K=1$ (unmodulated).
\end{remark}

\begin{remark}[Complex vs. real asymptotic SE]\label{rem:complex_real_asymp_se}
The only difference between the asymptotic state evolution in \eqref{eq:se_phi_asymp}--\eqref{eq:se_psi_asymp} for modulated complex SPARCs and that for unmodulated real-valued SPARCs \cite[Lemma 1]{hsieh2018spatially} is that $R$ within the indicator function in \eqref{eq:se_psi_asymp} is replaced by $2R$ in \cite{hsieh2018spatially}. This matches our intuition since a rate of $R$ nats over a complex AWGN channel corresponds to a rate of $R/2$ nats per dimension (cf. Sec. \ref{subsec:real_vs_complex}).
\end{remark}

\subsection{PSK-modulated SPARCs are asymptotically capacity achieving}\label{sec:proof_mod_scsparc}
The asymptotic state evolution equations in \eqref{eq:se_phi_asymp}--\eqref{eq:se_psi_asymp} have been analyzed  for two choices of base matrix $\W$:  for a suitable $(\omega, \Lambda, \rho)$ base matrix  in\cite{hsieh2018spatially, rush2020capacity}, and for the base matrix corresponding to an exponentially decaying power allocation in \cite{rush2017capacity}. These results show that in both cases, for any fixed $R < \mc{C}$ we have $\bar{\psi}_\sfc^T=0$ for $\sfc \in [\Lc]$, where the number of iterations $T$ is a finite value depending on the rate gap from capacity. That is,  the state evolution equations predict reliable decoding in the large system limit for $R < \mc{C}$.

From Corollary \ref{corr:asymp_se} we know that the asymptotic equations \eqref{eq:se_phi_asymp}--\eqref{eq:se_psi_asymp} hold for $K$-PSK modulated complex SPARCs, for any $K \geq 1$ (Remarks \ref{rem:mod_unmod_asymp_se} and \ref{rem:complex_real_asymp_se}). We can therefore directly use the asymptotic state evolution results in \cite{rush2017capacity} and  \cite{rush2020capacity} to argue  that  $K$-PSK modulated SPARCs are asymptotically capacity achieving for any finite $K$, with the base matrices used in  \cite{rush2017capacity} and \cite{rush2020capacity}.  
Theorem \ref{thm:sc_mod_sparcs}  gives this result for modulation applied to spatially coupled SPARCs defined via an $(\omega, \Lambda, \rho)$ base matrix. Theorem \ref{thm:pa_mod_sparcs} gives a  similar result for SPARCs with exponentially decaying power allocation.  We require some definitions to state the theorems.

For an $(\omega,\Lambda, \rho)$ base matrix , let $\vartheta = 1 + \frac{\omega-1}{\Lambda}$,  and
\be R^\star  =  \frac{1}{\vartheta} \ln (1 + \vartheta \, \snr), 
\label{eq:R_star} \ee
where $\snr = \frac{P}{\sigma^2}$. We will consider rates $R < R^\star$ (in nats), noting that  
$\mc{C} >   R^\star  > \frac{\mc{C}}{\vartheta}$, where
\be
\mc{C} =  \ln (1 +  \snr).
\ee
Observe that $\vartheta \to 1$ as $\frac{\omega}{\Lambda} \to 0$, and hence $R^\star$ can be made arbitrarily close to $\mc{C}$ for any fixed $\omega$ by choosing $\Lambda$ to be sufficiently large.  Finally, let
\begin{align}
\omega^\star &=  \frac{\vartheta\, \snr^2}{(1+\vartheta\,\snr)(R^\star - R)}. \label{eq:omeg_star}
\end{align}

\begin{theorem}[$K$-PSK modulated SPARCs with spatial coupling are capacity achieving]
For any  $R<\mc{C}$, let $\W$ be an $(\omega,\Lambda, \rho)$ base matrix with parameters chosen such that $R^\star > R$, $\omega > \omega^\star$ and $\rho = \min\{ \frac{1}{2}, \frac{R^\star - R}{3 \snr} \}$, where $R^\star, \omega^\star$ are defined in \eqref{eq:R_star} and \eqref{eq:omeg_star}.
Fix  $K$ to be a power of $2$. Let $\{ \mc{S}_n \}$ be a sequence  of rate $R$, $K$-PSK modulated SPARCs (indexed by code length $n$), with $\mc{S}_n$ defined via an $n\times \L\M$ design matrix constructed from the base matrix $\W$.
Let $\text{SER}(\mc{S}_n) := \frac{1}{\L}\sum_{\ell=1}^L \mathbbm{1}\{\widehat{\bbeta}^T_{\text{sec}(\ell)} \neq \bbeta_{\text{sec}(\ell)} \}$  denote the section error rate of the AMP decoder after $T$ iterations where $T = \lceil \frac{\Lambda \omega^\star}{2 \omega} \rceil$. Then 
$$\lim_{n \to \infty} \text{SER}(\mc{S}_n)  = 0 \text{ almost surely}, $$ 
where
the limit is taken with  $(\L,\M, n)$ all tending to infinity such that $R = \L\ln(\K\M)/n$.
\label{thm:sc_mod_sparcs}
\end{theorem}
\begin{IEEEproof}
Recall the definition of  $\bar{\psi}_\sfc^{T}$ from Corollary \ref{corr:asymp_se}, iteratively computed according to
\eqref{eq:se_phi_asymp}--\eqref{eq:se_psi_asymp}. Under the stated conditions on the parameters of the $(\omega,\Lambda, \rho)$ base matrix, 
\cite[Prop. 4.1]{rush2020capacity} shows that 
\begin{align}
\bar{\psi}_\sfc^{T} =0, \quad \sfc \in [\Lc].
\label{eq:psicT0}
\end{align} 
Furthermore, \cite[Theorem 2]{rush2020capacity} implies that 
\be \lim \frac{\| \bbeta^T - \bbeta \|^2}{L} = \frac{1}{\Lc} \sum_{\sfc \in[\Lc]} \bar{\psi}^{T}_{\sfc} \quad \text{almost surely}, \label{eq:MSE conv0}
\ee where the limit is taken with  $(\L,\M, n)$ all tending to infinity such that $R = \L\ln(\K\M)/n$.  The concentration result in \eqref{eq:MSE conv0} is shown for unmodulated real-valued SPARCs in \cite{rush2020capacity}, and the proof for the complex-valued case is essentially the same. Combining \eqref{eq:psicT0} and \eqref{eq:MSE conv0} with Lemma \ref{lem:ser_nmse} yields the statement of the  theorem.
\end{IEEEproof}

For any rate $R<\mc{C}$, we can choose base matrix parameters $\omega$ and $\Lambda$ as follows to  satisfy the conditions of the theorem.
Letting $\vartheta_0 = \frac{\mc{C}}{R}$, first choose  $\omega > \omega^\star$ (with $\vartheta$ replaced by $\vartheta_0$). Then  choose $\Lambda$ large enough that $\vartheta = 1 + \frac{\omega-1}{\Lambda} \leq \vartheta_0$. This ensures that  $R^\star > R$ and $\omega > \omega^\star$.

\begin{theorem}[$K$-PSK modulated SPARCs with exponentially decaying power allocation are capacity  achieving]
\label{thm:pa_mod_sparcs}
For any  $R<\mc{C}$, let $\W$ be a $1\times \L$ base matrix corresponding to the exponentially decaying power allocation, i.e.,
\be
W_{1\ell} = \L P\cdot \frac{e^{\mc{C}/\L}-1}{1-e^{-\mc{C}}}\cdot e^{-\mc{C}\ell/\L}, \quad \ell\in[\L].
\label{eq:exp_pow_alloc}
\ee
Fix $K$ to be a power of $2$. Let $\{ \mc{S}_n \}$ be a sequence  of rate $R$, $K$-PSK modulated SPARCs (indexed by code length $n$), with $\mc{S}_n$ defined via an $n\times \L\M$ design matrix constructed from the base matrix $\W$.
Let $\text{SER}(\mc{S}_n) := \frac{1}{\L}\sum_{\ell=1}^L \mathbbm{1}\{\widehat{\bbeta}^T_{\text{sec}(\ell)} \neq \bbeta_{\text{sec}(\ell)} \}$  denote the section error rate of the AMP decoder after $T$ iterations where 
$T = \lceil \frac{\mc{C}}{\ln(\mc{C}/R)} \rceil$. 
Then $$\lim_{n \to \infty} \text{SER}(\mc{S}_n)  = 0 \text{ almost surely}, $$ where
the limit is taken with  $(\L,\M, n)$ all tending to infinity such that $R = \L\ln(\K\M)/n$.
\end{theorem}

\begin{IEEEproof}
With the exponentially decaying allocation in \eqref{eq:exp_pow_alloc}, using  $\Lr =1$, $\Lc=\L$, the asymptotic state evolution recursion in Corollary \ref{corr:asymp_se} reduces to
\be
\bar{\psi}_\ell^{t+1} =  \mathbbm{1}\left\{ W_{1\ell} \leq R\Big(\sigma^2 + \frac{1}{L}\sum_{j=1}^L W_{1j} \, \bar{\psi}_{j}^{t} \Big) \right\}, \quad \ell \in [L].
\ee
This recursion is analyzed in \cite[Lemmas 1 and 2]{rush2017capacity}, where it is shown that 
\be \lim \frac{1}{L} \sum_{\ell=1}^L W_{1\ell} \, \bar{\psi}_{\ell}^{T} =0,  \label{eq:pow_alloc_SE}\ee
for $T = \lceil \frac{\mc{C}}{\ln(\mc{C}/R)} \rceil$. Furthermore, \cite[Theorem 2]{rush2020capacity}  implies that 
\be 
\lim \,  \left \vert \frac{\| \bbeta^T - \bbeta \|^2}{L} - \frac{1}{\L} \sum_{\ell\in[\L]} \bar{\psi}^{T}_{\ell} \right \vert =0 \quad \text{ almost surely}. \label{eq:MSE_conv1}
\ee
In both \eqref{eq:pow_alloc_SE} and \eqref{eq:MSE_conv1}, is the limit is taken with  $(\L,\M, n)$ all tending to infinity such that $R = \L\ln(\K\M)/n$. As $L \to \infty$, the base matrix entries $\{W_{1\ell}\}_{\ell \in [L]}$ in  \eqref{eq:exp_pow_alloc} are bounded above and below by strictly positive universal constants. Therefore, \eqref{eq:pow_alloc_SE} implies that $\lim \frac{1}{\L} \sum_{\ell \in[\L]} \bar{\psi}^{T}_{\ell} =0$. Using this in \eqref{eq:MSE_conv1} and then invoking Lemma \ref{lem:ser_nmse} yields the statement of the  theorem.
\end{IEEEproof}

Theorems \ref{thm:sc_mod_sparcs} and \ref{thm:pa_mod_sparcs} correspond to two different choices of base matrices for which $\bar{\psi}_\sfc^T=0$ for $\sfc \in [\Lc]$. The key requirement for a PSK-modulated complex SPARC design to be capacity achieving is that the underlying base matrix satisfies $\frac{1}{\Lc} \sum_{\sfc =1}^\Lc \bar{\psi}_\sfc^{T} =0$, for all fixed $R < \mc{C}$. Here $T$ is the final iteration, which in the examples above  is determined by the gap from capacity $(\mc{C} -R)$. Whenever we have a base matrix whose asymptotic state evolution recursion (in Corollary \ref{corr:asymp_se}) satisfies the above property, \cite[Theorem 2]{rush2020capacity}  and Lemma \ref{lem:ser_nmse} together imply that the design is capacity achieving in the large system limit.


\section{Numerical results}\label{sec:numerical_results}

\begin{figure}[!t]
\centering
\subfloat[Bit error rate]{\includegraphics[width=0.85\columnwidth]{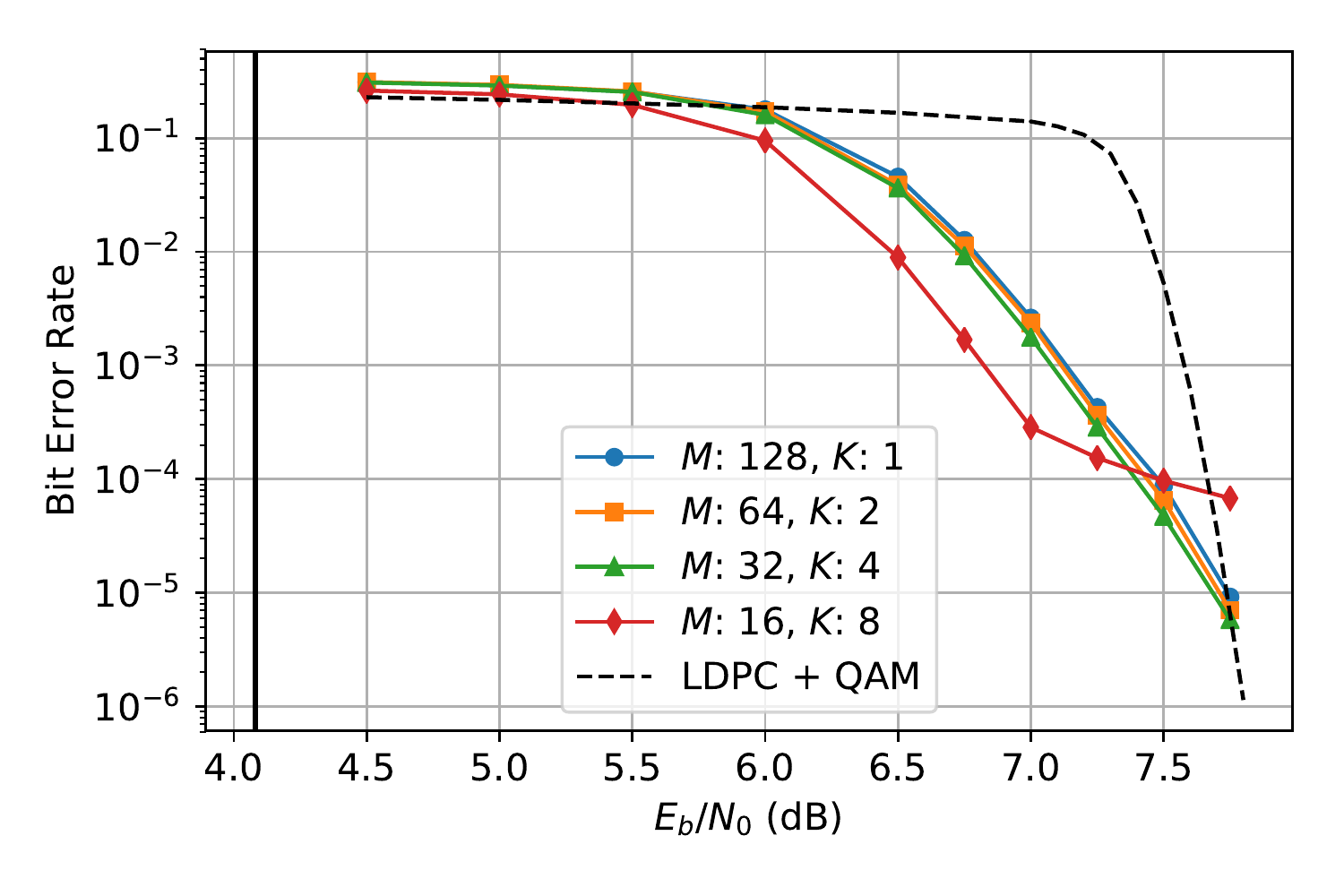}%
\label{fig:low_complex_R160_SC_ber}}
\hfil
\subfloat[Frame error rate]{\includegraphics[width=0.85\columnwidth]{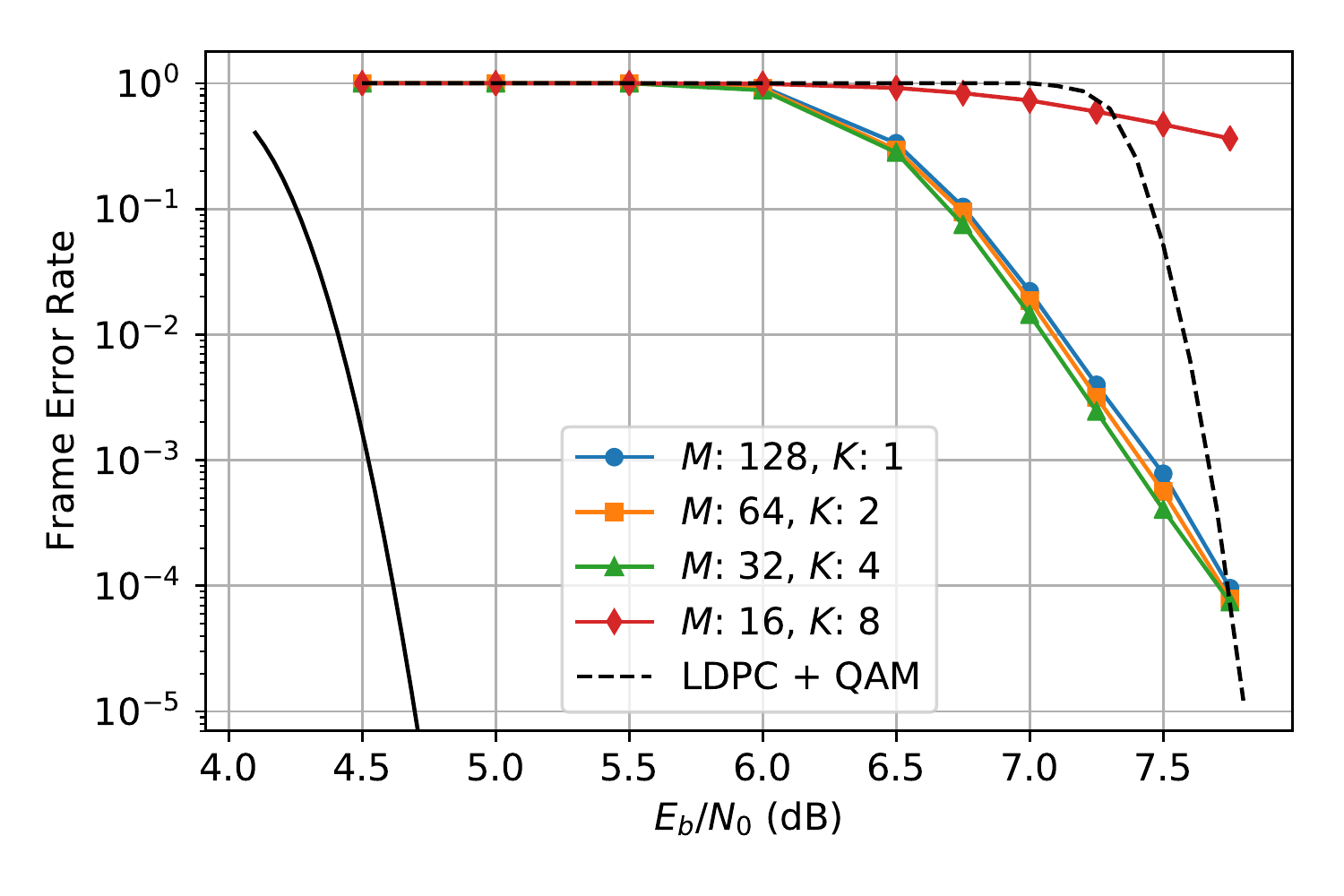}%
\label{fig:low_complex_R160_SC_fer}}
\caption{Error performance of modulated complex SPARCs defined via a $(\omega=6, \Lambda=32, \rho=0)$ base matrix. $R=1.593$ bits/dimension, $\L=960$, code length $n=2109$. Each curve represents a $\K$ and $\M$ pair with fixed $\K\M=128$. The dashed lines show the performance of coded modulation: (6480, 16200) DVB-S2 LDPC + 256 QAM, frame length $=2025$. The solid black line in subplot (a) is the AWGN Shannon limit for $R=1.6$ bits/dimension, and in subplot (b) it is the normal approximation to AWGN finite length error bound in \cite{polyanskiy2010channel}.}
\label{fig:low_complex_R160_SC}
\end{figure}

\begin{figure}
\centering
\subfloat[Bit error rate]{\includegraphics[width=0.83\columnwidth]{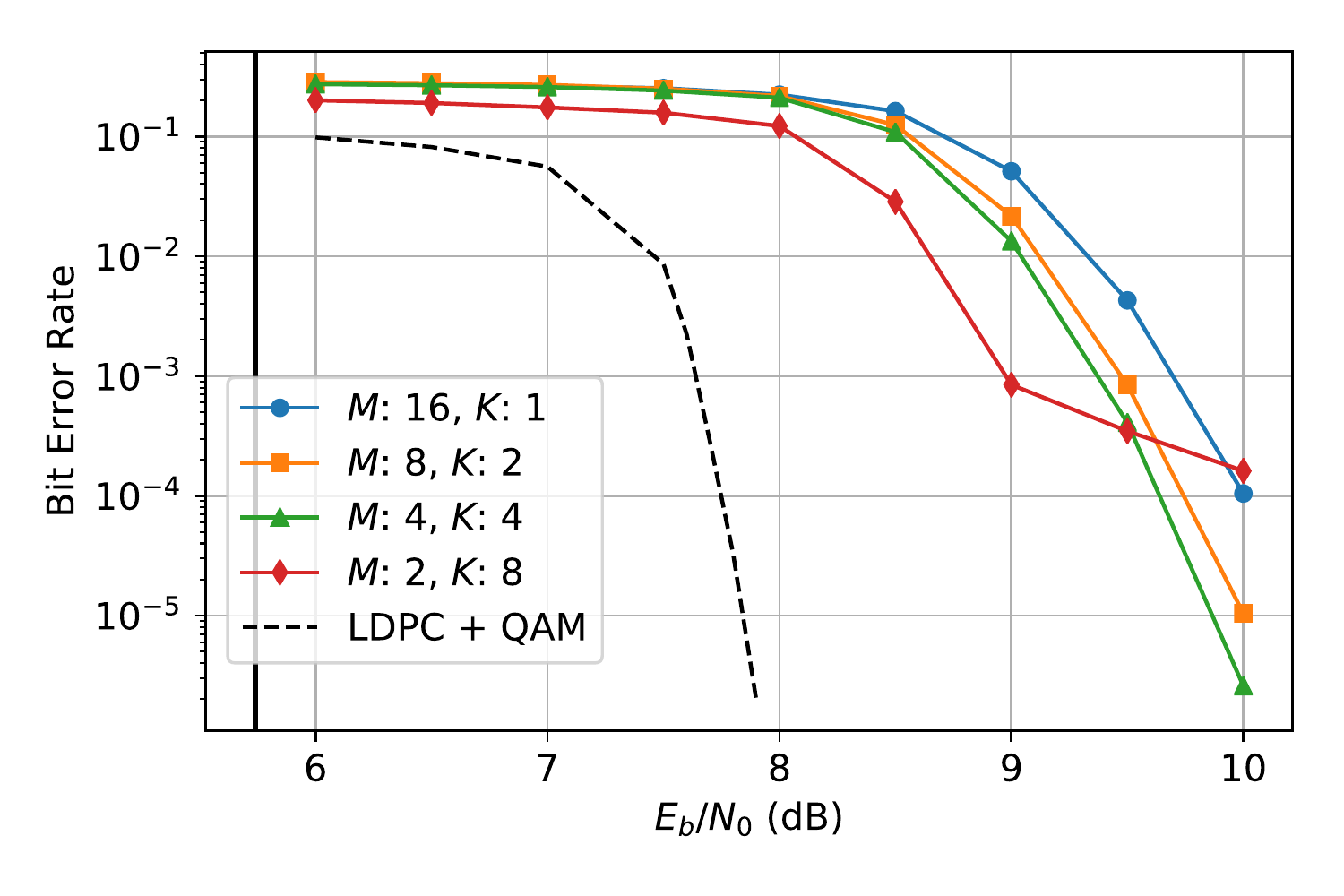}%
\label{fig:low_complex_R2_SC_ber}}
\hfil
\subfloat[Frame error rate]{\includegraphics[width=0.83\columnwidth]{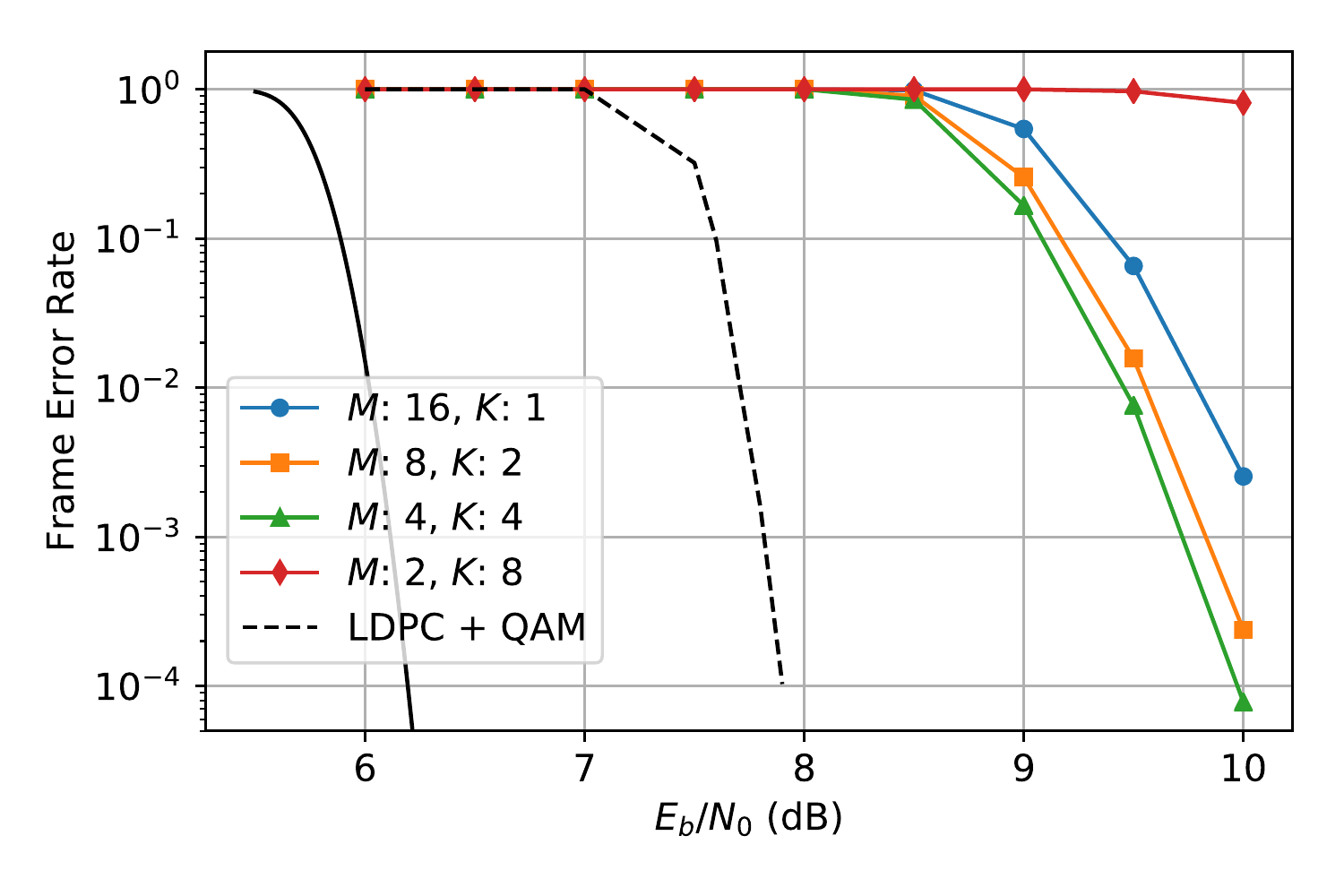}%
\label{fig:low_complex_R2_SC_fer}}
\hfil
\subfloat[Location error rate]{\includegraphics[width=0.83\columnwidth]{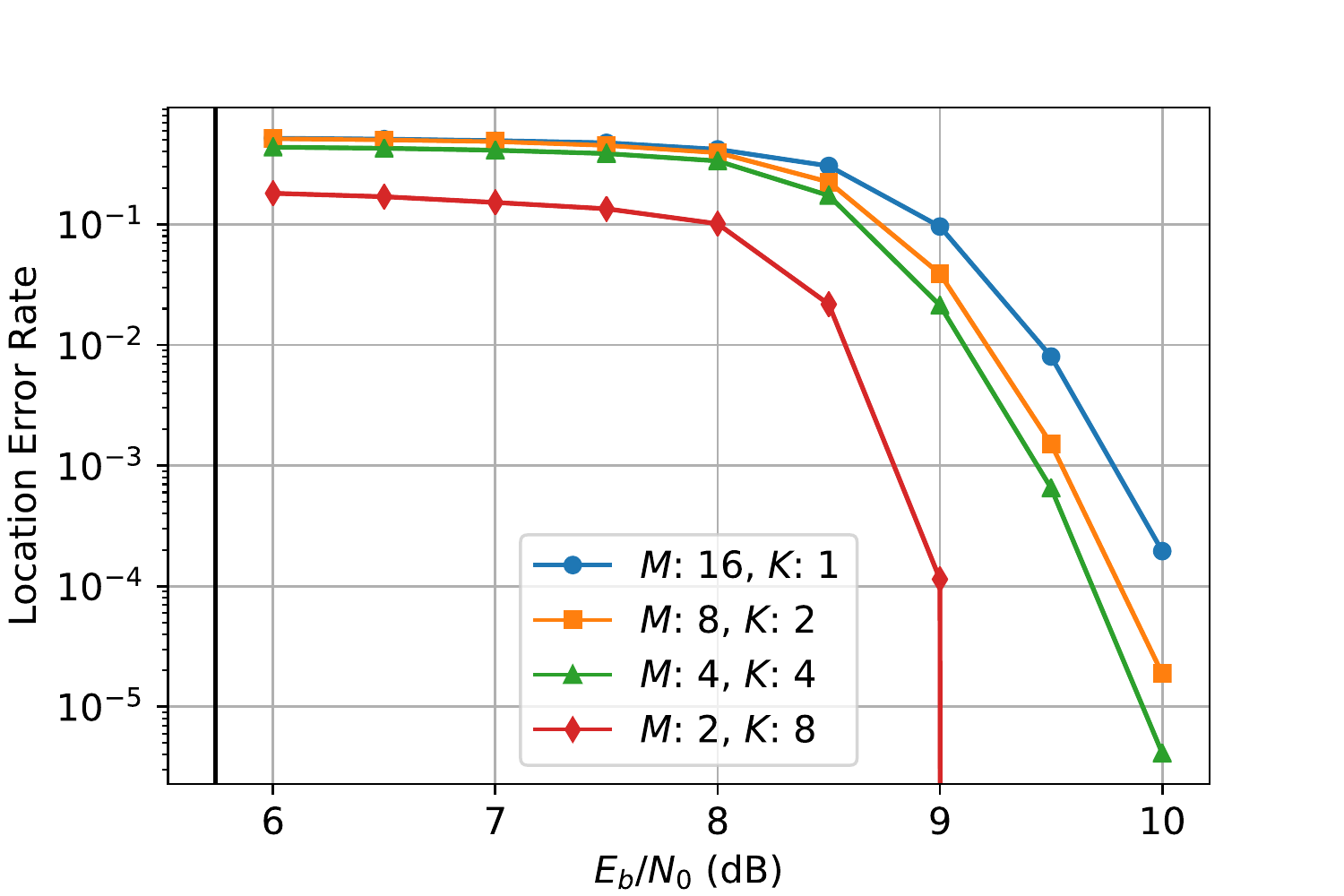}%
\label{fig:low_complex_R2_SC_ler}}
\hfil
\subfloat[Value error rate]{\includegraphics[width=0.83\columnwidth]{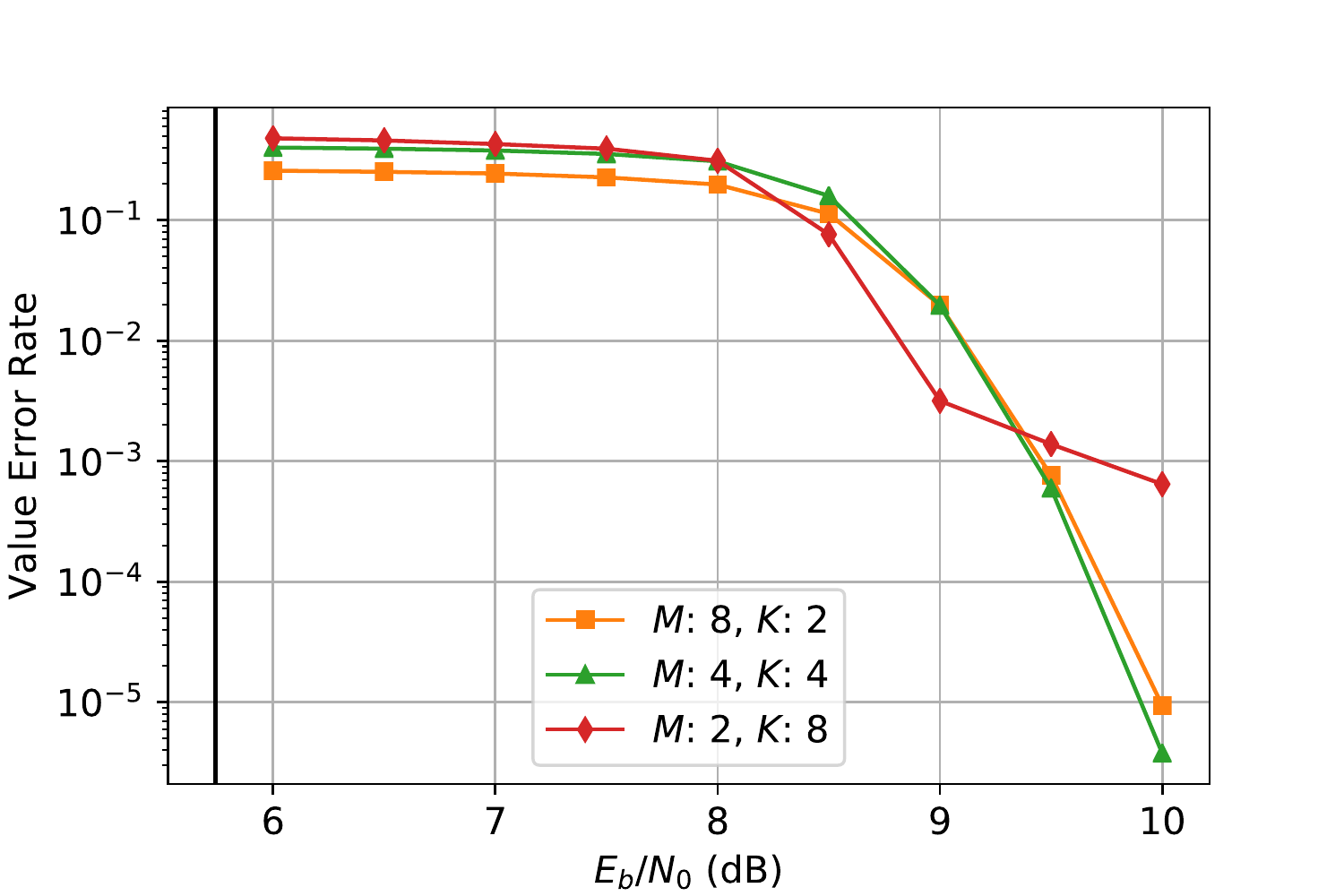}%
\label{fig:low_complex_R2_SC_ver}}
\caption{Error performance of modulated complex SPARCs defined via a $(\omega=6, \Lambda=32, \rho=0)$ base matrix. $R=1.99$ bits/dimension, $\L=2688$, code length $n=2701$. Each curve represents a $\K$ and $\M$ pair with fixed $\K\M=16$. The dashed lines show the performance of coded modulation: (10800, 16200) DVB-S2 LDPC + 64 QAM, frame length $=2700$. The solid black lines in subplots (a), (c) and (d) are the AWGN Shannon limit for $R=2.0$ bits/dimension, and in subplot (b) it is the normal approximation to AWGN finite length error bound in \cite{polyanskiy2010channel}.}
\label{fig:low_complex_R2_SC}
\end{figure}

In this section, we investigate the finite length error performance of modulated complex SPARCs  via numerical simulations.  Figures \ref{fig:low_complex_R160_SC} and \ref{fig:low_complex_R2_SC} show the finite length error performance of complex PSK-modulated SPARCs with  AMP decoding.
The error performance is evaluated using bit error rate (BER) and frame error rate (FER) as they are common performance metrics of interest. 
For reference, we also simulate and plot the error performance of coded modulation schemes (LDPC + QAM) using the AFF3CT toolbox \cite{cassagne2017fast}. The LDPC codes used are from the DVB-S2 standard \cite{dvb-s2}.

Within each figure, the SPARCs have the same rate $R$, code length $n$, and number of sections $\L$. We vary the code parameters $\K$ and $\M$ while keeping their product $\K \M$ constant, recalling that  $R = \frac{\L\log(\K\M)}{n}$ bits.
We observe that as $\K$ increases from $1$ (unmodulated) to $4$, both the BER and the FER improve.
At $\K=8$, the BER continues to improve at low values of $E_b/N_0$, but an error floor starts to appear at high $E_b/N_0$. 
However, the FER at $\K=8$ is significantly worse than that of $\K=1,2,4$ for all values of $E_b/N_0$.
We expect that a much larger $\M$ is required to achieve low FER with $\K=8$, but more investigation is required.

From Figs. \ref{fig:low_complex_R160_SC} and \ref{fig:low_complex_R2_SC}, we observe that the error performance of PSK-modulated SPARCs is better than the LDPC + QAM scheme for $R \approx 1.6$, but  worse for $ R \approx 2$. One reason for this is that the class of $(\omega, \Lambda, \rho)$ base matrices are capacity-achieving in the large system limit, but may not be optimal for finite length error performance.   Designing base matrices for modulated SPARCs that can achieve state of the art  error performance for a wider range of rates is an interesting direction for future work.  One idea is to design base matrices that use a combination of spatial coupling and power allocation.

Recall that in a modulated SPARC, a section error occurs when either the location or value (or both) of the non-zero entry in the section is decoded incorrectly. Figs. \ref{fig:low_complex_R2_SC_ler} and \ref{fig:low_complex_R2_SC_ver} show the location error rate (fraction of sections where the location of the non-zero entry was decoded in error) and the value error rate (fraction of sections where the value of the non-zero entry was decoded in error) from the same set of simulations used to plot Figs. \ref{fig:low_complex_R2_SC_ber} and  \ref{fig:low_complex_R2_SC_fer}.
We notice that with $\K\M$ held fixed,  the location error rate consistently improves as $\K$ increases and $\M$ decreases. We also notice that when $E_b/N_0 > 9$ dB, all errors in decoding the $\M=2, \K=8$ modulated complex SPARC were due to value errors.
Since bit errors arise from a combination of both location and value errors (see beginning of Section \ref{sec:err_perf_analysis}), the location error rate and value error rate curves help us understand the shape of the BER curve in Fig. \ref{fig:low_complex_R2_SC_ber}.


\emph{Implementation details}: For the simulations, a Discrete Fourier Transform (DFT) based design matrix was used instead of a Gaussian one. This  enables the matrix-vector multiplications in the AMP decoder \eqref{eq:amp_decoder} to be computed via the Fast Fourier Transform (FFT), which significantly lowers the decoding complexity and memory requirement. 
The error performance of DFT based design matrices were found to be similar to that of Gaussian matrices for large matrix sizes. Our approach is similar to that of \cite{barbier2017approximate, rush2017capacity} where Hadamard-based design matrices were used for unmodulated real-valued SPARCs. 

The AMP decoder for SPARCs was run for a maximum of 100 iterations 
whereas the belief propagation decoder for LDPC codes was run for 50 iterations.
The AMP decoder used the online estimates of the state evolution parameters described in \eqref{eq:amp_decoder_gamma}--\eqref{eq:amp_decoder_tau}. 
Since those parameters are estimates of certain noise variances related to the decoding error in each iteration of the AMP, we chose to stop the AMP decoder early if the change in one of those estimates fell below a prescribed threshold over consecutive iterations.

The code used for the simulations is available at \cite{KuanPyScripts}. This public Github repository contains a Python implementation of modulated SPARCs (both real-valued and complex, power allocated and spatially coupled) with AMP decoding.

\emph{Decoding complexity}: The complexity of the AMP decoder is dominated by the two matrix-vector multiplications (which are replaced by FFTs) in \eqref{eq:amp_decoder}, and the $\eta$ function in \eqref{eq:eta_function}. The complexity of the FFTs are $O(\L\M\log(\L\M))$ and the complexity of the $\eta$ function is $O(\L\M\K)$. Therefore, the overall complexity per iteration is $O(\L\M (\log(\L\M) + \K))$.
In each set of simulations, as $K$ increases and $M$ decreases, with the product $KM$ kept constant, the overall decoding complexity decreases. 

To compare the decoding complexities of unmodulated and modulated SPARCs with $KM$ held fixed, denote the values of $M$ for the two cases by $\M_{\text{unmod}}$ and $\M_{\text{mod}}$, so that $\M_{\text{unmod}} = \K \M_{\text{mod}}$. Then 
the ratio of decoding complexities is
\begin{align}
&\frac{\text{complexity for unmodulated SPARC}}{\text{complexity for modulated SPARC}}\nonumber\\[0.4em]
&= \K \cdot \frac{\log(\L \M_{\text{unmod}}) + 1}{\log(\L \M_{\text{unmod}}) + \K - \log \K}.
\end{align}
If $\K \ll \log(\L\M)$, then modulation can reduce decoding complexity by nearly $\K$ times. For example, in Fig. \ref{fig:low_complex_R160_SC} the decoding complexity is reduced by approximately 3.8 times from $(\K=1, \M=128)$ to $(\K=4, \M=32)$ while the error performance improves.

The simulations above indicate that for a fixed code length $n$, rate $R$, and number of sections $\L$, modulation can significantly reduce decoding complexity without sacrificing error performance. 
Modulation also allows more flexibility in the code design of SPARCs.
For example, due to the rate equation $R =\frac{\L\ln(\K\M)}{n}$, for a fixed code length $n$, number of sections $\L$, and section size $\M$, one can increase the rate of the SPARC by increasing the modulation parameter $\K$, and this increase of $\K$ will not affect the decoding complexity if $\K \ll \log(\L\M)$.


\section{Proof of Proposition \ref{prop:se_psi_bound}}\label{sec:proof_prop_se_psi_bound}

We will prove the proposition assuming that $K$, the size of the PSK constellation, is a power of $2$. The proof can be extended to all integer values $K$ using additional technical arguments, which we omit. 

We first obtain the lower and upper bounds of $\nu_\sfc^t$ given in \eqref{eq:se_nu_lb_ub}.
To do this, we first show that the absolute value of $\mathcal{E}(\tau)$ defined in \eqref{eq:se_E} is bounded. Indeed,
\begin{align}
	&|\mathcal{E}(\tau)|\nonumber\\
	&\leq \mathbb{E}\left[ \left| \frac{\sum_{k=1}^{K} \Re[\p_k]\cdot e^{\frac{1}{\tau}\Re\left[(1+\sqrt{\tau}U_1)\, \overline{\p_k}\right]}}{\sum_{a=1}^{K} e^{\frac{1}{\tau}\Re\left[(1+\sqrt{\tau}U_1)\, \overline{\p_{a}}\right]} + \sum_{j=2}^{M}\sum_{b=1}^{K} e^{\frac{1}{\sqrt{\tau}} \Re[U_j \overline{\p_b}]}}\right| \right]\nonumber\\
	&\stackrel{\text{(i)}}{\leq}\mathbb{E}\left[ \frac{\sum_{k=1}^{K} e^{\frac{1}{\tau}\Re\left[(1+\sqrt{\tau}U_1)\, \overline{\p_k} \right]}}{\sum_{a=1}^{K} e^{\frac{1}{\tau}\Re\left[(1+\sqrt{\tau}U_1)\, \overline{\p_{a}}\right]} + \sum_{j=2}^{M}\sum_{b=1}^{K} e^{\frac{1}{\sqrt{\tau}} \Re[U_j \overline{\p_b}]}} \right]\nonumber\\
	&\leq 1, \label{eq:se_E_bound_pm1}
\end{align}
where (i) is obtained by $|\Re[\p_k]\,|\leq1$ for all $k\in[\K]$. Using the above result in \eqref{eq:se_psi} we deduce that 
\begin{equation}\label{eq:se_psi_lb_ub}
0 \leq \psi_\sfc^t \leq 2.
\end{equation}
Using this, we obtain the upper and lower bounds for $\nu_\sfc^t$:
\begin{align}
	\nu_\sfc^t 
	& \stackrel{\text{(i)}}{=}  \frac{2}{R\,\Lr}\sum_{\sfr=1}^{\Lr}\frac{W_{\sfr \sfc}}{\sigma^2 + \frac{1}{\Lc}\sum_{\sfc'} W_{\sfr \sfc'} \psi_{\sfc'}^t}
	\stackrel{\text{(ii)}}{\leq} \frac{2}{R\, \Lr}\sum_{\sfr=1}^{\Lr}\frac{W_{\sfr \sfc}}{\sigma^2}\nonumber\\
	 &\hspace{13.5em}\stackrel{\text{(iii)}}{\leq}  \frac{2}{R} \cdot \frac{\xi_2}{\sigma^2}, \label{eq:se_nu_ub}\\
	\nu_\sfc^t 
	&\stackrel{\text{(i)}}{=} \frac{2}{R\, \Lr}\sum_{\sfr=1}^{\Lr}\frac{W_{\sfr \sfc}}{\sigma^2 + \frac{1}{\Lc}\sum_{\sfc'} W_{\sfr \sfc'} \psi_{\sfc'}^t} 
	 \stackrel{\text{(ii)}}{\geq}  \frac{2}{R\, \Lr}\sum_{\sfr=1}^{\Lr}\frac{W_{\sfr \sfc}}{\sigma^2 + 2\xi_2} \nonumber\\
	&\hspace{13.5em}\stackrel{\text{(iii)}}{\geq} \frac{2}{R} \cdot \frac{\xi_1}{\sigma^2 + 2\xi_2}.
\end{align}
The labelled steps can be obtained as follows: (i) using the definition of $\nu_\sfc^t$ and $\phi_\sfr^t$ in \eqref{eq:se_nu} and \eqref{eq:se_phi}, (ii) using \eqref{eq:se_psi_lb_ub} and $\xi_1 \leq \frac{1}{\Lc} \sum_\sfc W_{\sfr\sfc} \leq \xi_2$, and (iii) using $\xi_1 \leq \frac{1}{\Lr} \sum_\sfr W_{\sfr\sfc} \leq \xi_2$. Note that $W_{\sfr\sfc} \geq 0$ for $\sfr\in[\Lr], \sfc\in[\Lc]$ and $\sigma^2 > 0$.


In the remainder of this proof, we obtain the upper and lower bounds on $\psi_\sfc^{t+1}$ given in \eqref{eq:psi_c_ub} and \eqref{eq:psi_c_lb}. 
We do this by obtaining lower and upper bounds on $\mathcal{E}(\tau_\sfc^t)$ since $\psi_\sfc^{t+1} = 1 - \mathcal{E}(\tau_\sfc^t)$.
We will often drop the dependencies on the column block index $\sfc$ and the iteration index $t$ in our notation for brevity, e.g., $\tau_\sfc^t$ will often be denoted by $\tau$.

We will first rewrite $\mathcal{E}(\tau)$ in order to simplify the notation, assuming that $K\geq 4$. 
The proofs for other cases ($K=1, 2$) use similar arguments and will be discussed afterwards.
Recalling the definition of $\mathcal{E}(\tau)$ in \eqref{eq:se_E} and letting
\be\label{eq:mu_def}
\mu \coloneqq \frac{1}{\tau} = \nu \ln(KM),
\ee
where the equality is obtained using the definition of $\nu_\sfc^t$ in \eqref{eq:se_nu},
we have
%
\begin{align}\label{eq:se_E_mu}
	&\mathcal{E}(\tau) = \nonumber\\
	&\mathbb{E}\left[\frac{\sum_{k=1}^{K} \Re[\p_k]\cdot e^{\mu \Re[\p_k] + \sqrt{\mu}\Re[U_1 \overline{\p_k}]}}{\sum_{a=1}^{K} e^{\mu \Re[\p_a] + \sqrt{\mu}\Re[U_1\overline{\p_a}]} + \sum_{j=2}^{M}\sum_{b=1}^{K} e^{\sqrt{\mu}\Re[U_j \overline{\p_b}]}}\right],
\end{align}
where $U_1, \ldots,U_M \stackrel{\text{i.i.d}}{\sim} \mathcal{CN}(0,2)$, and $\p_k = e^{\imag 2\pi k/\K}$, for $k \in [\K]$.  Furthermore, we note that the set of PSK symbols
$\{\p_k\}_{k=1,\ldots,\K}$ can equally be represented as $\{\p_k\}_{k=i,\ldots,i+\K-1}$ for any integer $i$, and we have $\p_i = \p_{i \!\!\mod \!K}$. We now introduce some notation to simplify \eqref{eq:se_E_mu}:
\begin{align}
	\cos_a &= \cos(2\pi a/K) = \Re[\p_a] \quad \text{for integer } a, \label{eq:se_new_not1}\\
	\sin_a  &= \sin(2\pi a/K)  = \Im[\p_a]  \quad \text{for integer } a,\\
	\cot_a  &= \cot(2\pi a/K)  \quad \text{for integer } a,\\
	U_j^R & = \Re[U_j] \quad \text{for} \ j=1,\ldots,M,\\
	U_j^I &= \Im[U_j] \quad \text{for} \ j=1,\ldots,M.
\end{align}
Using the above notation, we have
\begin{align}
	&\mathcal{E}(\tau)
	\stackrel{\text{(i)}}{=} \mathbb{E}\left[\frac{\sum_{k=1}^{K} \cos_k \cdot \, e^{\mu \cos_k + \sqrt{\mu}\Re\left[U_1 \overline{\p_k}\right]}}{X + \sum_{a=1}^{K} e^{\mu \cos_a + \sqrt{\mu}\Re\left[U_1 \overline{\p_a}\right]}}\right] \nonumber\\
	%
	%
	&\stackrel{\text{(ii)}}{=} \mathbb{E}\left[\frac{\sum_{k=-K/4+1}^{K/4}  \cos_k  \cdot \, 2 \sinh(\mu \cos_k + \sqrt{\mu}\Re\left[U_1 \overline{\p_k}\right])}{X + \sum_{a=-K/4+1}^{K/4} 2 \cosh(\mu \cos_a + \sqrt{\mu}\Re\left[U_1 \overline{\p_a}\right])}\right] \nonumber\\
	&\stackrel{\text{(iii)}}{=} \mathbb{E} \left[\frac{\sum_{k=-K/4+1}^{K/4} 2 \cos_k \cdot \sinh Y_k}{X + \sum_{a=-K/4+1}^{K/4} 2\cosh Y_a}\right] \nonumber\\
	&= \mathbb{E}_{U_1} \mathbb{E}_X\left[\frac{\sum_{k=-K/4+1}^{K/4} 2 \cos_k \cdot \sinh Y_k}{X + \sum_{a=-K/4+1}^{K/4} 2\cosh Y_a} \, \Bigg | \, U_1\right], \label{eq:se_E_nu_1}
\end{align}
where (i) is obtained by substituting
\begin{align}
	X  \coloneqq  \sum_{j=2}^{M}\sum_{b=1}^{K} e ^{\sqrt{\mu}\Re\left[U_j \overline{\p_b}\right]} 
	     = \sum_{j=2}^{M}\sum_{b=1}^{K} e ^{\sqrt{\mu}\left[U_j^R\cos_b+U_j^I\sin_b \right]},\label{eq:se_X_def}
\end{align}
(ii) is obtained using $\p_a = - \p_{(a+\frac{K}{2}) \!\!\!\mod \!K}$ and $\cos_a = - \cos_{(a+\frac{K}{2})\!\!\!\mod \!K}$, noting that $\K$ is a multiple of 4, 
and (iii) is obtained with the following substitutions: 
for $k=-\frac{\K}{4}+1, \ldots, \frac{\K}{4}$,
\begin{align}
	Y_k &=  \mu \cos_k + \sqrt{\mu}\Re\left[U_1 \overline{\p_k}\right]  \nonumber\\
	         &=  \mu \cos_k + \sqrt{\mu}\left[U_1^R\cos_k + U_1^I\sin_k \right]. \label{eq:se_Y_def}
\end{align}

Starting from \eqref{eq:se_E_nu_1}, 
we prove the required upper and lower bounds for $\psi_\sfc^{t+1}$ in the following two subsections.


\subsection{Proof of the upper bound \eqref{eq:psi_c_ub}}\label{sec:proof_prop_se_psi_bound_ub}

We obtain the required upper bound on $\psi_\sfc^{t+1}$ given in \eqref{eq:psi_c_ub} by obtaining a lower bound on $\mathcal{E}(\tau_\sfc^t)$ since $\psi_\sfc^{t+1} = 1 - \mathcal{E}(\tau_\sfc^t)$.
The result will first be proven for the $K\geq8$ case. The other cases ($K=1, 2, 4$) use similar arguments and will be discussed afterwards.

From \eqref{eq:se_E_bound_pm1}, we know that $-1 \leq \mathcal{E}(\tau) \leq 1$. Furthermore, the same arguments used to obtain \eqref{eq:se_E_bound_pm1} shows that the expectation over $X$ in \eqref{eq:se_E_nu_1} is bounded as follows,
\begin{equation}
	-1 \leq \mathbb{E}_X\left[\frac{\sum_{k=-K/4+1}^{K/4} 2 \cos_k \cdot \sinh Y_k}{X + \sum_{a=-K/4+1}^{K/4} 2\cosh Y_a} \, \Bigg | \, U_1\right] \leq 1.\label{eq:se_Ex_giv_U_bound}
\end{equation}

To lower bound $\mc{E}(\tau)$, we first identify when the expression inside the inner expectation  in \eqref{eq:se_E_nu_1} is non-negative. Observe that $\cos_k$, $X$, and $\cosh Y_a$ are all non-negative for the values of $a$ and $k$ being considered, with $\cos_k= 0$ when $k=\frac{K}{4}$.
Furthermore, from the definition of $Y_k$ in \eqref{eq:se_Y_def}, for $k\in\{-K/4+1,\ldots, K/4-1\}$, we have $\sinh Y_k \geq 0$ if and only if
\be\label{eq:se_E_postive_ineq}
	U_1^R\cos_k + U_1^I \sin_k \geq - \sqrt{\mu} \cos_k.
\ee
 It can be easily verified that a sufficient condition for \eqref{eq:se_E_postive_ineq} to hold for $k\in\{-K/4+1,\ldots, K/4-1\}$ is that both $U_1^R$ and $U_1^I$ are greater than or equal to $\frac{-\sqrt{\mu}}{1 + \tan(\frac{\pi}{2} - \frac{2\pi}{K})}$.


Using this knowledge, we can split the expectation over $U_1$ in \eqref{eq:se_E_nu_1} into integrals over four regions such that in at least one of the regions the integrand is non-negative. For any $\underline{u} <0$, we have
\begin{align}
	&\mathcal{E}(\tau) = \nonumber\\
	&\int_{\underline{u}}^{\infty} \! \int_{\underline{u}}^{\infty} \!\phi(u^R)  \phi(u^I) \, \mathbb{E}_X \left[\ldots  \mid  U_1^R=u^R, U_1^I=u^I \right] du^R  du^I \nonumber \\
	&\qquad + \int_{-\infty}^{\underline{u}} \int_{-\infty}^{\underline{u}} \phi(u^R)  \phi(u^I) \, \mathbb{E}_X [\ldots ] \, du^R  du^I \nonumber\\
	&\qquad + \int_{\underline{u}}^{\infty} \int_{-\infty}^{\underline{u}} \phi(u^R)  \phi(u^I) \, \mathbb{E}_X [\ldots ] \, du^R  du^I \nonumber \\
	&\qquad + \int_{-\infty}^{\underline{u}} \int_{\underline{u}}^{\infty} \phi(u^R)  \phi(u^I) \, \mathbb{E}_X [\ldots ] \, du^R  du^I \nonumber \\
	&\stackrel{\text{(i)}}{=}  I_1 - Q(|\underline{u}|)^2 - 2[1 - Q(|\underline{u}|)] \,Q(|\underline{u}|) \nonumber\\
	&\geq I_1 - 2 Q(|\underline{u}|), \label{eq:se_E_geq_I1_2Q}
\end{align}
where $\phi(\cdot)$ denotes the standard Gaussian density and $I_1$ is the integral over the first region. Step (i) is obtained using $\mathbb{E}_X [\ldots]\geq -1$ from \eqref{eq:se_Ex_giv_U_bound}, 
using $U_1^R, U_1^I \stackrel{\text{i.i.d}}{\sim}\mathcal{N}(0, 1)$, 
defining  $Q(x) = \int_{x}^{\infty} \frac{1}{\sqrt{2 \pi}} e^{-z^2/2}\, dz$ to be the upper tail probability of the standard Gaussian distribution,
and using $1-Q(|x|) = Q(x)$ for $x<0$.

From  \eqref{eq:se_E_postive_ineq} and the discussion surrounding it, when $\underline{u} = \frac{-\sqrt{\mu}}{1 + \tan(\frac{\pi}{2} - \frac{2\pi}{K})} = \frac{-\sqrt{\mu}}{1 + \cot(\frac{2\pi}{K})} $, the integrand of $I_1$  is non-negative, and we therefore have the following lower bound for $\mathcal{E}(\tau)$:
\begin{align}
	\mathcal{E}(\tau) 
	\geq - 2 Q(|\underline{u}|) 
	\geq - \frac{2 (1+\cot(\frac{2\pi}{K}))}{\sqrt{2\pi\nu\ln(KM)}} \,  (KM)^{-\frac{\nu}{2 (1+\cot(\frac{2\pi}{K}))^2}}, \label{eq:se_E_lb_all_nu}
\end{align}
where the second inequality is obtained using \eqref{eq:mu_def} and the bound $Q(x) \leq \frac{1}{x\sqrt{2\pi}} e^{-x^2/2}$ for $x>0$.

The lower bound of $\mathcal{E}(\tau)$ in \eqref{eq:se_E_lb_all_nu} applies for all $\nu>0$. We now show that for $\nu>2$, one can obtain a better lower bound which shows that $\mathcal{E}(\tau)$ approaches the upper bound of 1 with growing $\M$. We do this by using a different choice for $\underline{u}$ to split the expectation over $U_1$ in  \eqref{eq:se_E_nu_1} into integrals over four different regions. Let
\be\label{eq:u_underline}
\underline{u} 
= \frac{-\alpha(\frac{\nu}{2}-1)}{\nu} \frac{\sqrt{\mu}}{1 + \cot(\frac{2\pi}{K})} \,.
\ee
Then, for any $\alpha \in (0,1)$ and $\nu > 2$, when $U_1^R\geq \underline{u}$ and $U_1^I\geq \underline{u}$ we have
\begin{align}
	U_1^R\cos_k + U_1^I \sin_k 
	 &\geq  -\sqrt{\mu}  \cos_k  \frac{\alpha (\frac{\nu}{2} -1) }{\nu} \, \frac{1 + \tan(\frac{2\pi k}{\K})}{1 + \tan(\frac{\pi}{2} - \frac{2\pi}{K})} \nonumber\\
	 &\geq  -\sqrt{\mu}  \cos_k , 
	  \label{eq:se_E_postive_ineq2}
\end{align}
for $\frac{-K}{4}+1 \leq k \leq \frac{K}{4}-1$. Thus, under these conditions, \eqref{eq:se_E_postive_ineq} holds and the integrand of $I_1$ in \eqref{eq:se_E_geq_I1_2Q} is non-negative.
In the following lemma, we obtain a stronger lower bound on $I_1$ for $\nu >2$.

\begin{lemma}\label{lem:I1_lb}
When $\underline{u} = \frac{-\alpha(\frac{\nu}{2}-1)}{\nu} \frac{\sqrt{\mu}}{1 + \cot(\frac{2\pi}{K})}$
for any $\alpha \in (0,1)$, $\nu>2$ and $\K\geq 4$, the $I_1$ term in \eqref{eq:se_E_geq_I1_2Q} can be lower bounded as follows:
\begin{align}
I_1 &\geq 1 - 3Q(|\underline{u}|) - 2(KM)^{-2z} \nonumber\\
      &\hspace{1.9em}	 - (KM)^{1+\frac{\nu}{2} - z} - (\K - 2) (KM)^{-(z-z^{\star})}, \label{eq:se_E_I1_lb3}
\end{align}
where
\begin{align}
z &= \nu - \frac{\alpha(\frac{\nu}{2}-1)}{1+\cot(\frac{2\pi}{\K})}, \label{eq:se_I1_z_def1}\\
z^\star &= z \cos \Big(\frac{2\pi}{\K}\Big) + (\nu-z) \sin\Big(\frac{2\pi}{\K}\Big). \label{eq:se_I1_z_star1}
\end{align}
\end{lemma}
The proof is given in Appendix \ref{appendix:proof_I1_lb}.


Applying the result of Lemma \ref{lem:I1_lb} in \eqref{eq:se_E_geq_I1_2Q} 
and
using the notation $\cos_1 = \cos(2\pi/K)$, $\sin_1 = \sin(2\pi/K)$, and $\cot_1 = \cot(2\pi/K)$,
we have, for any $\alpha \in (0,1)$ and $\nu>2$,
\begin{align}
	&\mathcal{E}(\tau) 
	\geq  1 - 5Q(|\underline{u}|) - 2(KM)^{-2z} - (KM)^{1+\frac{\nu}{2} - z} \nonumber\\
	&\hspace{14.2em} - (\K - 2) (KM)^{-(z-z^{\star})}\nonumber\\
	&\stackrel{\text{(i)}}{=}
		1 - 5Q(|\underline{u}|)
		- 2(KM)^{-2 \big(\nu - \frac{\alpha(\frac{\nu}{2}-1)}{1+\cot_1}\big)} \nonumber\\
		&\hspace{2em}- (KM)^{-\big(1-\frac{\alpha}{1+\cot_1}\big)(\frac{\nu}{2}-1)} \nonumber\\
		&\quad\quad - (\K - 2) (KM)^{-\big(\nu - \frac{\alpha(\frac{\nu}{2}-1)}{1+\cot_1}\big)\left(1-\cos_1\right)+ \frac{\alpha(\frac{\nu}{2}-1)}{1 + \cot_1}  \sin_1} \nonumber\\
	&\geq
		1 - 5Q(|\underline{u}|)
		- (KM)^{-\big(1-\frac{\alpha}{1+\cot_1}\big)(\frac{\nu}{2}-1)}\nonumber\\
		&\hspace{2em}- \K  (KM)^{-\big(\nu - \frac{\alpha(\frac{\nu}{2}-1)}{1+\cot_1}\big)\left(1-\cos_1\right)+ \frac{\alpha(\frac{\nu}{2}-1)}{1 + \cot_1}  \sin_1} \nonumber\\
	&\stackrel{\text{(ii)}}{\geq}
		1 - \frac{5\sqrt{\nu} (1+\cot_1) (KM)^{- \frac{\alpha^2(\frac{\nu}{2}-1)^2}{2\nu(1+\cot_1)^2}}}{(\nu/2-1)\alpha \sqrt{2\pi \ln(KM)} } \nonumber\\
		&\quad - (KM)^{-\big(1-\frac{\alpha}{1+\cot_1}\big)(\frac{\nu}{2}-1)} \nonumber\\
		&\quad - \K  (KM)^{ -  (1 - \cos_1) \left[ \nu (1 - \frac{\alpha}{2} (1 + \frac{1}{\sin_1 + \cos_1}) ) + \alpha (1 + \frac{1}{\sin_1 + \cos_1} ) \right]}, \label{eq:se_E_nu2_lb}
\end{align}
where (i) is obtained using the substitutions \eqref{eq:se_I1_z_def1} and \eqref{eq:se_I1_z_star1}, and (ii) from using the bound $Q(x) \leq \frac{1}{x\sqrt{2\pi}} e^{-x^2/2}$ for $x>0$ (noting that $\underline{u} <0$ for $\nu > 2$), and rearranging the exponent of the last term as follows:
\begin{align*}
& -\bigg(\nu - \frac{\alpha(\frac{\nu}{2}-1)}{1+\cot_1}\bigg)\left(1-\cos_1 \right)+ \frac{\alpha(\frac{\nu}{2}-1)}{1 + \cot_1}  \sin_1\nonumber \\
&  = - (1- \cos_1) \bigg[ \nu - \frac{\alpha(\nu/2-1)}{1+\cot_1} - \frac{ \alpha( \nu/2-1)}{1+\cot_1} \frac{\sin_1}{1- \cos_1} \bigg] \\
& = - (1- \cos_1)\bigg[ \nu \bigg( 1- \frac{\alpha/2}{(1+\cot_1)} \Big(1 + \frac{\sin_1}{1-\cos_1} \Big) \bigg) \nonumber\\
&\hspace{11em} + \frac{\alpha}{(1+\cot_1)} \Big(1 + \frac{\sin_1}{1-\cos_1} \Big)  \bigg] \\
& = - (1- \cos_1)\bigg[ \nu \bigg( 1- \frac{\alpha}{2} \Big(1 + \frac{1}{\sin_1+\cos_1} \Big) \bigg) \nonumber\\
&\hspace{14em}+ \alpha \Big(1 + \frac{1}{\sin_1+\cos_1} \Big)  \bigg].
\end{align*}

For any $\delta \in (0,\frac{1}{2})$, we consider the case $\nu > 2 + \delta$ and choose $\alpha = 1-\delta$.  With this choice, the exponent of the last term in \eqref{eq:se_E_nu2_lb} can be bounded as follows:
\begin{align}
& \nu \Big[1 - \frac{\alpha}{2} \big(1 + (\sin_1 + \cos_1)^{-1} \big) \Big] 
 + \alpha \big(1 + (\sin_1 + \cos_1)^{-1}\big)  \nonumber\\
& > 2 + \delta\left[ 1 - \frac{\alpha}{2}\left( 1 + (\sin_1 + \cos_1)^{-1} \right) \right] \nonumber \\
& \geq 2 + \delta(1-\alpha) = 2+ \delta^2, \label{eq:last_term_lb}
\end{align}
where the second inequality holds because $\sin(\frac{2\pi}{\K}) + \cos(\frac{2\pi}{\K}) \geq 1$ for $\K \geq 8$. 

Using  \eqref{eq:last_term_lb}  in \eqref{eq:se_E_nu2_lb} along with $\frac{\nu}{2}-1 > \frac{\delta}{2}$ and  $\alpha = 1-\delta$, we obtain the following bound on $\mathcal{E}(\tau)$ for sufficiently large $\M$ (dropping constant terms):
\begin{align}
	\mathcal{E}(\tau) & \geq
	1 - \frac{10\sqrt{\nu}\, (1+\cot(\frac{2\pi}{\K})) (KM)^{- \frac{(1-\delta)^2 \delta^2 }{8\nu (1+\cot(\frac{2\pi}{\K}))^2}}}{(1-\delta)\, \delta \sqrt{2\pi \ln(KM)}}\nonumber\\
	&\hspace{1.8em} - (KM)^{-\frac{\delta(\delta + \cot(\frac{2\pi}{\K}))}{2(1+\cot(\frac{2\pi}{\K}))}} 
	- K (KM)^{- (2+\delta^2)(1-\cos(\frac{2\pi}{K}))} \nonumber\\
	&\geq 1 - \frac{(1+\cot(\frac{2\pi}{\K})) (KM)^{- \frac{\kappa \delta^2}{(1+\cot(\frac{2\pi}{\K}))^2}} }{\delta \sqrt{\ln(KM)}}\nonumber\\
	&\hspace{1.8em} - K (KM)^{- 2(2+\delta^2) \sin^2(\frac{\pi}{\K})}, \label{eq:se_E_lb_final}
\end{align}
where $\kappa < \frac{(1-\delta)^2}{8\nu}$ is a suitably chosen universal positive constant.
For the second inequality we used $1-\delta > \frac{1}{2}$ and the fact that $\nu$ can be upper bounded by a positive constant \eqref{eq:se_nu_ub}. Comparing the exponents of the last two terms of \eqref{eq:se_E_lb_final}, using $\kappa < \frac{(1-\delta)^2}{8\nu}$, 
$\delta \in (0, \frac{1}{2})$ and $\nu >2$, we have
\begin{align}
	\frac{\kappa \delta^2}{(1+\cot(\frac{2\pi}{\K}))^2}
	&<\frac{(1-\delta)^2 \delta^2 }{8\nu (1+\cot(\frac{2\pi}{\K}))^2}
	< \frac{1}{256 \, (1+\frac{\cos_1}{\sin_1})^2} \nonumber\\
	&= \frac{(1-\cos_1)(1+\cos_1)}{256 \, (\sin_1+\cos_1)^2} 
	< \frac{2 (1-\cos_1)}{256}\nonumber \\
	&< \frac{2 (2+\delta^2)\sin^2({\pi}/{\K})}{256},\nonumber
\end{align}
where we have used the notation $\cos_1 = \cos(2\pi/K)$ and $\sin_1 = \sin(2\pi/K)$. The second last inequality above is  obtained using $\cos_1<1$, and $(\sin_1+\cos_1)^2 = [\sqrt{2} \sin(\frac{\pi}{4} + \frac{2\pi}{K})]^2 \geq 1$ for $K\geq8$. Therefore, the exponent of third term is more than 256 times larger than that of second term (in absolute value).

Summing up, we now have two lower bounds on $\mathcal{E}(\tau)$ for different values of $\nu$. When $\nu \leq 2+\delta$ we have \eqref{eq:se_E_lb_all_nu}, and when $\nu > 2+\delta$ we have \eqref{eq:se_E_lb_final}. By applying them to the state evolution equation $\psi_\sfc^{t+1} = 1 - \mathcal{E}(\tau_\sfc^t)$, we obtain the required result for the $\K \geq 8$ case. The results for the other cases ($\K=1,2,4$) use similar arguments and are explained below.

\textbf{$\boldsymbol{K=1}$:}
There is only one constellation symbol $\p_0 = 1$ which has no imaginary part. Therefore, all imaginary parts in the expression of $\mathcal{E}(\tau)$ disappear as shown below.
\begin{equation}\label{eq:se_E_K1}
	\mathcal{E}(\tau) = \mathbb{E}\left[\frac{e^{\nu\ln(M)  + \sqrt{\nu\ln(M)} U_1^R }}{e^{\nu\ln(M) + \sqrt{\nu\ln(M)}U_1^R} + \sum_{j=2}^{M} e^{\sqrt{\nu\ln(M)}U_j^R}}\right], \nonumber
\end{equation}
where $U_1^R,\ldots, U_M^R \stackrel{\text{i.i.d.}}{\sim}\stdnorm$. The above expression is the same as that given in \cite[(A.2)]{rush2019theerror}. Therefore, we can obtain the result from following the steps in \cite[Appendix~A]{rush2019theerror}.

\textbf{$\boldsymbol{K=2}$:}
In this case there are two constellation symbols $\p_0=1$ and $\p_1=-1$, which are both real. 
We follow steps similar to those used to obtain \eqref{eq:se_E_geq_I1_2Q} in the $\K\geq8$ case to obtain the following lower bound on $\mathcal{E}(\tau)$ for any $\underline{u} \leq 0$
\begin{align}
\mathcal{E}(\tau)
	&= \mathbb{E} \left[
		\frac{ 2\sinh ( \mu + \sqrt{\mu} U_1^R )}{2\cosh( \mu + \! \sqrt{\mu}  U_1^R ) + \sum_{j=2}^{M} (e^{\sqrt{\mu}  U_j^R} \! +  e^{-\sqrt{\mu} U_j^R})}\right] \nonumber\\
	&= \mathbb{E}_{U_1^R} \mathbb{E}_X\bigg[\frac{ 2\sinh( \mu + \sqrt{\mu} \, U_1^R )}{2\cosh( \mu + \sqrt{\mu} \, U_1^R ) + X}  \bigg|  U_1^R\bigg] \nonumber\\
	&=\int_{\underline{u}}^{\infty} \phi(u) \, \mathbb{E}_X \left[ \, \ldots \, | \, U_1^R=u \right] du \nonumber\\
	 &\hspace{8.2em} + \int_{-\infty}^{\underline{u}} \phi(u) \, \mathbb{E}_X \left[\, \ldots \, | \, U_1^R=u \right] du \nonumber\\
	&\stackrel{\text{(i)}}{\geq} \int_{\underline{u}}^{\infty} \phi(u) \, \mathbb{E}_X \left[ \, \ldots \, | \, U_1^R=u \right] du 
	+ \int_{-\infty}^{\underline{u}} \phi(u) (-1)  du \nonumber\\
	&\stackrel{\text{(ii)}}{=} I_1 - Q(|\underline{u}|), \label{eq:se_E_equals_I1_I2_K2}
\end{align}
where $\mu = \nu\ln(2M)$, 
$X = \sum_{j=2}^{M} (e^{\sqrt{\mu} \, U_j^R} + e^{-\sqrt{\mu} \,U_j^R})$,
and $I_1$ is the integral from $\underline{u}$ to $\infty$.
Step (i) is obtained using $|\frac{ 2\sinh(x)}{2\cosh(x) + X}| \leq 1$ for any $x$, and
step (ii) holds since $\underline{u} \leq 0$.

Similar to what was done in \eqref{eq:se_E_lb_all_nu} for the $\K\geq 8$ case, we choose $\underline{u} = -\sqrt{\mu} = -\sqrt{\nu \ln(2\M)}$ so that the integrand of $I_1$  is non-negative, and obtain the following lower bound on $\mathcal{E}(\tau)$ for all $\nu>0$.
\begin{align}
	\mathcal{E}(\tau) 
	&\stackrel{\text{(i)}}{\geq} - Q\left(\sqrt{\nu\ln(2M)}\right) 
	 \stackrel{\text{(ii)}}{\geq} \ - \frac{(2M)^{-\frac{\nu}{2}}}{\sqrt{2\pi \nu\ln(2M)}}, \label{eq:E_lb_allnu_K2}
\end{align}
where the labelled inequalities are obtained as follows: 
(i) $I_1\geq 0$ when $\underline{u} = -\sqrt{\mu}$, and
(ii) using the bound on the tail probability of a standard normal.

The lower bound on $\mathcal{E}(\tau)$ in \eqref{eq:E_lb_allnu_K2} applies for all values of $\nu>0$. To obtain a better bound for $\nu > 2$ which shows that $\mathcal{E}(\tau)$ approaches 1 with growing $\M$,  we choose
\be\label{eq:underline_u_nu2_K2}
\underline{u} 
=  \frac{-\alpha(\frac{\nu}{2}-1)}{\nu} \sqrt{\mu} 
= - \alpha \Big(\frac{\nu}{2}-1\Big)\sqrt{\frac{\ln(2M)}{\nu}}
\ee
for any $\alpha \in (0,1)$, and then using this choice of $\underline{u}$ in \eqref{eq:se_E_equals_I1_I2_K2} to obtain a lower bound for $I_1$.

We obtain the following lower bound on $I_1$ by following similar steps to the proof of Lemma \ref{lem:I1_lb} given in Appendix \ref{appendix:proof_I1_lb}. Recall that Lemma \ref{lem:I1_lb} obtains a lower bound on $I_1$ for $\nu>2$ in the $\K\geq4$ case.

\begin{align}
	I_1
	\geq 1 - Q(|\underline{u}|)
		- 2\, (2M)^{-((2-\alpha)\nu+2\alpha)} - (2M)^{-(1- \alpha)(\frac{\nu}{2}-1)}. \label{eq:E_I1_lb_K2}
\end{align}
Using \eqref{eq:E_I1_lb_K2} in \eqref{eq:se_E_equals_I1_I2_K2}, we have the following lower bound on $\mathcal{E}(\tau)$ for $\nu>2$ and $\underline{u}$ taking the value in \eqref{eq:underline_u_nu2_K2}.
\begin{align}\label{eq:se_E_lb_nu2_K2}
	\mathcal{E}(\tau)
	&\geq 1 - \frac{2\sqrt{\nu} (2M)^{-\frac{\alpha^2 (\frac{\nu}{2} -1)^2}{2\nu }}}{\alpha (\frac{\nu}{2} -1)\sqrt{2\pi \ln(2M)}} \nonumber\\
		&\hspace{1.8em} - 2 (2M)^{-((2-\alpha)\nu+2\alpha)} - (2M)^{-(1- \alpha)(\frac{\nu}{2}-1)},
\end{align}
where we used the bound $Q(x) \leq \frac{1}{x\sqrt{2\pi}} e^{-x^2/2}$ for $x>0$.

For any $\delta \in (0,\frac{1}{2})$, we consider the case $\nu > 2 + \delta$ and choose $\alpha=1-\delta$. Plugging these into \eqref{eq:se_E_lb_nu2_K2}, we have
\begin{align}
	\mathcal{E}(\tau)
	&\geq
	1 - \frac{4\sqrt{\nu} (2M)^{- \frac{\delta^2 (1-\delta)^2 }{8\nu}}}{\delta(1-\delta)\sqrt{2\pi \ln(2M)}} 
	- 2(2M)^{-(4+\delta + \delta^2)} \nonumber\\
	&\hspace{12.3em}- (2M)^{-\frac{\delta^2}{2}}\nonumber\\
	&\geq 1 - \frac{(2M)^{- \kappa_2 \delta^2}}{\delta\sqrt{\ln(2M)}}, \label{eq:se_E_lb_final_K2}
\end{align}
for sufficiently large $\M$ and a suitably chosen universal positive constant $\kappa_2$. The second inequality is obtained by noting that $1-\delta > \frac{1}{2}$ and that $\nu$ can be upper bounded by a positive constant \eqref{eq:se_nu_ub}.

By applying the two lower bounds \eqref{eq:E_lb_allnu_K2} and \eqref{eq:se_E_lb_final_K2} to the state evolution equation $\psi_\sfc^{t+1} = 1 - \mathcal{E}(\tau_\sfc^t)$ for the $\nu\leq 2+\delta$ and $\nu>2+\delta$ case respectively, we obtain the required result for $\K = 2$.

\textbf{$\boldsymbol{K=4}$:}
The arguments for this case are essentially the same as that for the $\K\geq 8$ case, noting that $\tan(\frac{\pi}{2} - \frac{2\pi}{\K}) = \cot(\frac{2\pi}{\K}) = 0$ and $\cos(\frac{2\pi}{\K})=0$ for $\K=4$.

By following the same arguments from \eqref{eq:mu_def}--\eqref{eq:se_E_lb_all_nu}, we obtain the following lower bound for $\mathcal{E}(\tau)$ for any $\nu>0$,
\begin{equation}
	\mathcal{E}(\tau) \geq  \frac{-2(4M)^{-\frac{\nu}{2}}}{\sqrt{2\pi\nu\ln(4M)}}.\label{eq:se_E_lb_all_nu_K4}
\end{equation}

To obtain the improved lower bound for the $\nu > 2$ case, we follow the same arguments from \eqref{eq:se_E_postive_ineq2} to \eqref{eq:se_E_lb_final} and arrive at the following. For any $\delta \in (0,\frac{1}{2})$ and $\nu > 2 + \delta$, we have
\begin{align}
	&\mathcal{E}(\tau) \nonumber\\
	&\geq
	1 - \frac{10\sqrt{\nu} (4M)^{-\frac{(1-\delta)^2 \delta^2 }{8\nu}}}{(1 - \delta)\delta \sqrt{2\pi \ln(4M)}}  
	- (4M)^{-\frac{\delta^2}{2}} 
	- 4(4M)^{- 2 - \delta^2 }\nonumber\\
	&\geq 
	1 - \frac{(4M)^{- \kappa_3 \delta^2}}{\delta \sqrt{\ln(4M)}}, \label{eq:se_E_lb_final_K4}
\end{align}
for sufficiently large $\M$ and a suitably chosen universal positive constant $\kappa_3$. The last inequality is obtained by noting that $1-\delta > \frac{1}{2}$ and that $\nu$ can be upper bounded by a positive constant \eqref{eq:se_nu_ub}.

By applying the two lower bounds \eqref{eq:se_E_lb_all_nu_K4} and \eqref{eq:se_E_lb_final_K4} to the state evolution equation $\psi_\sfc^{t+1} = 1 - \mathcal{E}(\tau_\sfc^t)$ for the $\nu\leq 2+\delta$ and $\nu>2+\delta$ case respectively, we obtain the required result for $\K = 4$. 


\subsection{Proof of lower bound \eqref{eq:psi_c_lb}}\label{sec:proof_prop_se_psi_bound_lb}

We obtain the required lower bound on $\psi_\sfc^{t+1}$ given in \eqref{eq:psi_c_lb} by obtaining an upper bound on $\mathcal{E}(\tau_\sfc^t)$ since $\psi_\sfc^{t+1} = 1 - \mathcal{E}(\tau_\sfc^t)$.
The result will be proven for the $K\geq4$ case. The $\K=1$ case was proved in \cite[Appendix~A]{rush2020capacity}.
The $\K=2$ case uses similar arguments as the $\K\geq 4$ case, so we focus on the latter.

We will use the following concentration inequality for the maximum of $N$ \iid\ standard Gaussian random variables $Z_1,\ldots, Z_N$.
For any $\e\in(0,1)$, 
\be\label{eq:max_gauss_bound}
\prob\left(\max_{1\leq j\leq N} Z_j < \sqrt{2\ln N}(1-\e)\right) 
\leq \exp\bigg(\frac{-\lambda N^{\e(2-\e)}}{\sqrt{\ln N}} \bigg),
\ee
where $\lambda$ is a universal positive constant.

Using the same steps and notations in \eqref{eq:mu_def}--\eqref{eq:se_Y_def}, 
we have that
\be
\mathcal{E}(\tau)
= \mathbb{E}_{U_1} \mathbb{E}_X\left[\frac{\sum_{k=-K/4+1}^{K/4} 2 \cos_k \cdot \sinh Y_k}{X + \sum_{a=-K/4+1}^{K/4} 2\cosh Y_a} \, \Bigg | \, U_1\right]. \label{eq:se_E_nu_1_1}
\ee

We will obtain an upper bound for $\mathcal{E}(\tau)$ 
using the fact that the term inside the expectation in \eqref{eq:se_E_nu_1_1} is strictly increasing in $U_1^R$ (the real part of $U_1\sim \mc{CN}(0,2)$). 
(Recall from \eqref{eq:se_Y_def} that $Y_k$ is a function of $U_1^R$ for each $k$.) 
To see that the term is increasing in $U_1^R$, we write 
\be
f \coloneqq \frac{\sum_{k=-K/4+1}^{K/4} 2 \cos_k \cdot \sinh Y_k}{X + \sum_{a=-K/4+1}^{K/4} 2\cosh Y_a} = \frac{f_\text{num}}{f_\text{den}},
\ee
and show that $\partial f/\partial U_1^R > 0$. Since
\be
\frac{\partial f}{\partial U_1^R}
= \Big(\frac{\partial f_\text{num}}{\partial U_1^R} \cdot f_\text{den} - f_\text{num} \cdot \frac{\partial f_\text{den}}{\partial U_1^R} \Big) \Big/ \Big(f_\text{den}^2\Big),
\ee
we show below  that $\frac{\partial f_\text{num}}{\partial U_1^R} \cdot f_\text{dem} - f_\text{num} \cdot \frac{\partial f_\text{dem}}{\partial U_1^R} > 0$. Indeed,
\begin{align}
	&\frac{\partial f_\text{num}}{\partial U_1^R} \cdot f_\text{den} - f_\text{num} \cdot \frac{\partial f_\text{den}}{\partial U_1^R} \nonumber \\
	&=(\sum_{k} 2\sqrt{\mu}  \cos_k^2   \cosh Y_k)
	(X + \sum_{a} 2\cosh Y_a) \nonumber\\
	&\hspace{2em}- (\sum_{k} 2\cos_k  \sinh Y_k )
	(\sum_{a} 2\sqrt{\mu}\cos_a   \sinh Y_a)\nonumber\\
	&= 4\sqrt{\mu} \bigg[
	\frac{X}{2} (\sum_{k} \cos_k^2  \cosh Y_k )\nonumber\\
	&\hspace{4em}+ (\sum_{k} \cos_k^2  \cosh Y_k )(\sum_{a} \cosh Y_a)\nonumber\\
	&\hspace{12.5em}- (\sum_{k} \cos_k  \sinh Y_k )^2 \bigg] \nonumber\\
	&\stackrel{\text{(i)}}{\geq} 4\sqrt{\mu} \bigg[
	\frac{X}{2} (\sum_{k} \cos_k^2  \cosh Y_k)
	+ (\sum_{k} \cos_k  \cosh Y_k )^2\nonumber\\
	&\hspace{12.5em}- (\sum_{k} \cos_k  \sinh Y_k )^2 \bigg] \nonumber\\
	&\stackrel{\text{(ii)}}{>}  4\sqrt{\mu} \cdot \frac{X}{2} 
	(\sum_{k} \cos_k^2  \cosh Y_k ) 
	\, > \, 0, \nonumber
\end{align}
where (i) is obtained using the Cauchy-Schwarz inequality and (ii) from $x^2 - y^2=(x+y)(x-y)$ and $\cosh(x) > |\sinh(x)|$ for all $x$.

Since the term inside the expectation in \eqref{eq:se_E_nu_1_1} is strictly increasing in $U_1^R$
and upper bounded by 1 (see \eqref{eq:se_E_bound_pm1}), 
we have the following upper bound on $\mathcal{E}(\tau)$.
Recall that $U_1^R, U_1^I$ $\stackrel{\iid}{\sim}\stdnorm$
and that $\tilde{\delta}\in (0,1)$ is the constant defined in Proposition \ref{prop:se_psi_bound}.
Let $\tilde{u} >0$ and $\alpha \in (0,1)$ be deterministic scalars to be specified later.
The summations over $k$ and $a$ below are from $(-\frac{K}{4}+1)$ to $\frac{K}{4}$.
\begin{align}
&\mathcal{E}(\tau) 
 \stackrel{\text{(i)}}{\leq}  \prob(U_1^R > \tilde{u}) \times 1
+\prob(U_1^R \leq \tilde{u}) \ \times \nonumber\\
&\hspace{1.25em}  \expec_{U_1^I}\mathbb{E}_X\left[\frac{\sum_{k} 2 \cos_k \, \sinh ( (\mu + \sqrt{\mu}\tilde{u})\cos_k + \sqrt{\mu}U_1^I \sin_k)}{X + \sum_{a} 2\cosh ( (\mu + \sqrt{\mu}\tilde{u})\cos_a + \sqrt{\mu}U_1^I \sin_a)} \right]\nonumber\\
& \stackrel{\text{(ii)}}{\leq}  Q(\tilde{u}) \nonumber\\
	&\hspace{1em}+ \mathbb{E} \Bigg[\frac{\sum_{k} 2 \cos_k \, \sinh ( \ldots)}{\sum_{j=2}^{M}\sum_{b=1}^{K} e^{\sqrt{\mu}\left[U_j^R\cos_b+U_j^I\sin_b \right]} + \sum_{a} 2\cosh (\ldots)} \Bigg]\nonumber\\
& \stackrel{\text{(iii)}}{\leq}  Q(\tilde{u}) \nonumber\\
	&\quad+ \mathbb{E} \left[\frac{\sum_{k} 2 \cos_k \, \sinh ( \ldots)}{\big(\max_{2\leq j \leq \M} e^{\sqrt{\nu\ln(\K\M)} U_j^R}\big) + \sum_{a} 2\cosh (\ldots)} \right]\nonumber\\
&\leq  Q(\tilde{u}) 
+ \prob\Big( \max_{2\leq j \leq \M} U_j^R < \sqrt{2\ln \M} (1-\alpha\tilde{\delta}) \Big) \times 1 \nonumber\\
&\hspace{3.5em}+ \prob\Big( \max_{2\leq j \leq \M} U_j^R \geq \sqrt{2\ln \M} (1-\alpha\tilde{\delta}) \Big) \, \times \nonumber\\
&\hspace{8em}   \mathbb{E}_{U_1^I}\left[\frac{\sum_{k} 2 \cos_k \, \sinh ( \ldots)}{ {\M}^{\sqrt{2\nu}(1-\alpha\tilde{\delta})} + \sum_{a} 2\cosh (\ldots)} \right]\nonumber\\
& \stackrel{\text{(iv)}}{\leq}  Q(\tilde{u}) 
+ e^{-\lambda (\sqrt{\ln \M})^{-1} \M^{\alpha\tilde{\delta}(2-\alpha\tilde{\delta})}} + \nonumber\\
&\hspace{0.5em}
\mathbb{E}\Bigg[\frac{\sum_{k} 2 \cos_k \, \sinh ( (\mu \! + \! \sqrt{\mu}\tilde{u})\! \cos_k + \sqrt{\mu} U_1^I \! \sin_k)}{ {\M}^{\sqrt{2\nu}(1-\alpha\tilde{\delta})} +  \sum_{a}  2 \cosh ((\mu \!+ \! \sqrt{\mu}\tilde{u}) \! \cos_a \!+ \sqrt{\mu} U_1^I \! \sin_a)} \Bigg]\label{eq:se_E_ub_0}
\end{align}
where 
$Q(x) = \int_{x}^{\infty} \frac{1}{\sqrt{2 \pi}} e^{-z^2/2}\, dz$
is the upper tail probability of the standard Gaussian distribution
and the labelled steps are obtained as follows:
(i) expanding $Y_k$ according to \eqref{eq:se_Y_def};
(ii) using the definition of $X$ in \eqref{eq:se_X_def} (note that the expectation at this step is over $U_2^R,\ldots, U_\M^R, U_1^I,\ldots, U_\M^I \stackrel{\iid}{\sim} \stdnorm$);
(iii) using $\mu = \nu\ln(\K\M)$ and noting that $\cos_{b=\K} = 1$, $\sin_{b=\K}=0$;
and (iv) using \eqref{eq:max_gauss_bound} with $\lambda$ being a universal positive constant.

We further bound $\mathcal{E}(\tau)$ by noting that the term inside the expectation term on the RHS of  \eqref{eq:se_E_ub_0} is bounded by 1. 
The summations over $k$ and $a$ below are from $(-\frac{K}{4}+1)$ to $\frac{K}{4}$.
\begin{align}
&\mathcal{E}(\tau)
 \leq  Q(\tilde{u}) 
+ e^{-\lambda (\sqrt{\ln \M})^{-1} \M^{\alpha\tilde{\delta}(2-\alpha\tilde{\delta})}} \nonumber\\
&\hspace{5.4em} + \prob(U_1^I < - \tilde{u}) \cdot 1 + \prob(U_1^I > \tilde{u})  \cdot 1 \ + \nonumber\\
&\hspace{1em} \int_{-\tilde{u}}^{\tilde{u}} \frac{\sum_{k} 2 \cos_k \, \sinh ( (\mu + \sqrt{\mu}\tilde{u})\cos_k + \sqrt{\mu}u \sin_k)}{{\M}^{\sqrt{2\nu}(1-\alpha\tilde{\delta})} \! + \! \sum_{a} \! 2\cosh ( (\mu \!+ \! \sqrt{\mu}\tilde{u})\cos_a \! + \sqrt{\mu}u \sin_a)} \nonumber\\
&\hspace{20.9em} \times \phi(u) du\nonumber\\
& \stackrel{\text{(i)}}{\leq} 3 Q(\tilde{u}) 
+ e^{-\lambda (\sqrt{\ln \M})^{-1} \M^{\alpha\tilde{\delta}(2-\alpha\tilde{\delta})}} \nonumber\\
&\ \ + \int_{-\tilde{u}}^{\tilde{u}} \frac{\sum_{k} 2  \sinh ( (\mu + \sqrt{\mu}\tilde{u})\cos_k + \sqrt{\mu} |u| |\sin_k|)}{{\M}^{\sqrt{2\nu}(1-\alpha\tilde{\delta})} + 2 \cosh (\mu + \sqrt{\mu}\tilde{u})}  \phi(u)\,du\nonumber\\
& \leq 3 Q(\tilde{u}) 
+ e^{-\lambda (\sqrt{\ln \M})^{-1} \M^{\alpha\tilde{\delta}(2-\alpha\tilde{\delta})}} \nonumber\\
&\quad  + (1 - 2 Q(\tilde{u})) \left[\frac{\sum_{k}  \exp ( (\mu + \sqrt{\mu}\tilde{u})\cos_k + \sqrt{\mu} \tilde{u} |\sin_k|)}{{\M}^{\sqrt{2\nu}(1-\alpha\tilde{\delta})} + \exp (\mu + \sqrt{\mu}\tilde{u})} \right]\nonumber\\
& \stackrel{\text{(ii)}}{\leq}  3 Q(\tilde{u}) 
	+ e^{-\lambda (\sqrt{\ln \M})^{-1} \M^{\alpha\tilde{\delta}(2-\alpha\tilde{\delta})}}\nonumber\\
	&\quad+ \frac{1 + \Delta \cdot \exp(-(\mu + \sqrt{\mu}\tilde{u}))}{1 + {\M}^{\sqrt{2\nu}(1-\alpha\tilde{\delta})} \cdot \exp(-(\mu + \sqrt{\mu}\tilde{u}))}, \label{eq:se_E_ub_1}
\end{align}
where $\phi(\cdot)$ denotes the standard Gaussian density
and the labelled steps are obtained as follows:
(i) noting that $\cos_k \leq 1$, $\cosh(\cdot)>0$, $\cos_{a=0} = 1$, $\sin_{a=0}=0$, and that $\sinh(\cdot)$ is an increasing function; and
(ii) substituting
\be
\Delta \coloneqq \sum_{k\in\{-\frac{\K}{4}+1,\ldots, \frac{\K}{4}\}\backslash 0}  \exp ( (\mu + \sqrt{\mu}\tilde{u})\cos_k + \sqrt{\mu} \tilde{u} |\sin_k|).
\ee

Since the RHS of \eqref{eq:se_E_ub_1} is increasing in $\Delta$, 
we obtain an upper bound on $\Delta$ to further bound $\mathcal{E}(\tau)$.
\begin{align}
\Delta 
&\leq \left(\frac{\K}{2} - 1\right) \max_{k\in\{-\frac{\K}{4}+1,\ldots, \frac{\K}{4}\}\backslash 0}  e^{ (\mu + \sqrt{\mu}\tilde{u})\cos_k + \sqrt{\mu} \tilde{u} |\sin_k|} \nonumber\\
&\stackrel{\text{(i)}}{=} \left(\frac{\K}{2} - 1\right) \max_{k\in\{1,\ldots, \frac{\K}{4}\}}  e^{ (\mu + \sqrt{\mu}\tilde{u})\cos_k + \sqrt{\mu} \tilde{u} \sin_k}, \label{eq:se_E_Delta_ub}
\end{align}
where step (i) is obtained by noting that taking the maximum over $k \in \{-K/4+1,\ldots,K/4\} \backslash 0$ is the same as taking the maximum over $k \in \{1,\ldots,K/4\}$ since $\cos_k = \cos_{-k}$ and $|\sin_k| = |\sin_{-k}|$.

We now show that for a certain choice of $\tilde{u}$, 
the maximum in \eqref{eq:se_E_Delta_ub} is achieved with $k=1$.
For any $\tilde{\alpha} \in (0,1)$, let 
\be\label{eq:tilde_u_1}
\tilde{u} = \frac{\tilde{\kappa}\tilde{\alpha}\tilde{\delta}}{2 (1+\cot(2\pi/K))} \sqrt{\frac{\ln(\K\M)}{\nu}},
\ee
where $\tilde{\kappa}>0$ is the positive constant defined in \eqref{eq:se_nu_lb_ub} that lower bounds $\nu$.
In order to show that $k=1$ achieves the maximum in \eqref{eq:se_E_Delta_ub} with this choice of $\tilde{u}$, we show that the derivative of the $\exp(\ldots)$ term with respect to $k$ is negative for all $k\in\{1,\ldots, \K/4\}$. 
Since $\exp(\cdot)$ is an increasing function, we only need to show this negative derivative result for its argument, which we denote by
\[
f_1(k) 
= (\mu + \sqrt{\mu}\tilde{u})\cos_k + \sqrt{\mu} \tilde{u} \sin_k.
\]
Recalling that $\cos_k = \cos(2\pi k /\K)$ and $\sin_k = \sin(2\pi k/\K)$, its derivative with respect to $k$ is
\begin{align}
f_1'(k) 
&= \frac{2\pi}{\K} \left[\sqrt{\mu} \tilde{u} \cos_k - (\mu + \sqrt{\mu}\tilde{u})\sin_k \right]\nonumber\\
&= \frac{2\pi}{\K} \mu \sin_k \left[\frac{\tilde{u}}{\sqrt{\mu}} \cot\left(\frac{2\pi k }{\K}\right) - \left(1 + \frac{\tilde{u}}{\sqrt{\mu}}\right)\right]\nonumber\\
& = \frac{2\pi}{\K} \mu \sin_k \left[\frac{\tilde{u}}{\sqrt{\mu}} \left( \cot\left(\frac{2\pi k}{\K}\right) -1\right) - 1 \right]\nonumber\\
&= \frac{2\pi}{\K} \mu \sin_k \left[\frac{\tilde{\kappa}\tilde{\alpha}\tilde{\delta}}{2\nu} \
\cdot \frac{\cot\left(\frac{2\pi k}{\K}\right) -1}{\cot\left(\frac{2\pi}{\K}\right) +1} - 1 \right]
< 0, \nonumber
\end{align}
where we use \eqref{eq:tilde_u_1}, 
$\mu=\nu\ln{\K\M}$, 
recalling that $\K\geq4$,
and noting that $\sin_k > 0$ for $k\in\{1,\ldots, \K/4\}$,
and also that $\cot(x)$ is decreasing for $x\in(0,\frac{\pi}{2})$.

Therefore, using the fact that $k=1$ achieves the maximum in \eqref{eq:se_E_Delta_ub} 
when $\tilde{u}$ is chosen according to \eqref{eq:tilde_u_1}, 
we have an upper bound on $\Delta$ 
which we use in \eqref{eq:se_E_ub_1} to further bound $\mathcal{E}(\tau)$.
In the following we use $\cot_1$ to denote $\cot(2\pi/\K)$.
\begin{align}
&\mathcal{E}(\tau)
\leq 3 Q(\tilde{u}) 
	+ e^{-\lambda (\sqrt{\ln \M})^{-1} \M^{\alpha\tilde{\delta}(2-\alpha\tilde{\delta})}}\nonumber\\
	&\quad+ \frac{1 + \left(\frac{\K}{2} - 1\right)  e^{ (\mu + \sqrt{\mu}\tilde{u})\cos_1 + \sqrt{\mu} \tilde{u} \sin_1} \cdot e^{-(\mu + \sqrt{\mu}\tilde{u})}}{1 + {\M}^{\sqrt{2\nu}(1-\alpha\tilde{\delta})} \cdot e^{-(\mu + \sqrt{\mu}\tilde{u})}}\nonumber\\
&\stackrel{\text{(i)}}{=} 3 Q(\tilde{u}) 
	+ e^{-\lambda (\sqrt{\ln \M})^{-1} \M^{\alpha\tilde{\delta}(2-\alpha\tilde{\delta})}}\nonumber\\
	&\quad+ \frac{1 + \left(\frac{\K}{2} - 1\right)  (\K\M)^{  -  (1-\cos_1) \big(\nu - \frac{\tilde{\kappa}\tilde{\alpha}\tilde{\delta}}{2} \cdot \frac{\frac{\sin_1}{1-\cos_1}-1}{1+\cot_1}\big)}}{1 + \K^{-\left(\nu + \frac{\tilde{\kappa}\tilde{\alpha}\tilde{\delta}}{2 (1+\cot_1)} \right)} \cdot \M^{\sqrt{2\nu} \big(1- \big(\alpha + \frac{\tilde{\kappa}\tilde{\alpha}}{2\sqrt{2\nu}(1+\cot_1)}\big) \tilde{\delta}\big) - \nu} }\nonumber\\
&\stackrel{\text{(ii)}}{\leq} 3 Q(\tilde{u}) 
	+ e^{-\lambda (\sqrt{\ln \M})^{-1} \M^{\alpha\tilde{\delta}(2-\alpha\tilde{\delta})}}\nonumber\\
	&\quad+ \frac{1 + \left(\frac{\K}{2} - 1\right)  (\K\M)^{  -  (1-\cos_1) (\nu - \tilde{\kappa}\tilde{\alpha}\tilde{\delta} )}}{\K^{-\left(\nu + \frac{\tilde{\kappa}\tilde{\alpha}\tilde{\delta}}{2 (1+\cot_1)} \right)} \cdot \M^{\sqrt{2\nu} \big(1- \big(\alpha + \frac{\tilde{\kappa}\tilde{\alpha}}{2\sqrt{2\nu}(1+\cot_1)}\big) \tilde{\delta}\big) - \nu} }\nonumber\\
&\stackrel{\text{(iii)}}{\leq} 3 (\K\M)^{-\frac{(\tilde{\kappa}\tilde{\alpha}\tilde{\delta})^2}{8\nu(1+\cot_1)^2}}
	+ e^{-\lambda (\sqrt{\ln \M})^{-1} \M^{\alpha\tilde{\delta}(2-\alpha\tilde{\delta})}}\nonumber\\
	&\quad+ \frac{\K^{\big(\nu + \frac{\tilde{\kappa}\tilde{\alpha}\tilde{\delta}}{2 (1+\cot_1)} \big)} \cdot
	\big(1 + \K  (\K\M)^{- (1-\cos_1) (\nu - \tilde{\kappa}\tilde{\alpha}\tilde{\delta} )}\big)}{ \M^{\sqrt{2\nu} \big(1- \big(\alpha + \frac{\tilde{\alpha}}{2}\big) \tilde{\delta}\big) - \nu}},
\label{eq:se_E_ub_2}
\end{align}
where step (i) is obtained by substituting in \eqref{eq:tilde_u_1};
(ii) is obtained by noting that for $\K\geq 4$,
\begin{align}
&\frac{\frac{\sin_1}{1-\cos_1}-1}{1+\cot_1}
= \frac{\frac{1-\cos_1^2}{1-\cos_1}-\sin_1}{\cos_1+\sin_1}
= \frac{1 + \cos_1-\sin_1}{\cos_1+\sin_1}\nonumber\\
&\hspace{6em}= \frac{1}{\cos_1+\sin_1} + \tan\Big(\frac{\pi}{4} - \frac{2\pi}{\K}\Big)
\leq 2,
\end{align}
and (iii) by using the bound $Q(x) \leq e^{-x^2/2}$ for $x>0$,
and from 
\begin{align}
\frac{\tilde{\kappa}}{2\sqrt{2\nu}(1+\cot_1)} 
\leq \frac{\tilde{\kappa}}{2\nu} \leq \frac{1}{2},
\end{align}
where the inequalities are obtained using $\tilde{\kappa} < \nu < 2$ and $\cot_1 \geq 0$ for $\K\geq4$.

Finally, from the final steps of the proof of the unmodulated ($\K=1$) case in \cite[Appendix~A.A]{rush2020capacity},
we know that for $\alpha=1/64$, $\tilde{\alpha}=10/32$ and $\nu \in [\tilde{\kappa}, 2 -\tilde{\delta}]$, we have
\be\label{eq:sqrt_nu_lb}
\sqrt{2\nu} \Big(1- \Big(\alpha + \frac{\tilde{\alpha}}{2}\Big) \tilde{\delta}\Big) - \nu 
\geq \frac{1}{32}\tilde{\delta} + \frac{55}{512}\tilde{\delta}^2.
\ee
By using \eqref{eq:sqrt_nu_lb} in \eqref{eq:se_E_ub_2}
and noting that $\nu$ can be upper bounded by a positive constant (see \eqref{eq:se_nu_lb_ub}), we obtain the required result:
for sufficiently large $\M$, any $\tilde{\delta} \in (0,1)$, and $\nu < 2-\tilde{\delta}$, we have
\[
\mathcal{E}(\tau) \leq \M^{-\alpha_{K}\tilde{\delta}^2},
\]
where $\alpha_{K}$ is a positive constant depending only on $\K$ and the bounds of $\nu$ in \eqref{eq:se_nu_lb_ub}.

\section{Conclusion} \label{sec:conc}

We proposed a generalization of sparse superposition codes for the complex AWGN channel where information is encoded in both the locations and the values of the non-zero entries of the message vector. This generalization introduces more flexibility in the SPARC code design, allowing us to reduce decoding complexity without affecting error performance (at a given rate), or equivalently, to increase the rate without increasing decoding complexity. 

The values of the non-zero entries in the modulated SPARC were chosen from a $K$-PSK constellation to ensure that they do not counteract the effect of power allocation or spatial coupling. We proposed an AMP decoding algorithm, and analyzed its performance by obtaining analytical bounds on a key state evolution parameter which predicts the NMSE in each iteration (Proposition \ref{prop:se_psi_bound}). 
These bounds showed that in the large system limit, the NMSE for $K$-PSK modulated complex SPARCs does not depend on the value of $K$, which allowed us to reduce the asymptotic analysis of $K$-PSK modulated complex SPARCs to the unmodulated case ($K=1$). This equivalence was used to establish that $K$-PSK modulated SPARCs are capacity-achieving, with suitable power allocation or spatial coupling.

In the future, we would like to better understand the effect of the modulation parameter $\K$ on the error performance  when the code parameters $\L, \M, n$ are finite. For the unmodulated case, an exponential bound on the probability of excess section error rate was  obtained in \cite{rush2019theerror, rush2020capacity}; it would be of interest to obtain a similar bound for modulated SPARCs which highlights how the error probability depends on $\K$.
Another key question is how to design base matrices that give improved finite length error performance for PSK-modulated SPARCs. The base matrix designs used in this paper were the same as those used for unmodulated SPARCs. While these designs are capacity-achieving in the large system limit, we expect that finite length error performance can be significantly improved by using a combination of spatial coupling and power allocation tailored to modulated SPARCs. Another direction for research is to explore the use of modulated SPARCs in unsourced random access schemes \cite{fengler2019sparcs,fengler2020unsourced, amalladinne2020approximate}.


%

\appendices

\section{Proof of Lemma \ref{lem:ser_nmse}}\label{appendix:proof_lemma_se_nmse}

Recall that $\text{sec}(\ell)\coloneqq\{(\ell-1)\M+1,\ldots,\ell \M\}$ for $\ell \in [\L]$ and $\bbeta_{\text{sec}(\ell)}\in\mathbb{C}^{\M}$ denotes the $\ell$-th section of the message vector.
We will show that if $\bbeta_{\text{sec}(\ell)}$ is decoded in error after $T$ iterations of AMP decoding, then the squared error of that section is lower bounded by a positive number that is a function of the modulation parameter $\K$. More precisely, we show that 
\begin{equation}
\label{eq:SER_to_MSE}
	\widehat{\bbeta}^T_{\text{sec}(\ell)} \neq \bbeta_{\text{sec}(\ell)} 
	\ \Rightarrow \
	\|\bbeta^{T}_{\text{sec}(\ell)} - \bbeta_{\text{sec}(\ell)} \|^2 \, \geq \, h(\K),
\end{equation}
where 
\begin{equation}\label{eq:hK_def}
	h(\K) \, = \,
	\begin{cases}
	\frac{1}{4} \quad &\text{if} \ \K=1,2, 4,\\
	\sin^4(\frac{\pi}{\K}) \quad &\text{if} \ \K \geq 8.
	\end{cases}
\end{equation}
Then \eqref{eq:SER_to_MSE} implies
\begin{align}
\text{SER} 
&=   \frac{1}{\L}\sum_{\ell=1}^\L \mathbbm{1}\{\widehat{\bbeta}^T_{\text{sec}(\ell)} \neq \bbeta_{\text{sec}(\ell)} \} \nonumber\\
&\leq  \frac{1}{h(\K)} \frac{1}{\L} \sum_{\ell=1}^{\L} \|\bbeta^{T}_{{\text{sec}(\ell)}} - \bbeta_{{\text{sec}(\ell)}}\|^2  \nonumber\\
&=  \frac{1}{h(\K)} \cdot \frac{\|\bbeta^{T} - \bbeta \|^2}{\L}, \nonumber
 \end{align}
which is the required result after substituting in \eqref{eq:hK_def}. We now prove the statements given in \eqref{eq:SER_to_MSE}, \eqref{eq:hK_def}.

We denote the location index of the non-zero entry of $\bbeta_{\text{sec}(\ell)}$ as $\text{sent}(\ell)$ (let's say it is in column block $\sfc$). By symmetry of the PSK constellation, we can assume without loss of generality that the value of the non-zero entry is $\p_\K=+1$. Thats is, for $j\in \text{sec}(\ell)$,
\begin{equation}
	\beta_j \, = \,
	\begin{cases}
	1 \quad &\text{if} \ j=\text{sent}(\ell),\\
	0 \quad &\text{otherwise}.
	\end{cases}
\end{equation}
Thus,
\begin{align}
&\|\bbeta^{T}_{\text{sec}(\ell)} - \bbeta_{\text{sec}(\ell)} \|^2
\geq |\beta^{T}_{\text{sent}(\ell)} - \beta_{\text{sent}(\ell)} |^2 
= |\beta^{T}_{\text{sent}(\ell)} - 1 |^2 \nonumber\\
&\stackrel{\text{(i)}}{=} \bigg| \sum_{k=1}^\K \p_k \cdot \frac{e^{\Re(\overline{s^{T-1}_{\text{sent}(\ell)}} \p_k)/\tau^{T-1}_\sfc}}{\sum_{j'\in \text{sec}(\ell)} \sum_{k'=1}^{\K} e^{\Re(\overline{s^{T-1}_{j'}} \p_{k'})/\tau^{T-1}_\sfc}} - 1 \bigg|^2 \nonumber\\
&\stackrel{\text{(ii)}}{=} \bigg[ \sum_{k=1}^\K \cos\Big(\frac{2\pi k}{\K}\Big) \cdot p_k - 1 \bigg]^2 + \bigg[\sum_{k=1}^\K \sin\Big(\frac{2\pi k}{\K}\Big) \cdot p_k \bigg]^2 \nonumber\\
&\geq \bigg[ \sum_{k=1}^\K \cos\Big(\frac{2\pi k}{\K}\Big) \cdot p_k - 1 \bigg]^2, \label{eq:sq_err_beta_sec_ell_lb}
\end{align}
where (i) is obtained using the expression for $\beta^T_{\text{sent}(\ell)}$ derived from \eqref{eq:amp_decoder}, \eqref{eq:eta_function} and \eqref{eq:se_tilde_tau}, and (ii) from substituting in $\p_k = e^{\imag 2\pi k/\K}$ and defining 
\be
p_k \coloneqq \frac{e^{\Re(\overline{s^{T-1}_{\text{sent}(\ell)}} \p_k)/\tau^{T-1}_\sfc}}{\sum_{j'\in \text{sec}(\ell)} \sum_{k'=1}^{\K} e^{\Re(\overline{s^{T-1}_{j'}} \p_{k'})/\tau^{T-1}_\sfc}}, \qquad k\in[\K].
\ee

Towards proving \eqref{eq:SER_to_MSE}, assume that $\widehat{\bbeta}^T_{\text{sec}(\ell)} \neq \bbeta_{\text{sec}(\ell)}$. Then from \eqref{eq:amp_hard_decision} we know that
$\Re(\overline{s^{T-1}_{\text{sent}(\ell)}} \p_{\K}) \leq \Re(\overline{s^{T-1}_{j^*}} \p_{k^*})$ for some 
$(j^*, k^*) \neq ( \text{sent}(\ell), K) $.
This gives us the following inequality.
\begin{align}
p_{\K} 
&= \frac{e^{\Re(\overline{s^{T-1}_{\text{sent}(\ell)}} \p_{\K})/\tau^{T-1}_\sfc}}{\sum_{j'\in \text{sec}(\ell)} \sum_{k'=1}^{\K} e^{\Re(\overline{s^{T-1}_{j'}} \p_{k'})/\tau^{T-1}_\sfc}} \nonumber\\
& \leq  \frac{e^{\Re(\overline{s^{T-1}_{j^*}} \p_{k^*})/\tau^{T-1}_\sfc}}{\sum_{j'\in \text{sec}(\ell)} \sum_{k'=1}^{\K} e^{\Re(\overline{s^{T-1}_{j'}} \p_{k'})/\tau^{T-1}_\sfc}} \nonumber \\
&\leq 1 - \frac{e^{\Re(\overline{s^{T-1}_{\text{sent}(\ell)}} \p_{\K})/\tau^{T-1}_\sfc}}{\sum_{j'\in \text{sec}(\ell)} \sum_{k'=1}^{\K} e^{\Re(\overline{s^{T-1}_{j'}} \p_{k'})/\tau^{T-1}_\sfc}} \nonumber\\
&= 1-p_{\K}, 
\label{eq:pk_leq_1_pk}
\end{align}
where the second inequality is obtained by noting that
\be
\sum_{j\in\text{sec}(\ell)} \sum_{k=1}^\K \frac{e^{\Re(\overline{s^{T-1}_{j}} \p_{k})/\tau^{T-1}_\sfc}}{\sum_{j'\in \text{sec}(\ell)} \sum_{k'=1}^{\K} e^{\Re(\overline{s^{T-1}_{j'}} \p_{k'})/\tau^{T-1}_\sfc}} = 1. \nonumber
\ee
From \eqref{eq:pk_leq_1_pk} and the fact that $p_\K\geq 0$, we deduce $0\leq p_\K \leq \frac{1}{2}$.

Continuing from \eqref{eq:sq_err_beta_sec_ell_lb}, we obtain the required lower bounds on the squared error of section $\ell$ when it is decoded in error.
\begin{align}
	&\bigg[ \sum_{k=1}^\K \cos\Big(\frac{2\pi k}{\K}\Big) \cdot p_k - 1 \bigg]^2\nonumber\\
	&\quad=
	\begin{cases}
		(1\cdot p_1 - 1)^2 \quad &\text{if} \ \K=1,\\
		(-1\cdot p_1 + 1\cdot p_2 - 1)^2 \quad &\text{if} \ \K=2,\\
		(-1\cdot p_2 + 1\cdot p_4 - 1)^2 \quad &\text{if} \ \K=4,
	\end{cases} \nonumber\\
	&\quad\geq \frac{1}{4},
	\label{eq:K4_bound}
\end{align}
where the last inequality is obtained using $0\leq p_\K \leq \frac{1}{2}$ and $p_k \geq 0$ for $k\in[\K]$.

For the $\K\geq 8$ case, first notice that $|\sum_k \cos(2\pi k/\K) \cdot p_k| \leq 1$ since $p_k\geq0$ for $k\in[\K]$ and $\sum_k p_k \leq 1$. Moreover,
\begin{align}
&\sum_{k=1}^\K \cos\Big(\frac{2\pi k}{\K}\Big)\cdot p_k \nonumber\\
&\leq 
\cos\Big(\frac{2\pi \K}{\K}\Big) \cdot p_\K 
+ \bigg ( \max_{k\in\{1,\ldots,\K-1\}} \cos\Big(\frac{2\pi k}{\K}\Big) \bigg)\Big(\sum_{k=1}^{\K-1} p_k \Big) \nonumber \\
&\leq p_\K + \cos\Big(\frac{2\pi}{\K}\Big) \cdot (1-p_\K).
\end{align}
Using this together with $|\sum_k \cos(2\pi k/\K) \cdot p_k| \leq 1$, we obtain
\begin{align}
&\bigg[ \sum_{k=1}^\K \cos\Big(\frac{2\pi k}{\K}\Big) \cdot p_k - 1 \bigg]^2 \nonumber\\
& \geq \bigg[ (1-p_\K) \Big(\cos(2\pi/\K) -1 \Big) \bigg]^2  \nonumber\\
&\geq  \frac{1}{4} \Big(1- \cos(2\pi/\K) \Big)^2 = \sin^4\left({\pi}/{\K}\right).
\label{eq:K8_bound}
\end{align}
Here the second inequality is obtained using $0\leq p_\K\leq \frac{1}{2}$.  Using  \eqref{eq:K4_bound} and \eqref{eq:K8_bound} in \eqref{eq:sq_err_beta_sec_ell_lb} completes the proof of \eqref{eq:SER_to_MSE}, and hence the lemma.


\section{Proof of Lemma \ref{lem:I1_lb}}\label{appendix:proof_I1_lb}

From \eqref{eq:se_E_geq_I1_2Q}, we have the following expression for $I_1$ (the summations over $k$ and $a$ below are from $(-\frac{K}{4}+1)$ to $\frac{K}{4}$):
\begin{align}
I_1 &= 
\int_{\underline{u}}^{\infty} \int_{\underline{u}}^{\infty} \phi(u^R) \phi(u^I) \, \times \nonumber\\
& \hspace{2em}  \mathbb{E}_X \Bigg[\frac{\sum_{k} 2 \cos_k \, \sinh Y_k}{X + \sum_{a} 2\cosh Y_a} \, \bigg | \,  U_1^R \! = \! u^R, U_1^I \!=\! u^I \Bigg] du^R  du^I,
\label{eq:I1_app}
\end{align}
where $\cos_k$, $Y_k$ for $k\in\{-\K/4+1,\ldots, \K/4\}$ and $X$ are defined in \eqref{eq:se_new_not1}, \eqref{eq:se_Y_def} and \eqref{eq:se_X_def} respectively, $\K\geq 4$, and $\phi(\cdot)$ denotes the standard Gaussian density.
We aim to lower bound $I_1$ with
\be\label{eq:u_underline_appendix}
	\underline{u} 
	= \frac{-\alpha(\frac{\nu}{2}-1)}{\nu} \frac{\sqrt{\mu}}{1 + \cot(\frac{2\pi}{K})} 
\ee
for any $\alpha \in (0,1)$ and $\nu > 2$.
Recall from the arguments around equation \eqref{eq:u_underline} and \eqref{eq:se_E_postive_ineq2} that the integrand of $I_1$ is non-negative for $\nu>2$ with this choice of $\underline{u}$.

Using Jensen's inequality for the expectation in \eqref{eq:I1_app}, we obtain
\begin{align}
	&I_1 
	\geq \int_{\underline{u}}^{\infty} \! \int_{\underline{u}}^{\infty} \phi(u^R)  \phi(u^I)  \frac{\sum_{k} 2 \cos_k \, \sinh Y_k}{\mathbb{E}X + \sum_{a} 2\cosh Y_a} du^R  du^I \nonumber\\
	&\stackrel{\text{(i)}}{\geq} \int_{\underline{u}}^{\infty} \!\int_{\underline{u}}^{\infty} \phi(u^R)  \phi(u^I)  \frac{\sum_{k} 2 \cos_k \, \sinh Y_k}{(KM)^{1+\frac{\nu}{2}} + \sum_{a} 2\cosh Y_a} du^R  du^I \nonumber\\
	&\stackrel{\text{(ii)}}{\geq} \int_{\underline{u}}^{\infty} \!\int_{\underline{u}}^{\infty} \phi(u^R)  \phi(u^I) \frac{ 2 \sinh Y_0}{(KM)^{1+\frac{\nu}{2}} + \sum_{a} 2\cosh Y_a}  du^R  du^I.\label{eq:se_E_I1_lb}
\end{align}
Inequality (i) is obtained as follows using the definition of $X$ in \eqref{eq:se_X_def} and the moment generating function of a standard normal:
\begin{align*}
	\mathbb{E}X 
	&= \sum_{j=2}^{M}\sum_{b=1}^{K}\mathbb{E}\big[e^{\sqrt{\mu}\cos_b U_j^R}\big] \mathbb{E}\big[e^{\sqrt{\mu}\sin_b U_j^I}\big] \nonumber\\
	&= \sum_{j=2}^{M}\sum_{b=1}^{K} e^{\frac{\mu}{2} (\cos_k^2 + \sin_k^2)} \nonumber\\
	&= \sum_{j=2}^{M}\sum_{b=1}^{K} (KM)^{\frac{\nu}{2}}
	< (KM)^{1+\frac{\nu}{2}}, \label{eq:EX_ub}
\end{align*}
where we have used $\mu = \nu \ln(KM)$.  Inequality (ii) in \eqref{eq:se_E_I1_lb} is obtained by taking only the $k=0$ term in the numerator's summation, recalling that $\cos_k$ and  $\sinh Y_k$ for $k\in\{-K/4+1,\ldots, K/4\}$ are all non-negative when $U_1^R \geq \underline{u}$ and $U_1^I \geq \underline{u}$.

Next, we further lower bound $I_1$ using the fact that the term $\frac{2 \sinh Y_0}{(KM)^{1+\frac{\nu}{2}} + \sum_{a} 2\cosh Y_a}$ in \eqref{eq:se_E_I1_lb} is a strictly increasing function of $U_1^R$. (Recall from \eqref{eq:se_Y_def} that $Y_k$ is a function of $U_1^R$ for each $k$.) 
To see that the term is increasing in $U_1^R$, we write $ f := \frac{2 \sinh Y_0}{(KM)^{1+\frac{\nu}{2}} + \sum_{a} 2\cosh Y_a}  = \frac{f_\text{num}}{f_\text{den}}$, and show that $\frac{\partial f}{\partial U_1^R} > 0$. Since
\be
\frac{\partial f}{\partial U_1^R}
= \Big(\frac{\partial f_\text{num}}{\partial U_1^R} \cdot f_\text{den} - f_\text{num} \cdot \frac{\partial f_\text{den}}{\partial U_1^R} \Big) \Big/ \Big(f_\text{den}^2\Big),
\ee
we show below  that $\frac{\partial f_\text{num}}{\partial U_1^R} \cdot f_\text{dem} - f_\text{num} \cdot \frac{\partial f_\text{dem}}{\partial U_1^R} > 0$. Indeed,
\begin{align}
	&\frac{\partial f_\text{num}}{\partial U_1^R} \cdot f_\text{den} - f_\text{num} \cdot \frac{\partial f_\text{den}}{\partial U_1^R} \nonumber  \\
	&= (2\sqrt{\mu}  \cos_0 \cosh Y_0)
	\big((KM)^{1+\frac{\nu}{2}} + \sum_{a} 2\cosh Y_a\big) \nonumber\\
	&\hspace{9.5em} - (2 \sinh Y_0 )
	\big(\sum_{a} 2\sqrt{\mu}\cos_a   \sinh Y_a\big)\nonumber\\
	&= 2\sqrt{\mu} \cosh Y_0 \,
	\Big[(KM)^{1+\frac{\nu}{2}}\nonumber\\
	&\hspace{7.5em} + 2\sum_{a} \big(\cosh Y_a
	- \frac{\sinh Y_0}{\cosh Y_0} \cos_a  \sinh Y_a \big) \Big] \nonumber\\
	&= 2\sqrt{\mu} \cosh Y_0 \,
	\Big[(KM)^{1+\frac{\nu}{2}} \nonumber\\
	&\hspace{4.5em}+ 2 \sum_{a} \cosh Y_a \cdot
	\big(1 - \tanh(Y_0) \tanh(Y_a) \cos_a \big) \Big]\nonumber\\
	&>0. \label{eq:derivative}
\end{align}
In the above, the summations over $a$ go from $-K/4+1$ to $K/4$.
The inequality in  \eqref{eq:derivative} holds because
$0\leq \cos_a \leq 1$ for $a\in\{-K/4+1,\ldots, K/4\}$, and $\abs{\tanh(x)} <1$ for all $x \in \mathbb{R}$.

Using the fact that the integrand in \eqref{eq:se_E_I1_lb} is strictly increasing in $U_1^R$, we further bound $I_1$  from below using the minimum value of $U_1^R$ in the range, i.e., $U_1^R = \underline{u}$ where $\underline{u}$ is given by \eqref{eq:u_underline_appendix}.
Using the expressions for $Y_0$ and $Y_a$ from \eqref{eq:se_Y_def}, and the substitution
\be
z 
= \frac{(\mu + \sqrt{\mu} \, \underline{u})}{\ln (KM)}
= \nu - \frac{\alpha(\frac{\nu}{2}-1)}{1+\cot(\frac{2\pi}{K})}, \label{eq:se_I1_z_def}
\ee
the RHS of \eqref{eq:se_E_I1_lb} is lower bounded as
\begin{align}
	&I_1  
	\geq \int_{\underline{u}}^{\infty} \! \int_{\underline{u}}^{\infty} 
	 \phi(u^R) \, \phi(u^I) \, \times \nonumber\\
	&\hspace{1em} \frac{2 \sinh[ z \ln(KM)]  \quad du^R  du^I}{(KM)^{1+\frac{\nu}{2} \ } \! + \! \sum_{a} \! 2\cosh [z \ln(KM) \cos_a \! + \sqrt{\mu} u^I \sin_a]}
	  \nonumber\\
	&\stackrel{\text{(i)}}{\geq} Q(\underline{u}) \int_{\underline{u}}^{-\underline{u}}  \phi(u) \, \times\nonumber\\
	&\hspace{1em}\frac{2 \sinh[ z \ln(KM)] }
	{(KM)^{1+\frac{\nu}{2}} + \sum_{a} 2\cosh [z \ln(KM) \cos_a + \sqrt{\mu} u\sin_a]} du \nonumber\\
	&\stackrel{\text{(ii)}}{=} Q(\underline{u}) \int_{\underline{u}}^{-\underline{u}}  
	\frac{\phi(u) \, 2 \sinh[ z \ln(KM)] }
	{(KM)^{1+\frac{\nu}{2}} + 2\cosh [z \ln(KM)] + \Delta(u)} 
	du\nonumber\\
	&=  Q(\underline{u}) \int_{\underline{u}}^{-\underline{u}} \phi(u) \, \times \nonumber\\
	&\hspace{2.0em}\frac{1 - (KM)^{-2z}}
	{1 + (KM)^{-2z} + (KM)^{1+\frac{\nu}{2} - z} + (KM)^{-z} \Delta(u) } \, du \nonumber\\
	&\stackrel{\text{(iii)}}{\geq}  Q(\underline{u}) \int_{\underline{u}}^{-\underline{u}} \phi(u) \,
	[1 - (KM)^{-2z}] \, \times \nonumber\\
	&\hspace{2.25em}[1 - (KM)^{-2z} - (KM)^{1+\frac{\nu}{2} - z} - (KM)^{-z} \Delta(u)] \, du \nonumber\\
	&\geq  Q(\underline{u}) \int_{\underline{u}}^{-\underline{u}} \phi(u) \,  
	\big[1 - 2(KM)^{-2z} - (KM)^{1+\frac{\nu}{2} - z} \nonumber\\
	&\hspace{15.25em}- (KM)^{-z} \Delta(u)\big] \, du \nonumber\\
	&\stackrel{\text{(iv)}}{\geq}  Q(\underline{u}) \int_{\underline{u}}^{-\underline{u}} \phi(u) \,
	\big[1 - 2(KM)^{-2z} - (KM)^{1+\frac{\nu}{2} - z} \nonumber\\
	&\hspace{12.0em}- (\K - 2) (KM)^{-(z-z^{\star})}\big] \, du \nonumber\\
	&\stackrel{\text{(v)}}{=}  [1-Q(|\underline{u}|)] 
	\cdot [1-2Q(|\underline{u}|)] 
	\cdot \big[1 - 2(KM)^{-2z} \nonumber\\
	&\hspace{6.75em}- (KM)^{1+\frac{\nu}{2} - z} - (\K - 2) (KM)^{-(z-z^{\star})}\big] \nonumber\\
	&\geq  1 - 3Q(|\underline{u}|)
	- 2(KM)^{-2z} - (KM)^{1+\frac{\nu}{2} - z} \nonumber\\
	&\hspace{11em}- (\K - 2) (KM)^{-(z-z^{\star})}. \label{eq:se_E_I1_lb2}
\end{align}
The labelled steps are obtained as follows: 
(i) noting that the integrand is non-negative and independent of $u^R$, 
recalling that $U_1^R\sim\stdnorm$ and that $Q(\cdot)$ is the upper tail probability of the standard Gaussian distribution, 
changing the limit of the integral for $u^I$,
and dropping the superscript on $u^I$ for brevity;
(ii) introducing the substitution
\begin{equation}\label{eq:se_I1_lb_Delta}
	\Delta(u) = \sum_{a\in \{-\frac{K}{4}+1,\ldots,\frac{K}{4}\} \backslash 0} \hspace{-1em}\cosh[z \ln(KM)\cos_a + \sqrt{\mu} u \sin_a ];
\end{equation}
(iii) using $\frac{1}{1+x} \geq 1-x$ for $x\geq0$, noting that $\Delta(u) > 0$;
(iv) from the upper bound on $\Delta(u)$ for $|u| \leq -\underline{u}$ shown below in \eqref{eq:se_I1_lb_Delta_ub};
and (v) from recalling that $\phi(\cdot)$ is the standard Gaussian density and noting that $Q(\underline{u}) = 1 - Q(|\underline{u}|)$ since $\underline{u}<0$.

To complete the proof of the lemma, it remains to show the upper bound on  $\Delta(u)$ that is used in step (iv) of \eqref{eq:se_E_I1_lb2}.
 Using the definition of $\Delta(u)$ in  \eqref{eq:se_I1_lb_Delta}, for $|u| \leq -\underline{u}$,
where $\underline{u}$ is given in \eqref{eq:u_underline_appendix},
we have
\begin{align}
	&\Delta(u) 
	\stackrel{\text{(i)}}{\leq} \sum_{a\in \{-\frac{K}{4}+1,\ldots,\frac{K}{4}\}  \backslash 0} \hspace{-2em} 2\cosh[z \ln(KM)\cos_a + \sqrt{\mu} |u| |\sin_a| ] \nonumber\\
	&= \sum_{a\in \{-\frac{K}{4}+1,\ldots,\frac{K}{4}\} \backslash 0}  
	\hspace{-2em} 2\cosh\Big[z \ln(KM) \cos_a \nonumber\\
	&\hspace{10em}+ \frac{\alpha(\frac{\nu}{2}-1)}{1+\cot(2\pi/\K)} \ln(KM)  |\sin_a| \Big] \nonumber\\
	&\stackrel{\text{(ii)}}{\leq} \Big(\frac{K}{2}-1\Big) \max_{a\in \{1,\ldots,\frac{K}{4}\}} 2\cosh \Big[z \ln(KM) \cos_a \nonumber\\
	&\hspace{14em}+ (\nu-z) \ln(KM)  \sin_a \Big] \nonumber\\
	&\stackrel{\text{(iii)}}{=} \Big(\frac{K}{2}-1\Big) \cdot 2 \cosh [z^{\star} \ln(KM) ] \nonumber\\
	&\stackrel{\text{(iv)}}{\leq} (K- 2) (KM)^{z^{\star}}, \label{eq:se_I1_lb_Delta_ub}
\end{align}
where the labelled steps are obtained as follows: 
(i) for $u\geq \underline{u}$, we know $z \ln(KM)\cos_a + \sqrt{\mu} u \sin_a \geq 0$ and that $\cosh(x)$ is an increasing function for $x\geq0 $;
(ii) using \eqref{eq:se_I1_z_def} and noting that taking the maximum over $a\in \{-K/4+1,\ldots,K/4\} \backslash 0$ is the same as taking the maximum over $a\in \{1,\ldots,K/4\}$ since $\cos_a = \cos_{-a}$ and $|\sin_a| = |\sin_{-a}|$;
(iii) holds because the maximum is achieved with $a=1$ (shown below) and defining $z^\star$ as
\begin{equation}\label{eq:se_I1_z_star}
z^\star 
= z \cos\Big(\frac{2\pi}{\K}\Big) + (\nu-z) \sin\Big(\frac{2\pi}{\K}\Big),
\end{equation}
where $z$ is defined in \eqref{eq:se_I1_z_def};
and (iv) using $(KM)^{z^\star} +(KM)^{-z^\star} \leq 2 (KM)^{z^\star}$ since $z^\star\geq0$.

In order to show that $a=1$ achieves the maximum of
\begin{align}
&f_1(a) 
:= \cosh [z \ln(KM) \cos_a + (\nu-z) \ln(KM)  \sin_a \nonumber\\
&= \cosh \bigg[\nu \ln(KM) \bigg(\Big(1 - \frac{\alpha(\frac{1}{2}-\frac{1}{\nu})}{1+\cot(2\pi/\K)}\Big) \cos_a \nonumber\\
&\hspace{10em}+ \frac{\alpha(\frac{1}{2}-\frac{1}{\nu})}{1+\cot(2\pi/\K)}  \sin_a \bigg) \bigg] \label{eq:se_I1_z_star_proof}
\end{align}
for $a\in \{1,\ldots,K/4\}$, we show that $f_1'(a) < 0$ for all $a$ in this region.
Since $\cosh(x)$ is increasing function for $x\geq 0$ and the argument of the $\cosh$ term in \eqref{eq:se_I1_z_star_proof} is non-negative, we only need to show that $f_2'(a) < 0$ for all $a\in \{1,\ldots,K/4\}$, where
\be
f_2(a) := \bigg(1 - \frac{\alpha(\frac{1}{2}-\frac{1}{\nu})}{1+\cot(2\pi/\K)}\bigg) \cos_a + \frac{\alpha(\frac{1}{2}-\frac{1}{\nu})}{1+\cot(2\pi/\K)}  \sin_a.
\ee
For $\alpha\in(0,1)$ and $\nu>2$, we have
\begin{align}
f_2'(a)
&=\frac{2\pi}{\K} \Bigg[ \frac{\alpha(\frac{1}{2}-\frac{1}{\nu})}{1+\cot(\frac{2\pi}{\K})} \cos\Big(\frac{2\pi a}{\K}\Big) \nonumber\\
&\hspace{5.5em}- \bigg(1 - \frac{\alpha(\frac{1}{2}-\frac{1}{\nu})}{1+\cot(\frac{2\pi}{\K})}\bigg) \sin\Big(\frac{2\pi a}{\K}\Big) \Bigg] \nonumber\\
&=\frac{2\pi}{\K} \sin\Big(\frac{2\pi a}{\K}\Big) \bigg[ \alpha\Big(\frac{1}{2}-\frac{1}{\nu}\Big)  \frac{1 + \cot(\frac{2\pi a}{\K})}{1 + \cot(\frac{2\pi}{\K})} - 1 \bigg], \nonumber
\end{align}
which is negative for $a\in \{1,\ldots,K/4\}$ since $\sin(\frac{2\pi a}{\K})>0$ and $\frac{1 + \cot(\frac{2\pi a}{\K})}{1 + \cot(\frac{2\pi}{\K})}\leq 1$ for $\K\geq4$ and $a\in \{1,\ldots,K/4\}$.  This completes the proof of \eqref{eq:se_I1_lb_Delta_ub}, and hence the lemma.

\section*{Acknowledgment}
We thank A. Greig and A. Dupuis for their work on early versions of modulated SPARCs, and A. Guill\'en i F\`abregas for helpful discussions.
We thank the Associate Editor and the anonymous reviewers for helpful comments and suggestions that led to an improved paper.

\ifCLASSOPTIONcaptionsoff
  \newpage
\fi



%
\bibliographystyle{IEEEtran}
\bibliography{mod_sparcs}

\begin{thebibliography}{10}
\providecommand{\url}[1]{#1}
\csname url@samestyle\endcsname
\providecommand{\newblock}{\relax}
\providecommand{\bibinfo}[2]{#2}
\providecommand{\BIBentrySTDinterwordspacing}{\spaceskip=0pt\relax}
\providecommand{\BIBentryALTinterwordstretchfactor}{4}
\providecommand{\BIBentryALTinterwordspacing}{\spaceskip=\fontdimen2\font plus
\BIBentryALTinterwordstretchfactor\fontdimen3\font minus
  \fontdimen4\font\relax}
\providecommand{\BIBforeignlanguage}[2]{{%
\expandafter\ifx\csname l@#1\endcsname\relax
\typeout{** WARNING: IEEEtran.bst: No hyphenation pattern has been}%
\typeout{** loaded for the language `#1'. Using the pattern for}%
\typeout{** the default language instead.}%
\else
\language=\csname l@#1\endcsname
\fi
#2}}
\providecommand{\BIBdecl}{\relax}
\BIBdecl

\bibitem{joseph2012least}
A.~Joseph and A.~R. Barron, ``Least squares superposition codes of moderate
  dictionary size are reliable at rates up to capacity,'' \emph{IEEE Trans.
  Inf. Theory}, vol.~58, no.~5, pp. 2541--2557, May 2012.

\bibitem{joseph2014fast}
A.~{Joseph} and A.~R. {Barron}, ``Fast sparse superposition codes have near
  exponential error probability for {$R<{\cal{C}}$},'' \emph{IEEE Trans. Inf.
  Theory}, vol.~60, no.~2, pp. 919--942, Feb. 2014.

\bibitem{cho2013approximate}
S.~Cho and A.~R. Barron, ``Approximate iterative {Bayes} optimal estimates for
  high-rate sparse superposition codes,'' in \emph{Sixth Workshop on Inf.
  Theoretic Methods in Sci. and Eng.}, 2013, pp. 35--42.

\bibitem{rush2017capacity}
C.~Rush, A.~Greig, and R.~Venkataramanan, ``Capacity-achieving sparse
  superposition codes via approximate message passing decoding,'' \emph{IEEE
  Trans. Inf. Theory}, vol.~63, no.~3, pp. 1476--1500, Mar. 2017.

\bibitem{venkataramanan19monograph}
R.~Venkataramanan, S.~Tatikonda, and A.~Barron, ``Sparse regression codes,''
  \emph{Foundations and Trends in Communications and Information Theory},
  vol.~15, no. 1-2, pp. 1--195, 2019.

\bibitem{polyanskiy2017perspective}
Y.~Polyanskiy, ``A perspective on massive random-access,'' in \emph{Proc. IEEE
  Int. Symp. Inf. Theory}, 2017.

\bibitem{fengler2019sparcs}
A.~Fengler, P.~Jung, and G.~Caire, ``{SPARCs} and {AMP} for unsourced random
  access,'' in \emph{Proc. IEEE Int. Symp. Inf. Theory}, 2019.

\bibitem{fengler2020unsourced}
------, ``Unsourced multiuser sparse regression codes achieve the symmetric
  {MAC} capacity,'' in \emph{Proc. IEEE Int. Symp. Inf. Theory}, 2020, pp.
  3001--3006.

\bibitem{amalladinne2020approximate}
V.~K. Amalladinne, A.~K. Pradhan, C.~Rush, J.-F. Chamberland, and K.~R.
  Narayanan, ``On approximate message passing for unsourced access with coded
  compressed sensing,'' in \emph{Proc. IEEE Int. Symp. Inf. Theory}, 2020, pp.
  2995--3000.

\bibitem{chen2017capacity}
X.~{Chen}, T.~{Chen}, and D.~{Guo}, ``Capacity of {G}aussian many-access
  channels,'' \emph{IEEE Trans. Inf. Theory}, vol.~63, no.~6, pp. 3516--3539,
  June 2017.

\bibitem{zadik2019improved}
I.~{Zadik}, Y.~{Polyanskiy}, and C.~{Thrampoulidis}, ``Improved bounds on
  {Gaussian MAC} and sparse regression via {Gaussian} inequalities,'' in
  \emph{Proc. IEEE Int. Symp. Inf. Theory}, July 2019, pp. 430--434.

\bibitem{ravi2019capacity}
J.~Ravi and T.~Koch, ``Capacity per unit-energy of {G}aussian many-access
  channels,'' in \emph{Proc. IEEE Int. Symp. Inf. Theory}, 2019.

\bibitem{barbier2015approximate}
J.~Barbier, C.~Sch{\"{u}}lke, and F.~Krzakala, ``Approximate message-passing
  with spatially coupled structured operators, with applications to compressed
  sensing and sparse superposition codes,'' \emph{J. Stat. Mech.: Theory Exp.},
  vol. 2015, no.~5, p. P05013, 2015.

\bibitem{barbier2017approximate}
J.~Barbier and F.~Krzakala, ``Approximate message-passing decoder and capacity
  achieving sparse superposition codes,'' \emph{IEEE Trans. Inf. Theory},
  vol.~63, no.~8, pp. 4894--4927, Aug. 2017.

\bibitem{barbier2019universal}
J.~{Barbier}, M.~{Dia}, and N.~{Macris}, ``Universal sparse superposition codes
  with spatial coupling and {GAMP} decoding,'' \emph{IEEE Trans. Inf. Theory},
  vol.~65, no.~9, pp. 5618--5642, Sept. 2019.

\bibitem{rush2020capacity}
\BIBentryALTinterwordspacing
C.~Rush, K.~Hsieh, and R.~Venkataramanan, ``Capacity-achieving spatially
  coupled sparse superposition codes with {AMP} decoding,'' {\em to appear in
  IEEE Trans. Inf. Theory}. [Online]. Available:
  \url{https://arxiv.org/abs/2002.07844}
\BIBentrySTDinterwordspacing

\bibitem{rush2019theerror}
C.~{Rush} and R.~{Venkataramanan}, ``The error probability of sparse
  superposition codes with approximate message passing decoding,'' \emph{IEEE
  Trans. Inf. Theory}, vol.~65, no.~5, pp. 3278--3303, May 2019.

\bibitem{barbier2016proof}
J.~{Barbier}, M.~{Dia}, and N.~{Macris}, ``Proof of threshold saturation for
  spatially coupled sparse superposition codes,'' in \emph{Proc. IEEE Int.
  Symp. Inf. Theory}, July 2016, pp. 1173--1177.

\bibitem{dvb-s2}
``{Digital Video Broadcasting (DVB); Second generation framing structure,
  channel coding and modulation systems for Broadcasting, Interactive Services,
  News Gathering and other broadband satellite applications (DVB-S2); Part 1:
  DVB-S2 },'' {European Telecommunications Standards Institute (ETSI) EN 302
  307-1 V1.4.1}, July 2014.

\bibitem{greig2017thesis}
A.~Greig, ``Design techniques for efficient sparse regression codes,'' Ph.D.
  dissertation, Dept. Eng., Cambridge Univ., Cambridge, UK, 2018.

\bibitem{liang2016ISTC}
C.~{Liang}, J.~{Ma}, and L.~{Ping}, ``Towards {G}aussian capacity, universality
  and short block length,'' in \emph{Proc. 9th Int. Symp. Turbo Codes Iterative
  Inf. Process.}, 2016, pp. 412--416.

\bibitem{liang2020cc}
S.~{Liang}, C.~{Liang}, J.~{Ma}, and L.~{Ping}, ``Compressed coding,
  {AMP}-based decoding, and analog spatial coupling,'' \emph{IEEE Trans.
  Commun.}, vol.~68, no.~12, pp. 7362--7375, 2020.

\bibitem{hsieh2018spatially}
K.~{Hsieh}, C.~{Rush}, and R.~{Venkataramanan}, ``Spatially coupled sparse
  regression codes: Design and state evolution analysis,'' in \emph{Proc. IEEE
  Int. Symp. Inf. Theory}, June 2018, pp. 1016--1020.

\bibitem{luby2001improved}
M.~G. {Luby}, M.~{Mitzenmacher}, M.~A. {Shokrollahi}, and D.~A. {Spielman},
  ``Improved low-density parity-check codes using irregular graphs,''
  \emph{IEEE Trans. Inf. Theory}, vol.~47, no.~2, pp. 585--598, Feb. 2001.

\bibitem{thorpe03}
J.~{Thorpe}, ``Low-density parity-check ({LDPC}) codes constructed from
  protographs,'' Jet Propuls. Lab., Pasadena, CA, USA, Tech. Rep., Aug. 2003.

\bibitem{divsalar2009capacity}
D.~{Divsalar}, S.~{Dolinar}, C.~R. {Jones}, and K.~{Andrews},
  ``Capacity-approaching protograph codes,'' \emph{IEEE J. Sel. Areas Commun.},
  vol.~27, no.~6, pp. 876--888, Aug. 2009.

\bibitem{mitchell2015spatially}
D.~G.~M. Mitchell, M.~Lentmaier, and D.~J. Costello, ``Spatially coupled {LDPC}
  codes constructed from protographs,'' \emph{IEEE Trans. Inf. Theory},
  vol.~61, no.~9, pp. 4866--4889, Sept. 2015.

\bibitem{donoho2009message}
D.~L. Donoho, A.~Maleki, and A.~Montanari, ``Message-passing algorithms for
  compressed sensing,'' \emph{Proc. Natl. Acad. Sci. U.S.A.}, vol. 106, no.~45,
  pp. 18\,914--18\,919, 2009.

\bibitem{krzakala2012statistical}
F.~Krzakala, M.~M\'ezard, F.~Sausset, Y.~F. Sun, and L.~Zdeborov\'a,
  ``Statistical-physics-based reconstruction in compressed sensing,''
  \emph{Phys. Rev. X}, vol.~2, p. 021005, May 2012.

\bibitem{bayati2011thedynamics}
M.~Bayati and A.~Montanari, ``The dynamics of message passing on dense graphs,
  with applications to compressed sensing,'' \emph{IEEE Trans. Inf. Theory},
  vol.~57, no.~2, pp. 764--785, Feb. 2011.

\bibitem{rangan2010generalized}
S.~{Rangan}, ``Generalized approximate message passing for estimation with
  random linear mixing,'' in \emph{Proc. IEEE Int. Symp. Inf. Theory}, July
  2011, pp. 2168--2172.

\bibitem{FletcherRanganIMA18}
A.~K. Fletcher and S.~Rangan, ``{Iterative reconstruction of rank-one matrices
  in noise},'' \emph{Information and Inference: A Journal of the IMA}, vol.~7,
  no.~3, pp. 531--562, 01 2018.

\bibitem{deshpande2014information}
Y.~Deshpande and A.~Montanari, ``Information-theoretically optimal sparse
  {PCA},'' in \emph{Proc. IEEE Int. Symp. Inf. Theory}, 2014, pp. 2197--2201.

\bibitem{maleki2013asymptotic}
A.~{Maleki}, L.~{Anitori}, Z.~{Yang}, and R.~G. {Baraniuk}, ``Asymptotic
  analysis of complex {LASSO} via complex approximate message passing
  ({CAMP}),'' \emph{IEEE Trans. Inf. Theory}, vol.~59, no.~7, pp. 4290--4308,
  July 2013.

\bibitem{anitori2013design}
L.~{Anitori}, A.~{Maleki}, M.~{Otten}, R.~G. {Baraniuk}, and P.~{Hoogeboom},
  ``Design and analysis of compressed sensing radar detectors,'' \emph{IEEE
  Trans. Signal Process.}, vol.~61, no.~4, pp. 813--827, Feb. 2013.

\bibitem{jeon2015optimality}
C.~Jeon, R.~Ghods, A.~Maleki, and C.~Studer, ``Optimality of large {MIMO}
  detection via approximate message passing,'' in \emph{Proc. IEEE Int. Symp.
  Inf. Theory}, 2015, pp. 1227--1231.

\bibitem{donoho2013information}
D.~L. Donoho, A.~Javanmard, and A.~Montanari, ``Information-theoretically
  optimal compressed sensing via spatial coupling and approximate message
  passing,'' \emph{IEEE Trans. Inf. Theory}, vol.~59, no.~11, pp. 7434--7464,
  Nov. 2013.

\bibitem{polyanskiy2010channel}
Y.~{Polyanskiy}, H.~V. {Poor}, and S.~{Verd{\'u}}, ``Channel coding rate in the
  finite blocklength regime,'' \emph{IEEE Trans. Inf. Theory}, vol.~56, no.~5,
  pp. 2307--2359, May 2010.

\bibitem{cassagne2017fast}
A.~Cassagne, O.~Hartmann, M.~L\'eonardon, T.~Tonnellier, G.~Delbergue,
  C.~Leroux, R.~Tajan, B.~{Le Gal}, C.~J\'ego, O.~Aumage, and D.~Barthou,
  ``Fast simulation and prototyping with {AFF3CT},'' in \emph{Proc. IEEE Int.
  Workshop on Signal Processing Systems}, Oct. 2017.

\bibitem{KuanPyScripts}
K.~Hsieh, ``Python implementation of sparse regression codes,''
  \url{https://github.com/kuanhsieh/sparc\_public}, 2020.

\end{thebibliography}

\end{document}